\documentstyle[12pt]{article}

\advance\textheight by \topskip
\begin{document}
\tolerance=100000
\def\Dir{\kern -6.4pt\Big{/}}
\def\DDir{\kern -7.6pt\Big{/}}
\def\DGir{\kern -6.0pt\Big{/}}
\def\Ord{\buildrel{\scriptscriptstyle <}\over{\scriptscriptstyle\sim}}
\def\simlt{\rlap{\lower 3.5 pt \hbox{$\mathchar \sim$}} \raise 1pt \hbox {$<$}}
\def\OOrd{\buildrel{\scriptscriptstyle >}\over{\scriptscriptstyle\sim}}
\def\simgt{\rlap{\lower 3.5 pt \hbox{$\mathchar \sim$}} \raise 1pt \hbox {$>$}}
\def\pl #1 #2 #3 {{\it Phys.~Lett.} {\bf#1} (#2) #3}
\def\np #1 #2 #3 {{\it Nucl.~Phys.} {\bf#1} (#2) #3}
\def\zp #1 #2 #3 {{\it Z.~Phys.} {\bf#1} (#2) #3}
\def\pr #1 #2 #3 {{\it Phys.~Rev.} {\bf#1} (#2) #3}
\def\prep #1 #2 #3 {{\it Phys.~Rep.} {\bf#1} (#2) #3}
\def\prl #1 #2 #3 {{\it Phys.~Rev.~Lett.} {\bf#1} (#2) #3}
\def\mpl #1 #2 #3 {{\it Mod.~Phys.~Lett.} {\bf#1} (#2) #3}
\def\rmp #1 #2 #3 {{\it Rev. Mod. Phys.} {\bf#1} (#2) #3}
\def\xx #1 #2 #3 {{\bf#1}, (#2) #3}
\def\preprint{{\it preprint}}
\def\sm{\ifmmode{{\cal {SM}}}\else{${\cal {SM}}$}\fi}
\def\mssm{\ifmmode{{\cal {MSSM}}}\else{${\cal {MSSM}}$}\fi}
\def\MH{\ifmmode{{M_{H^0}}}\else{${M_{H^0}}$}\fi}
\def\Mh{\ifmmode{{M_{h^0}}}\else{${M_{h^0}}$}\fi}
\def\MA{\ifmmode{{M_{A^0}}}\else{${M_{A^0}}$}\fi}
\def\MHpm{\ifmmode{{M_{H^\pm}}}\else{${M_{H^\pm}}$}\fi}
\def\tb{\ifmmode{\tan\beta}\else{$\tan\beta$}\fi}
\def\ctb{\ifmmode{\cot\beta}\else{$\cot\beta$}\fi}
\def\ta{\ifmmode{\tan\alpha}\else{$\tan\alpha$}\fi}
\def\cta{\ifmmode{\cot\alpha}\else{$\cot\alpha$}\fi}
\def\tba{\ifmmode{\tan\beta=1.5}\else{$\tan\beta=1.5$}\fi}
\def\tbb{\ifmmode{\tan\beta=30}\else{$\tan\beta=30$}\fi}
\def\cab{\ifmmode{c_{\alpha\beta}}\else{$c_{\alpha\beta}$}\fi}
\def\sab{\ifmmode{s_{\alpha\beta}}\else{$s_{\alpha\beta}$}\fi}
\def\cba{\ifmmode{c_{\beta\alpha}}\else{$c_{\beta\alpha}$}\fi}
\def\sba{\ifmmode{s_{\beta\alpha}}\else{$s_{\beta\alpha}$}\fi}
\def\ca{\ifmmode{c_{\alpha}}\else{$c_{\alpha}$}\fi}
\def\sa{\ifmmode{s_{\alpha}}\else{$s_{\alpha}$}\fi}
\def\cb{\ifmmode{c_{\beta}}\else{$c_{\beta}$}\fi}
\def\sb{\ifmmode{s_{\beta}}\else{$s_{\beta}$}\fi}
\def\be{\begin{equation}}
\def\ene{\end{equation}}
\def\ba{\begin{eqnarray}}
\def\ena{\end{eqnarray}}
\def\ar{\rightarrow}
\def\F{\ifmmode{\cal F}\else{$\cal F$}\fi}
\def\X{\ifmmode{\cal X}\else{$\cal X$}\fi}
\def\Y{\ifmmode{\cal Y}\else{$\cal Y$}\fi}
\def\Z{\ifmmode{\cal Z}\else{$\cal Z$}\fi}
\def\li{\ifmmode{p_i,\lambda}\else{$p_i,\lambda$}\fi}
\def\lj{\ifmmode{p_j,\lambda'}\else{$p_j,\lambda'$}\fi}
\def\l #1{\ifmmode{p_{#1},\lambda_{#1}}\else{$p_{#1},\lambda_{#1}$}\fi}
\def\m #1{\ifmmode{q_{#1},\lambda_{#1}}\else{$q_{#1},\lambda_{#1}$}\fi}
\def\r #1{\ifmmode{r_{#1},-}\else{$r_{#1},-$}\fi}
\thispagestyle{empty}
\setcounter{page}{0}

\begin{flushright}
{\large DFTT 62/94}\\
{\large DTP/94/108}\\
{\rm June 1995\hspace*{.5 truecm}}\\
\end{flushright}

\vspace*{\fill}

\begin{center}
{\Large \bf Intermediate mass Higgs bosons of the
Minimal Supersymmetric
Standard Model at the
proposed CERN LEP$\otimes$LHC
$ep$ collider\footnote{Work supported in part by Ministero
dell' Universit\`a e della Ricerca Scientifica (S.M.), by University of
Durham Studentship and by World Lab. Fellowship ICSC (G.A.L).}}\\[2.cm]
{\large Ghadir Abu Leil$^a$ and Stefano
Moretti$^{a,b,}$\footnote{Address
after September 1995: Cavendish Laboratory,
University of Cambridge,
Madingley Road,
Cambridge, CB3 0HE, U.K.}}\\[2cm]
{\it a) Department of Physics, University of Durham,}\\
{\it    South Road, Durham DH1 3LE, U.K.}\\[1cm]
{\it b) Dipartimento di Fisica Teorica, Universit\`a di Torino,}\\
{\it    and INFN, Sezione di Torino,}\\
{\it    V. Pietro Giuria 1, 10125 Torino, Italy.}\\[2cm]
\end{center}

\vspace*{\fill}

\begin{abstract}
{\normalsize
\noindent
The production of the \mssm\ Higgs bosons $H^0,h^0,A^0$
and $H^\pm$, in the intermediate mass range of the
$A^0$, at two different values of $\tan\beta$,
is studied at the possible CERN  LEP$\otimes$LHC $ep$ collider,
through $\gamma p$ interactions, by
photons generated via Compton back--scattering of laser
light. Signatures in which $H^0,h^0,A^0\ar b\bar b$
and $H^\pm\ar \tau\nu_\tau$ are considered.
Flavour identification on $b$--jets is assumed.
Backgrounds to Higgs signals are computed.
Explicit formulae for the helicity amplitudes
of the Higgs processes are given.}
\end{abstract}

\vspace*{\fill}

\newpage
\subsection*{Introduction}

We know that, despite its innumerable experimental successes,
the Standard Model (\sm) \cite{SM} cannot be a fundamental
theory valid up to an arbitrary\footnote{A scale
which has to be less than the Plank scale $M_{Planck}\sim
10^{19}$, where a description which includes quantum gravity is needed.}
energy scale $\Lambda$.
It should rather be regarded as an effective low energy model,
which has to be replaced at an energy close to the Fermi scale
$G_F^{-1/2}\approx300$ GeV by some more fundamental theory.
This can be seen from the fact that, for $\Lambda>>G_F^{-1/2}$, the one--loop
radiative corrections to the \sm\ Higgs mass $M_\phi$ are quadratically
divergent (naturalness or hierarchy problem) \cite{naturalness}.\par
Supersymmetric (${\cal {SUSY}}$) models can solve this.
The most intriguing among them
is probably the Minimal Supersymmetric Standard Model (\mssm)
\cite{MSSM}. It incorporates two complex Higgs doublets of fundamental scalar
fields $(H^0_1,H^-_1)$ and $(H^+_2,H^0_2)$, which, after a spontaneous
symmetry breaking, originate  five Higgs
bosons: the ${\cal {CP}}$--even neutral $H^0$ and $h^0$,
the ${\cal {CP}}$--odd neutral
 $A^0$ and the charged  $H^\pm$'s\footnote{The
three neutral Higgs states of the \mssm\ will be collectively indicated by the
symbol $\Phi^0$.}.
The attractions of the \mssm\ are numerous. It is a predictive model:
all masses and couplings in the Higgs sector can be expressed at tree--level
in terms of only two real parameters, the ratio of the vacuum expectation
values $v_1$ and $v_2$ of the two doublets (i.e., $\tan\beta=\frac{v_2}{v_1}$)
and the mass of one of the bosons (e.g., $M_{A^0}$),
and, at
the same time, the radiative corrections can be kept well under control.
It breaks the gauge symmetry close to the electroweak scale $G_F^{-1/2}$
and, if combined with Grand Unification Theories (${\cal {GUT}}$), it predicts
a value for the Weinberg angle $\theta_W$ in good agreement with the measured
one and a value for the Grand Unification Mass $M_{\cal {GUT}}$ which can
explain the not--observed proton decay \cite{protondecay}. It supplies a
natural candidate for the dark matter in terms of the Lightest Supersymmetric
Particle ({${\cal {LSP}}$), which is stable, neutral and weakly
interacting
(i.e., neutralino).
Finally, so far, it survived stringent experimental constrains: e.g.,
the most part of its parameter space has not yet been excluded by
LEP data \cite{limMSSM}.\par
While upper limits on the \mssm\ Higgs boson masses can be deduced by
arguments connected with the request of unitarity of the theory, which
implies that at least one neutral \mssm\ Higgs must have mass below
$\sim 1$ TeV \cite{guide,unitarityMSSM1,unitarityMSSM2},
lower limits can be extracted at present
colliders. From LEP I ($\sqrt s_{ee}=M_{Z^0}$) experiments,
as a result of searches for $e^+e^-\rightarrow Z^{0*}h^0$ and
$e^+e^-\rightarrow h^0A^0$ events, one obtains \cite{limMSSM}
\begin{equation}
M_{h^0}\OOrd 44.5~{\rm {GeV}}\quad\quad{\rm {and}}\quad\quad
M_{A^0}\OOrd
45~{\rm {GeV}}.
\end{equation}
Extensive studies have been carried out on the detectability of
\mssm\ Higgs particles by the next generation of high energy
machines, both at a $pp$ hadron collider \cite{guide,LHC,SSC} and at an
$e^+e^-$ Next Linear Collider (NLC) \cite{guide,LepII,NLC,ee500,LC92,JLC}.\par
The region $M_{A^0} < 80-90$ GeV will be studied at LEP II
($\sqrt s_{ee}=170-190$ GeV), by the Higgs decay channel $b\bar b$
\cite{LepII}, via one or both the processes
$e^+e^-\rightarrow Z^{0*}\rightarrow Z^{0}h^0$ (bremsstrahlung) and
$e^+e^-\rightarrow Z^{0*} \rightarrow h^0A^0$ (neutral pair production)
\cite{brenppMSSM}.\par
Higgses with larger masses will be searched for
at $pp$ colliders like the LHC\footnote{Since the most part of
the results on Higgs searches at the SSC can be transposed to the LHC, in the
following we will arbitrarily confuse
the two bibliographies on this argument,
even though we know that the SSC project
has been definitely set aside.}, with $\sqrt s_{pp}=10, 14$ TeV and
${\cal L}\approx 10-100$ fb$^{-1}$,
or at $e^+e^-$ NLCs, with $\sqrt s_{ee}=300-2000$ GeV and
${\cal L}\approx 10-20$ fb$^{-1}$.\par
At the LHC, because of the huge QCD background,
the mass range 80 GeV $\Ord M_{\Phi^0}
\Ord 130$ GeV is the most difficult to study since
in this case a neutral Higgs boson mainly
decays to $b\bar b$ pairs,
for a large choice of the \mssm\ parameters.
However, studies have shown that the discovery
of a neutral Higgs boson via the $\Phi^0\rightarrow\gamma\gamma$
mode at hadron colliders can be exploited for the discovery of $H^0$
for 80 GeV $\Ord M_{A^0} \Ord$ 100 GeV
and of $h^0$ for $M_{A^0}\OOrd$ 170 GeV, at all $\tan\beta$.
For heavier masses, the ``gold--plated'' decay channel ($\Phi^0\ar4\ell$)
is useful for the $H^0$
if $\tan\beta\Ord 7$ and 100 GeV $\Ord M_{A^0} \Ord$ 300 GeV, but not
for the $h^0$
because of its too light mass\footnote{For the $stop$ mass $m_{\tilde t}=1$ TeV
and all --ino masses greater than 200 GeV.}.
Recently, it has been also shown \cite{btagg} that with the $b$--tagging
capabilities \cite{SDC} of the LHC experiments\footnote{If
the higher luminosity and a large number of tracks per event can successfully
be dealt with.}, it might be possible to rely,
over a substantial portion of the parameter space, on the
$t\bar t\Phi^0$ production channel, with one $t$
decaying semileptonically and $\Phi^0\rightarrow b\bar b$,
for 80 GeV $\Ord M_{\Phi^0} \Ord$ 130 GeV,
for at least one of the ${\cal {MSSM}}$ Higgses $h^0$ or $H^0$,
removing the ``window of unobservability' for
100 GeV $\Ord M_{A^0} \Ord$ 170 GeV and  $\tan\beta\OOrd 2$,
which remained in previous analyses.
Moreover, it has been found \cite{noiA0} also that the reaction $bg\rightarrow
bZ^0\Phi^0$ is an excellent candidate for the discovery
of $A^0$ and at least one of the other two neutral Higgses
over the whole intermediate range of $M_{A^0}$ for
large values of $\tan\beta$, through the same decay channel $\Phi^0\rightarrow
b\bar b$.
With respect to charged Higgses, for low(high) values of $M_{H^\pm}$
the dominant production mechanism is $gg\rightarrow t\bar t\rightarrow
H^+H^-b\bar b$($bg\rightarrow tH^-$). Because of QCD backgrounds,
only the
low mass case gives a detectable signal over a non--negligible region
of $(M_{A^0},\tan\beta)$ \cite{kz}.\par
At NLC energies, other than via bremsstrahlung and neutral pair production
(this latter for $H^0A^0$ final states too \cite{brenppMSSM}),
\mssm\ Higgses can be produced also via the fusion processes
$e^+e^-\rightarrow
\bar\nu_e\nu_eW^{\pm*} W^{\mp*}(e^+e^-Z^{0*}Z^{0*})\rightarrow\bar\nu_e\nu_e
(e^+e^-)h^0/H^0$ \cite{fusionMSSM}
and the charged pair production $e^+e^-\rightarrow\gamma^*,Z^{0*}
\rightarrow H^+H^-$ \cite{cppMSSM}. The lightest ${\cal {CP}}$--even
Higgs $h^0$ can be detected
over the whole \mssm\ parameter space, independently of the top and
$squark$ masses.
Therefore, if the $h^0$ will not be found at the NLC, the \mssm\ is ruled out.
If the $H^0$ and $A^0$ boson
masses are less then $\approx 230$ GeV, there exists a very
large area in the parameter
space where all neutral Higgses can be contemporaneously
detected for $\sqrt s_{ee}=500$ GeV \cite{DHZ}.
A charged Higgs with $M_{H^\pm}<m_b+m_t$
mainly decays to
$\nu_\tau\tau^+(\bar\nu_\tau\tau^-)$
and $c\bar s(\bar c s)$ pairs (with the leptonic mode dominating
for $\tan\beta>1$). If kinematically allowed, a heavy $H^\pm$ decays via
the top mode $H^\pm\rightarrow t\bar b(\bar t b)$
(and in some part of the parameter space also
to $W^\pm h^0$). In both cases the signature is a cascade with a $\tau$ or
a $b$ in the final state: therefore, an extremely good
mass resolution is crucial in order
to reduce the backgrounds from top and boson pair production. For an
intermediate $H^\pm$, if $\tan\beta>1$,
a possible signature is an apparent breaking of the $\tau$ {\it vs}
$\mu$/$e$ universality.
At higher $e^+e^-$ energies, such as $\sqrt s_{ee}=$1--2 TeV, fusion
mechanisms become dominant over other production processes \cite{JLC,Bur}.\par
The conversion of
$e^+e^-$ NLCs into $\gamma\gamma$ and/or $e\gamma$ colliders, by
photons generated via Compton back--scattering of laser light,
provides new possibilities of detecting and studying Higgs bosons \cite{laser}.
For the \mssm, at a NLC with $\sqrt s_{ee}=500$ GeV,
$\gamma\gamma\rightarrow\Phi^0$ reactions are important
in searching for heavy $H^0$ and $A^0$ bosons: they
can be detected up to mass values of $\approx 0.8\sqrt s_{ee}$, for moderate
$\tan\beta$ and if a luminosity of 20 fb$^{-1}$, or more, can be
achieved \cite{gMSSM}.
For the $H^0$, the channels $H^0\rightarrow h^0h^0$, if $M_{H^0}
\Ord 2m_t$, and $H^0\rightarrow t\bar t$, for $M_{H^0}\OOrd 2m_t$, appear more
interesting than the decays $H^0\rightarrow b\bar b$ and
$H^0\rightarrow
Z^0Z^0$.
For the $A^0$, the feasible reactions are $\gamma\gamma\rightarrow
A^0\rightarrow
Z^0h^0/b\bar b$, if $M_{A^0}\Ord 2m_t$, and $\gamma\gamma\rightarrow
A^0\rightarrow t\bar t$,
if $M_{A^0}\OOrd 2m_t$. If $\tan\beta\Ord$ 20, only the $b\bar b$ channel is
useful for the $A^0$, with $M_{A^0}\Ord 250$ GeV\footnote{Since the
$h^0$
mass never becomes
large, the only important channel is $\gamma\gamma\rightarrow h^0
\rightarrow b\bar b$,
allowing its detection for $M_{h^0}\OOrd 60$ GeV ($M_{A^0}\OOrd 70$ GeV).}.
Recently, it has been shown that the intermediate mass $H^+H^-$ pair production
via $\gamma\gamma$ fusion is greater (e.g., at least by a factor
2 at
$\sqrt s_{ee}=500$ GeV)
than the corresponding $e^+e^-$ mode,
and charged Higgses can be detected using the three decay
modes $\nu_\tau\tau^+\bar\nu_\tau\tau^-$, $c\bar s\bar c s$ and
$c\bar s\bar\nu_\tau\tau^- +\nu_\tau\tau^+ \bar c s$
in a complementary way in order to cover all the intermediate mass region of
$H^\pm$ \cite{ggHH}.
The $e\gamma$ option at NLCs is quite interesting
in studying \mssm\ Higgs boson production via the processes
$e^-\gamma\rightarrow \nu_eW^-\Phi^0$,
$e^-\gamma\rightarrow \nu_eH^-\Phi^0$ and
$e^-\gamma\rightarrow e^-H^+H^-$,
in the intermediate mass range of $M_{A^0}$ and for a large choice
of $\tan\beta$'s \cite{ioPR}.\par
The option of $ep$ colliders in detecting and studying \mssm\ Higgs bosons
has been only marginally exploited, so far, with respect to the possibilities
of $pp$ and $e^+e^-$ accelerators.
The only presently operating $ep$ high energy machine is HERA \cite{epHERA},
which, however, has been primarily designed for providing accurate data on the
proton structure functions in the small--$x$ region, more than being devoted
to Higgs searches, which are almost impossible even for the more
favourable cases of $A^0$-- and $H^\pm$--production \cite{GGv}.
In fact, most of these searches rely on
very special conditions, which seem to be excluded by recent
limits on Higgs and top masses: e.g., very high $\tan\beta$ ($\approx40$)
in order to detect neutral Higgses $\Phi^0$ via $Z^0Z^0$-- and
$\gamma\gamma$--fusion processes \cite{BatesNg}, or very light
charged Higgses and/or top quark
for $H^\pm$--production via $\gamma\gamma$--
\cite{Korea} and $\gamma g$--fusion \cite{Munich}.
Furthermore, the $H^\pm$--production
mechanism via bremsstrahlung off
heavy quarks $\gamma q\ar q'H^\pm$ suffers from a strong
Cabibbo--Kobayashi--Maskawa or ${\cal O}({m_q\over M_{W^\pm}})$
suppression (where $q$ is the emitting initial light quark) \cite{HanLiu}.
Finally, the production of neutral \mssm\ Higgses through bremsstrahlung
off $b$--quarks, exploited in ref.~\cite{qqhHERA},
can hardly be useful,
since it depends not only on a good $b$-- and/or heavy lepton--tagging,
but also on the fact that only
large $\tan\beta$ ($\approx20$) and Higgs masses
$M_{\Phi^0}\Ord90$ GeV can give detectable signals\footnote{Region
that can be more easily covered by LEP II.}.\par
In the future, another $ep$ collider is contemplated to be operating,
the CERN LEP$\otimes$LHC machine, obtainable by combining an
electron/positron beam of LEP II and a proton beam of
the LHC \cite{LHC,epLEPLHC}. The detailed studies on the detectability
of an intermediate mass \sm\ Higgs boson $\phi$ at such a machine presented
in ref.~\cite{ZeppenfeldLHC} (using $W^\pm W^\mp$-- and
$Z^0Z^0$--fusion processes \cite{HanLiu,GGv,WWZZep,corrWWep},
with $\phi$ decaying
to $b\bar b$) can be  transposed to the case of ${\cal {CP}}$--even
neutral \mssm\ Higgs bosons, but increasing the requirements
on luminosity and/or on $b$--tagging identification, due to the
smaller $H^0$ and $h^0$ cross sections with respect to the \sm\ case.
Charged Higgs bosons  can be detected at
LEP$\otimes$LHC energies via the decay $t(\bar t)\ar H^+b(H^-\bar b)$,
if $M_{H^\pm}\Ord m_t-m_b$, while for
$M_{H^\pm}\OOrd m_t-m_b$, good sources of $H^\pm$ bosons are
the photo--production $\gamma b\ar H^- t$ (through bremsstrahlung
photons) and the $W^\pm$--mediated process $e^-b\ar\nu_eH^-b$, studied
in ref.~\cite{CruzSampayoLEPLHC}.\par
 Concerning photon--initiated processes,
only recently has the possibility of resorting
to back--scattered laser $\gamma$'s, also at the CERN $ep$ collider,
\cite{CheungLEPLHC} been suggested. This option has
been applied to the case of \sm\ Higgs production but, obviously,
it could turn out to be useful for \mssm\ Higgs bosons also.\par
It is the purpose of this paper to study at the LEP$\otimes$LHC
$ep$ collider the reactions
\be\label{proc1}
q\gamma\ar q'W^\pm\Phi^0,
\ene
\be\label{proc2}
q\gamma\ar q Z^0\Phi^0,
\ene
\be\label{proc3}
q\gamma\ar q'H^\pm\Phi^0,
\ene
\be\label{proc4}
q\gamma\ar q \Phi^0\Phi^{0{'}},
\ene
\be\label{proc5}
q\gamma\ar qH^+ H^-,
\ene
\be\label{proc6}
g\gamma\ar q\bar q'H^\pm,
\ene
\be\label{proc7}
g\gamma\ar q\bar q\Phi^0,
\ene
where $\Phi^{0(')}=H^0,h^0$ and $A^0$, in the intermediate mass range
of $A^0$, for all possible (anti)flavours of the (anti)quarks $q(q')$,
using laser back--scattered photons.
We discuss their relevance for the detection of the
\mssm\ Higgs bosons and the study of their parameters,
assuming $b$--tagging identification.\par
We did not study the processes
\be\label{noproc1}
q\gamma\ar q W^\pm H^\mp,
\ene
\be\label{noproc2}
q\gamma\ar q' Z^0H^\pm,
\ene
since here the \mssm\ Higgs bosons directly couple to the quark line
in each Feynman diagram at tree--level, so we expect that they are suppressed
through the Yukawa coupling
by the hadron structure functions, with  respect to  processes
(\ref{proc1})--(\ref{proc7}), where $\Phi^0$ and $H^\pm$ also couple
to the vector bosons $\gamma$, $Z^0$ and $W^\pm$ (see diagrams
(7), (8), (13), (14) of fig.~1 and diagrams (7), (8), and (15)--(17)
of fig.~2).\par
There are at least two important motivations for studying processes
(\ref{proc1})--(\ref{proc7}), and at the LEP$\otimes$LHC collider.
First, the CERN
$ep$ option could be operating before any NLC, so it would
constitute the first TeV energy environment partially free from the
enormous background arising from QCD processes (typical of the purely
hadron colliders),
which prevents the possibility of detailed
studies of the various parameters of an intermediate mass Higgs boson.
Second, even in the case that
LEP II and LHC can together cover
all the parameter space $(M_{A^0},\tan\beta)$,
nevertheless, processes (\ref{proc1})--(\ref{proc7})
offer the opportunity for studying a large variety
of \mssm\ interactions involving Higgs bosons:
in fact, all the vertices displayed in tabs.~A.I--A.IV
occur.
Moreover, the additional heavy particles
$t(\bar t)$, $W^\pm$, $Z^0$ and the second Higgs boson
can be used for tagging purposes, increasing the signal
versus background ratio.\par
The plan of the paper is as follows.
In Section II we give some details of the calculation
and the numerical values adopted for the various parameters.
Section III is devoted to the presentation of the results while
the Conclusions are in Section IV. Finally,
in the Appendix, we give the tree--level helicity amplitudes for
processes (\ref{proc1})--(\ref{proc7}).

\subsection*{Calculation}

In the unitary gauge  the Feynman diagrams which enter
in describing reactions (\ref{proc1})--(\ref{proc7})
at tree--level  are shown in figs.~1, 2 and 3. For the various
possible combinations of $(q,q',V,V^*,S^*,\Phi^0)$ in fig.~1,
$(q,q',V^*,S^*,S^{*'},\Phi,\Phi')$ in fig.~2, and $(q,q',S^*,\Phi)$ in fig.~3,
see details in the Appendix.
All quarks have been considered massive, so
diagrams with a direct coupling of $\Phi^0/H^\pm$
to fermion lines have been computed for each combination of
flavours.\par
The matrix elements have been evaluated by means of
the spinor techniques of refs.~\cite{ks,mana} and the
{\tt FORTRAN} codes have been compared with the corresponding ones
implemented by the method of ref. \cite{hz}.
The amplitudes have been tested for gauge invariance, and
it has been also verified that, with appropriate couplings,
hadronic distributions
and luminosity function of photons, our results
for the processes $q\gamma\rightarrow q'W^\pm\Phi^0$, $g\gamma\ar
q\bar q\Phi^0$ and $g\gamma\ar t\bar b H^-$
reproduce those of ref.~\cite{CheungLEPLHC} (for a ${\cal {SM}}$ Higgs),
of ref.\cite{qqhHERA} and of ref.~\cite{Munich}, respectively.
Furthermore, since a simple adaptation of the formulae given in the
Appendix (by
changing photon couplings from quarks into leptons and setting the
quark masses equal to zero) allows us to reproduce the computations
of ref.~\cite{ioPR}, we have checked, where possible, our helicity amplitudes
also in these limits.\par
As proton structure functions we adopted the recent set MRS(A) \cite{MRSA},
fixing the $\mu$--scale equal
to the Center
of Mass energy (CM)  at parton level (i.e.,
$\mu=\sqrt{\mathaccent 94{s}}_{q(g)\gamma}$).
The strong coupling constant $\alpha_s$, which appears in the the
gluon initiated
processes, has been evaluated at next--to--leading order, with
$\Lambda^4_{{QCD}_{\overline{MS}}}=230$ MeV
and a scale $\mu$ equal to the one used for the proton
structure functions, and consistent with the quark flavour entering
in the partonic subprocess. We are confident that
changing the energy scale and/or the distribution functions choice should not
affect our results by more than a factor of two.\par
For the energy spectrum of the back--scattered (unpolarized) photon
we have used \cite{backscattered}
\begin{equation}\label{backsc}
F_{\gamma/e} (x)= \frac{1}{D(\xi)}[1-x+\frac{1}{1-x}-\frac{4x}{\xi(1-x)}
      +\frac{4x^2}{\xi^2(1-x)^2}],
\end{equation}
where $D(\xi)$ is the normalisation factor
\begin{equation}
D(\xi)=\left(1-\frac{4}{\xi}-\frac{8}{\xi^2}\right)\ln(1+\xi)
+\frac{1}{2}+\frac{8}{\xi}-\frac{1}{2(1+\xi)^2},
\end{equation}
and $\xi=4E_0\omega_0/m_e^2$, where
$\omega_0$ is the incoming laser photon energy
and $E_0$ the (unpolarized) electron/positron one. In eq.~(\ref{backsc}),
$x=\omega/E_0$ is the fraction
of the energy of the incident electron/positron carried by the
back--scattered photon,
with a maximum value
\begin{equation}
x_{\rm {max}}=\frac{\xi}{1+\xi}.
\end{equation}
In order to maximise $\omega$ while avoiding $e^+e^-$ pair creation, one takes
$\omega_0$ such that $\xi=2(1+\sqrt 2)$. So, in the end, one gets the typical
values $\xi\simeq 4.8$, $x_{\rm {max}}\simeq 0.83$, $D(\xi)\simeq 1.8$.\par
In the case of $q(g)\gamma$ scatterings from $ep$ collisions,
the total cross section $\sigma$ is obtained by folding
the subprocess cross section $\hat\sigma$ with
the photon $F_{\gamma/e}$ and hadron $H_{q(g)/p}$ luminosities:
\begin{equation}
\sigma(s_{ep})=
\int_{x^\gamma_{\rm {min}}}^{x^\gamma_{\rm {max}}}dx^\gamma
\int_{x^{q(g)}_{\rm {min}}}^{1-x^\gamma}dx^{q(g)}
F_{\gamma/e}(x^\gamma)
H_{q(g)/p}(x^{q(g)})
\hat\sigma(\hat s_{q(g)\gamma}=x^\gamma x^{q(g)}s_{ep}),
\end{equation}
where $\hat s_{q(g)\gamma}$ is the CM
energy at parton (i.e., $q(g)\gamma$) level, while
\begin{equation}
x^\gamma_{\rm {min}}x^{q(g)}_{\rm {min}}=
\frac {(M_{\rm {final}})^2}{s_{ep}},
\end{equation}
with $M_{\rm {final}}$ the sum of the final state particle masses.\par
The total cross section has been then obtained numerically integrating
over the phase space using the Monte Carlo routine VEGAS
\cite{vegas}.\par
So far, to our knowledge, a detailed study, like in the case of $e\gamma$ and
$\gamma\gamma$ collisions \cite{backscattered}, on the efficiency of the laser
back--scattering method in converting
$e\ar \gamma$ at $ep$ colliders does not exist. In this work we assume the
effective $\gamma p$ luminosity to be equal to the $ep$ one (see
ref.~\cite{SMGAL}).
For the discussion of the results we have assumed an overall
total integrated luminosity ${\cal L}=3$ fb$^{-1}$, according to value
adopted in ref.~\cite{CheungLEPLHC}.\par
Within the \mssm, in order to simplify the discussion, we assume an universal
soft supersymmetry--breaking mass \cite{corrMH0iMSSM,corrMHMSSM}
\be
m_Q^2=m_U^2=m_D^2=m_{\tilde q}^2,
\ene
and a negligible mixing in the $stop$ and $sbottom$ mass matrices
\be
A_t=A_b=\mu=0.
\ene
If we also neglect the $bottom$ mass in the formulae
of refs.~\cite{corrMH0iMSSM,corrMHMSSM}, the one--loop corrected
masses of the ${\cal {MSSM}}$  neutral ${\cal {CP}}$--even
Higgs bosons can be expressed in terms
of a single parameter $\epsilon$ \cite{0pmLEPLHCSSC}, given
by\
\begin{equation}\label{m2}
\epsilon = \frac{3e^{2}}{8\pi^{2} M^{2}_{W^\pm}\sin^2\theta_W}
m_{t}^{4}
{\rm {ln}}\left( 1 +
\frac{{m}^{2}_{\tilde q}}{m_{t}^{2}} \right).
\end{equation}
Diagonalization of the mass squared matrix leads to the
expressions\footnote{Throughout this paper we use the notations
$s_x=\sin(x)$, $c_x=\cos(x)$,
$t_x=\tan(x)$ (with $x=\alpha,\beta,2\alpha$ and $2\beta$),
$s_{\alpha\beta}=\sin({\beta+\alpha})$,
$c_{\alpha\beta}=\cos({\beta+\alpha})$, $s_{\beta\alpha}=\sin({\beta-\alpha})$
and $c_{\beta\alpha}=\cos({\beta-\alpha})$.}
\begin{eqnarray}\label{m1}
M^{2}_{h^0,H^0}& = &
\frac{1}{2}[M^{2}_{A^0} + M_{Z^0}^{2} + \epsilon/s^2_{\beta}] \nonumber \\
&   & \pm \left\{ [ (M^{2}_{A^0} - M^{2}_{Z^0})c_{2\beta} + \epsilon/
s^{2}_{\beta}]^{2}
+(M^{2}_{A^0} + M^{2}_{Z^0})^{2}s^{2}_{2\beta} \right\}^{1/2},
\end{eqnarray}
while the mixing angle $\alpha$ in the ${\cal {CP}}$--even sector
is defined at one--loop by
\begin{equation}\label{m3}
t_{2\alpha} = \frac{(M_{A^0}^{2} +
M_{Z^0}^{2})s_{2\beta}}{(M_{A^0}^{2} -
M_{Z^0}^{2})
c_{2\beta} + \epsilon/s^{2}_\beta},
\quad\quad -\frac{\pi}{2}<\alpha\leq0.
\end{equation}
For the ${\cal {MSSM}}$ charged Higgs masses we have maintained the
tree--level relations
\begin{equation}
M_{H^\pm}^2=M_{A^0}^2+M_{W^\pm}^2,
\end{equation}
since the one--loop corrections are quite small if compared
with the corresponding ones for neutral Higgses \cite{corrMHMSSM}.\par
Concerning the numerical part of our work,
we have adopted
$\alpha_{em}= 1/128$ and  $\sin^2\theta_W\equiv s^2_W=0.23$, for
the e.m. coupling constant and the sine squared of the Weinberg angle,
respectively.
For the gauge boson masses and widths we have taken:
$M_{Z^0}=91.175$ GeV, $\Gamma_{Z^0}=2.5$ GeV,
$M_{W^\pm}=M_{Z^0}\cos\theta_W\equiv M_{Z^0}c_W\approx80$ GeV and
$\Gamma_{W^\pm}=2.2$ GeV;
while for the fermion masses: $m_e=0.511\times10^{-3}$ GeV,
$m_\mu=0.105$ GeV, $m_\tau=1.78$ GeV, $m_u=8.0\times10^{-3}$ GeV,
$m_d=15.0\times10^{-3}$ GeV, $m_s=0.3$ GeV, $m_c=1.7$ GeV, $m_b=5.0$ GeV
and, e.g., according to the CDF announcement \cite{CDF},
$m_t=175$ GeV, with all widths equal to zero except
the top one, which has been computed at tree--level
within the \mssm, using the expressions \cite{widthtopMSSM}:
\begin{eqnarray}
\frac{\Gamma(t\ar bH^+)}{\Gamma(t\ar bW^+)}&=&
\frac{\lambda(M_{H^\pm}^2,m_b^2,m_t^2)^{1/2}}
     {\lambda(M_{W^\pm}^2,m_b^2,m_t^2)^{1/2}}\times \\ \nonumber
&&\frac{(m_t^2+m_b^2-M_{H^\pm}^2)(m_t^2t^{-2}_\beta+m_b^2t^2_\beta)+4m_t^2
m_b^2}
       {M_{W^\pm}^2(m_t^2+m_b^2-2M_{W^\pm}^2)+(m_t^2-m_b^2)^2},
\end{eqnarray}
and \cite{widthtopSM}
\begin{eqnarray}
{\Gamma(t\ar bW^+)}&=&|V_{tb}|^2
\frac{G_Fm_t}{8\sqrt{2}\pi}\lambda(M_{W^\pm}^2,m_b^2,m_t^2)^{1/2}\times\\
\nonumber
&&     \left\{[1-\left(\frac{m_b}{m_t}\right)^2]^2
      +[1+\left(\frac{m_b}{m_t}\right)^2]
       \left(\frac{M_{W^\pm}}{m_t}\right)^2
      -2\left(\frac{M_{W^\pm}}{m_t}\right)^4\right\},
\end{eqnarray}
where $V_{tb}$ is the Cabibbo--Kobayashi-Maskawa mixing term (here set equal
to 1), $G_F={\sqrt{2}g^2}/{8M_{W^\pm}^2}$ the electroweak Fermi constant,
 $g=e/s_W$ with $-e$ the electron charge, and $\lambda^{1/2}$
the usual kinematic factor
\be
\lambda(M_a,M_b,M_c)^{1/2}=[M_a^2+M_b^2+M_c^2-2M_aM_b-2M_aM_c-2M_bM_c]^{1/2}.
\ene
All neutrino's have been considered
massless: i.e.,
$m_{\nu_e}=m_{\nu_\mu}=m_{\nu_\tau}=0$, with null corresponding
widths.\par
The widths of the \mssm\ Higgs bosons have been evaluated
for the same \mssm\ parameters we adopted in the cross
section analysis: for the numerical values as for
further details on their computation we refer to \cite{BRs}.\par
Finally, the universal supersymmetry--breaking squark mass has been fixed
in the numerical analysis to the value $m_{\tilde q}=1$ TeV, and at
the
same time, for simplicity, we have ignored the presence of not--Higgs
supersymmetric particles (i.e., squarks, sleptons, gauginos, higgsinos).\par
We have analysed processes (\ref{proc1})--(\ref{proc7}) in the mass
range 60 GeV $\Ord M_{A^0}\Ord$ 140 GeV, with tan$\beta=1.5$,
30, at the $ep$ CM energy $\sqrt s_{ep}=1.36$  TeV.

\subsection*{Results}

As it is unpractical to cover all regions of the
\mssm\ parameter space $(M_{A^0},\tan\beta)$ (for intermediate
masses of the pseudoscalar Higgs boson), we have chosen here,
as representative for
$\tan\beta$, the two extreme values
1.5 and 30, whereas $M_{A^0}$
spans in the range 60 to 140 GeV. Also, due to the huge
amount of computing time that otherwise would have been
necessary, and contrary to the \sm\ analysis
of ref.~\cite{SMGAL}, we concentrate here only on the energy
of the proposed CERN $ep$ collider ($\sqrt
s_{ep}=1.36$ TeV) \cite{LHC}.
At this CM energy the cross sections (summed over all possible
flavour combinations) for the processes:
\be\label{after1}
q\gamma\ar q'W^\pm\Phi^0,\quad\quad\quad \Phi^0=H^0,h^0,A^0,
\ene
\be\label{after2}
q\gamma\ar q Z^0\Phi^0,\quad\quad\quad \Phi^0=H^0,h^0,A^0,
\ene
\be\label{after3}
q\gamma\ar q'H^\pm\Phi^0,\quad\quad\quad \Phi^0=H^0,h^0,A^0,
\ene
\be\label{after4}
q\gamma\ar q \Phi^0\Phi^{0{'}},\quad\quad
(\Phi^0,\Phi^{0{'}})=(H^0,A^0),(h^0,A^0),
\ene
\be\label{after5}
q\gamma\ar qH^+ H^-,
\ene
\be\label{after6}
g\gamma\ar q\bar q'H^\pm,
\ene
\be\label{after7}
g\gamma\ar q\bar q\Phi^0,\quad\quad\quad \Phi^0=H^0,h^0,A^0,
\ene
are given in tabs.~Ia--VIIb.
Since the production rates for the reactions
\be\label{after4bis}
q\gamma\ar q \Phi^0\Phi^{0{'}},\quad\quad
(\Phi^0,\Phi^{0{'}})=(H^0,H^0),(H^0,h^0),(h^0,h^0),(A^0,A^0),
\ene
are generally\footnote{Apart from the cases $(h^0,h^0)$ and
$(A^0,A^0)$
for $M_{A^0}=60-80$  GeV, with $\tan\beta=1.5,30$ and $30$,
respectively,
which can reach cross sections of ${\cal O}(1)$ fb.}
never larger than ${\cal O}(10^{-2})$ fb and
are beyond any experimental possibility of detection,
we do not give their rates here and we will not consider them in the
forthcoming analysis either\footnote{Also, in some instances, results
given in tabs.~Ia--VIIb will be very small. Nevertheless, we present them
with the purpose of comparison, in order to facilitate the discussion
in terms of dependence on
masses, couplings, etc ...}.
Before proceeding further, a
few comments are in order now, concerning the characteristics of the
signals.\par
Process (\ref{after1}) gives quite large rates for the case
$\Phi^0=H^0$
and not too large values of $M_{A^0}$ ($\Ord 120$ GeV), both for
$\tan\beta=1.5$ and $\tan\beta=30$, with  the cross sections
corresponding to the last case being larger. Significantly large numbers occur
also in the case $\Phi^0=h^0$, more at small than at large
$\tan\beta$'s. Phase space effects due to the increase of \MH\
and \Mh\ lower down the cross sections,
whereas the strong change of trend at large $\tan\beta$'s and
$\MA\approx120-140$ GeV is due to the sudden steep decrease of the
$H^0W^+W^-$ coupling (proportional to \cba), and to the corresponding
increase of the $h^0W^+W^-$ one (proportional to \sba). Higgs
bremsstrahlungs diagrams (numbers 1--6 in fig.~1) are in fact drastically
suppressed because of the Yukawa coupling $\Phi^0q\bar q$,
proportional to $m_q$, since $q$ (due to the partonic distributions)
is most of the times a light quark. Because of this, and
since the $A^0$ does not couple at tree--level
to the $W^\pm$'s, the case $\Phi^0=A^0$ generally gives much smaller
rates. Only the case \tbb\ (i.e., large $\Phi^0D\bar D$ coupling to
down-type quarks $D$), for small enough phase space suppression
(i.e., if $\MA\approx60$ GeV), can  give cross sections of ${\cal
O}(1)$ fb.\par
Same considerations as the above mentioned apply to the case of reaction
 (\ref{after2}), even though the suppressed $\Phi^0Z^0Z^0$ couplings
(with respect to the case $\Phi^0W^+W^-$, being $\Phi^0=H^0,h^0$)
yield contributions which are in general an order of magnitude
smaller than in the previous case. At \tba\ only the $h^0$ seems
to be interesting, whereas at \tbb\ both the $H^0$ and the $h^0$ show
negligible numbers.
Finally, graphs with Higgs--strahlungs
off $b$--quarks contribute to keep the rates for the $A^0$ at \tbb\
at the level of ${\cal O}(1)$ fb, if \MA\ is not too large, whereas
at \tba\ numbers are completely negligible.\par
The coupling of the $A^0$
to the vertices $\Phi^0 W^\pm H^\mp(\gamma)$ (see tab.~A.I in the
appendix) does not suffer from angular factor suppression (there is no
dependence on $\alpha$ and $\beta$),
whereas $H^0$'s and $h^0$'s do. Therefore, the rates for the $A^0$ in
the case of reaction (\ref{after3}) are larger than the ones of the $CP$--even
scalars, both a \tba\ and \tbb. This latter observation is always true
apart from the case \tbb\ and $M_{A^0}\Ord 120$ GeV, where numbers for the
pseudoscalar and the light scalar are practically the same, as the
value of \cba\ approaches 1. This also proves that diagrams
with $\Phi^0H^+H^-$ couplings (graphs 13--14 in fig.~2, which
are zero for $\Phi^0=A^0$) do not count. The same can be affirmed for
neutral Higgs bremsstrahlung diagrams,
because they always occur in conjunction
with a $H^\pm q\bar q'$  Yukawa coupling
(see the practically unchanged rates for the $A^0$ at both
values of \tb).\par
The case $(\Phi^0,\Phi^{0'})=(H^0,A^0)$ in process (\ref{after4}) is
never interesting (and it has been shown for comparison purposes only,
against the combination $h^0A^0$). Due to the double Yukawa coupling,
diagrams 1--6 in fig.~2 essentially never enter. Diagrams 9--10
are strongly suppressed at \tba, whereas at \tbb\ they give
a small contribution (because of the $A^0 D\bar D$ vertex). However,
the largest rates come from diagrams 7--8, which are proportional
to $\sba^2$ and $\cba^2$, for the $H^0$ and the $h^0$, respectively.
As the second coupling is larger than the first one and $M_{H^0}>\Mh$
in our range of interest, it is clear
that $H^0$ rates are again smaller compared to the $h^0$ ones (especially
at \tbb).\par
Process (\ref{after5}) is one of those for which the production rates
are bigger, if \MA\ is not too large. The major partonic contributions
here come from the subprocess with resonant top--quarks
(i.e., $b\gamma\ar bH^+H^-$).
Diagrams with $\gamma^*(Z^{0*})H^+H^-$ couplings
(i.e., with a virtual photon or $Z^0$ splitting into $H^+H^-$--pairs)
 are dominant only in the other
cases (for $q=u,d,s,c$). The increase of the rates with \tb\
is due to the larger contribution of graphs 8--10 and 13--14, which
involve $\Phi^0D\bar D$ couplings ($\Phi^0=H^0$ and $h^0$).\par
For process (\ref{after6}), practically, the whole of the partonic
contribution comes from the combination $g\gamma\ar t\bar bH^+$ +
c.c., because of the $bt$ Yukawa couplings of the $H^\pm$ and
because of the top resonance.
Therefore, the increase of the rates with the increase of \tb\
in tabs.~VIa--b exclusively depends on and can be understood
in terms of the coupling
$H^\pm tb$. Graphs with $\gamma H^+H^-$ vertices
are generally suppressed in the $tbH^\pm$  case, and
phase space effects act
in such a way to strongly reduce the rates for increasing \MA\
(because of
the quite large value of $m_t$).
\par
In case of process (\ref{after7}) we can greatly appreciate
the benefits of $\Phi^0D\bar D$ Yukawa couplings
with large $\tan\beta$: in fact,
all the flavours $\Phi^0=H^0,h^0$ and $A^0$ have large cross
sections at \tbb. This happens especially
for the pseudoscalar (it has a $\sim\tb$ quark--coupling) and
the light scalar ($\sim\sa/\cb$ quark--coupling). The decrease of their
rates with an increasing \MA\  is due to a phase space effect in the
former case, whereas in the latter also a reduction due to the
diminishing of \sa\ occurs. Since the $H^0$ quark--coupling is
proportional to $\ca/\cb$, in this case things proceed
 the other way round. In addition, the suppression due to
phase space effects
is small here, as \MH\ varies by only $\approx10$ GeV in the
usual $M_{A^0}$ range, if \tbb. At \tba\ rates are generally much
smaller,
being noticeable only for $h^0$ (small mass and large \sa).
\par
The main lines of the analysis we will perform in order to select the
signal events out of the backgrounds
are the same ones already adopted for the
\sm\ case, in ref.~\cite{SMGAL}. In order to maximise
the event rates, we will consider the decay channels with highest
Branching Ratio (BR).
Therefore, we will look
for the Higgs decay channel
$\Phi^0\ar b\bar b$ for the neutral
Higgs flavours $h^0$, $A^0$ and $H^0$,
whereas, in case of charged Higgses,
we will concentrate on the decay $H^\pm\ar \tau\nu_\tau$.
 We know that for $\tan\beta=1.5(30)$ and 60 GeV $\leq M_{A^0}\leq$
140 GeV($\MA\approx60$ GeV),
corresponding
to 145 GeV $\Ord M_{H^0}\Ord$ 180 GeV($\MH\approx129$ GeV),
the BRs of the decay channels $H^0\ar
W^{+*}W^{-*}$($H^0\ar W^{+*}W^{-*}$)
and, for  $\MH\Ord150$ GeV,  $H^0\ar h^0h^0$($H^0\ar
h^0h^0$ and  $H^0\ar A^0A^0$),
are larger than BR($H^0\ar b\bar b)$
\cite{BRs}. Nevertheless, we concentrate here on the last decay only, for
various reasons. In the case of $W^{+*}W^{-*}$--decays, we
should first add, in any case, an additional reduction factor due to
the
$W^{\pm *}$--decay channels
(that we
should, in some how, identify). Second, we would end up considering signatures
of the type $jj(Y)$, $(\tau\nu_\tau)(Y)$ or $(b\bar b)(Y)$ (see later
on), where $Y=
4j,2j2\ell$, or $4\ell$, with the clear disadvantages of
dealing  either with a large number of jets (for
$Y=4j,2j\ell\nu_\ell$,
which would
have both a large QCD and combinatorial background) or with missing
energy/momenta (for $Y=\ell\nu_\ell\ell\nu_\ell$,
which would prevent from reconstructing
Higgs peaks by means of invariant mass spectra).
In the case of $h^0h^0$-- and $A^0A^0$--decays,
in order to keep high rates, we should consider
the channels  $h^0h^0,A^0A^0\ar
4b$, which would lead to the difficult requirement of
recognising with high efficiency at least four $b$'s in a single event.
Whereas the decay $H^0\ar b\bar b$ implies that the only reduction
factor
is the
$b\bar b$--BR, which ranges in the above interval
of $M_{H^0}$, for $\tan\beta=1.5(30)$, between $\approx4(3)$ and
$\approx20(90)\%$
\cite{BRs}.
In the case of charged Higgs decays, if \tba\ and
$\MHpm\OOrd150$ GeV, the $tb$ channel has a BR larger than the
one into $\tau\nu_\tau$--pairs. However, as BR$(H^\pm\ar tb)$ is not too
drastically larger (so the loss of statistics is not substantial) and
the decay chain $H^\pm\ar tb\ar b\bar bW^\pm+b\bar bH^\pm$ would lead to a
more complicated final state with additional backgrounds, for the
moment,
we consider the $H^\pm\ar\tau\nu_\tau$ channel only.\par
We will require hadronic decays of the massive vectors bosons
($W^\pm$ and $Z^0$) and, in order to select the $b\bar b$ Higgs decay out
of the QCD background, we will assume excellent
flavour identification of $b$--quarks \cite{btagging},
such that we can get rid of
the non--$b$
multi--jet photoproduction, $W^\pm$ + jets and
$Z^0$ + jets  background events \cite{CheungLEPLHC}, and
that a $M_{b\bar b}$ cut around the $\Phi^0$
 masses (see later
on) is sufficient in order to suppress
the above processes in the case of $\gamma^* / g^*\ar b\bar b$ splitting.
\par
Therefore, we expect the
following signatures:
\be\label{signature1}
q'W^\pm\Phi^0 \ar  (jj)(b\bar b)X,
\ene
\be\label{signature2}
q Z^0\Phi^0  \ar  (jj)(b\bar b)X,
\ene
\be\label{signature3}
q'H^\pm\Phi^0 \ar (\tau\nu_\tau)(b\bar b)X,
\ene
\be\label{signature4}
q H^0A^{0}, q h^0A^{0} \ar (b\bar b)(b\bar b)X,
\ene
\be\label{signature5}
qH^+ H^-\ar (\tau\nu_\tau)(\tau\nu_\tau)X,
\ene
\be\label{signature6}
q\bar q'H^\pm\ar jj(\tau\nu_\tau)X~({\rm{if}}~q\bar q'\ne
t\bar b)~~{\rm{or}}~~t\bar b(\tau\nu_\tau)X\ar b\bar
b(\tau\nu_\tau)X~({\rm{if}}~q\bar q'= t\bar b),
\ene
\be\label{signature7}
q\bar q\Phi^0 \ar (jj)(b\bar b)X~({\rm{if}}~q\ne
b,t)~~{\rm{or}}~~(b\bar b)
(b\bar b)X~({\rm{if}}~q= b,t),
\ene
where X represents the untagged particles in the final states.
\par
Concerning the expected backgrounds\footnote{These have been evaluated
with the help of Madgraph/HELAS \cite{Tim}.} to the above signatures,
in case of neutral scalar production (i.e.,
eqs.~(\ref{signature1})--(\ref{signature4}) and (\ref{signature7}))
we have to consider the same processes already analysed
in ref.~\cite{SMGAL} for the \sm: i.e., $ep\ar W^\pm Z^0 X$,
$ep\ar \bar t b X\ar b\bar bW^- X$, $ep\ar t\bar tX\ar b\bar bW^\pm
X$, $ep\ar Z^0Z^0 X$ and $ep\ar q\bar q Z^0$.
In the case of double and single
charged scalar production
(i.e.,
eqs.~(\ref{signature5})--(\ref{signature6}), respectively), we must add
the reactions $ep\ar W^+W^-X$ and $ep\ar tb W^\pm X$.
We also notice how the process $ep\ar Z^0 Z^0 X$ is a background to
$H^+H^-$--production when $Z^0Z^0\ar
(\tau^+\tau^-)(\nu_\tau\bar\nu_\tau)$
and that the double and single top--resonant backgrounds $t\bar tX$
and $tbX$, as in the
\mssm\ $t$--quarks can decay either to $bW^\pm$-- or $bH^\pm$--pairs,
are a potential background  for $W^\pm\Phi^0X\ar W^\pm(b\bar b)X$,
$H^\pm\Phi^0X\ar H^\pm(b\bar b)X$ and
$t\bar bH^-+\bar t bH^+\ar b\bar bH^\pm X$.\par
In tabs.~VIIIa--b we update the results given in \cite{SMGAL} for the
neutral scalar production backgrounds,
as we are using here a more recent
set of structure functions (compare to tab.~III in ref.~\cite{SMGAL}),
and, at the same time, we give the rates also for the additional cases
$ep\ar qW^+W^-$ ($q\ne b$) and $ep\ar q\bar q' W^\pm$ ($q\bar q'\ne t\bar b$).
In tab.~VIIIa, a sum over all possible combinations of flavours (not
involving top resonances) is implied everywhere. In particular, we
notice how in the
case of the subprocesses
$b\gamma\ar W^- Z^0 t$ + c.c. and $g\gamma\ar t\bar tZ^0$ there are
top quarks involved as well: however, as they are produced on--shell
in our computations, they do not have any dependence on the \mssm\
parameters.
On the contrary, in the case of the top resonant backgrounds $tbX$,
$t\bar tX$, $W^+W^-X$ (via $b$--initiated subprocesses)
and $q\bar q'W^\pm X$ (for $q\bar q'=t\bar b$ + c.c.)
there is such a dependence. Since $\Gamma_t^\mssm$ is function
of $M_{H^\pm}$ and \tb\
 (at tree--level), ten different cross sections appear in
tab.~VIIIb. The total top width in the \mssm\ (together with
the BRs of the top quark into $bW^\pm$ and $bH^\pm$),
for the two values
\tba\ and 30, is given in tab.~IX.\par
Also, we would like to stress here a few details concerning
the rates for
top production via $g\gamma$--fusion. The case
labeled $tbW^\pm X$ corresponds to top production via the
two--to--three body subprocess $g\gamma\ar t\bar bW^-$ + c.c.
(including all the 8
diagrams at tree--level giving a gauge invariant set),
whereas
$tbX\ar b\bar bW^+W^- X$ and $t\bar tX\ar b\bar bW^+W^- X$ correspond to
the rates obtained for
the subprocesses $g\gamma\ar b\bar bW^+W^-$
via graphs with one (12 diagrams)
or two (2 diagrams) top resonances, respectively. That is, in the case
of the two--to--four body process, we considered only the amplitudes
squared of two subsets of
the complete set of tree--level Feynman graphs, neglecting
their interference.
This clearly turns out to be an approximation (and not gauge invariant).
However,
as single and double top production in $g\gamma\ar b\bar b W^+W^-$
events are by far the dominant contributions we expect to reproduce
quite accurately the complete calculation.
In order to check the self--consistency of our results, e.g., one can
take, on the one hand,
the cross section for $t\bar bW^- X$ + c.c. in case of $tt$-- (2 diagrams,
yielding,
e.g., at \tba\ and $\MA=60$ GeV, $\approx1195$ fb) plus the one for single
$t$--production (6 diagrams, with $\sigma\approx406$ fb, for the same
choice of parameters $(\MA,\tb)$ as above)
and multiply these by the corresponding
BR$(t\ar bW^\pm)$ within the \mssm\ (see tab.~IX), after dividing by
two the
contribution of the $tt$--resonant part (thus avoiding problems
of double counting), and, on the other hand, the sum of the rates in
third and fourth column of tab.~VIIIb,
then he ends up
with numbers that are `roughly' the ones within the computational errors of the
others.
The above approximate procedure has been adopted in order to avoid
prohibitive CPU--time consumes
in calculating the complete $g\gamma\ar b\bar bW^+W^-$
process (52 Feynman graphs at tree--level, including
Higgs contributions and keeping the $W^\pm$'s on--shell).\par
In case of neutral Higgs production, we divide the backgrounds in
continuum and discrete. The first
are the ones in which the $b\bar b$--pair does not come
from a $Z^0$--resonance (i.e., $tbX$ and $t\bar tX$), and the second
the ones in which the $b$'s are the decay products of the $Z^0$.
Following the above distinction also in the case of double and single
$H^\pm$--production, it turns out that $H^\pm$--signals have only
discrete backgrounds, in which the $\tau\nu_\tau$--pair comes from a decaying
$W^\pm$.\par
Although the background rates are in some instances much larger
that the corresponding signals, one has to remember that the discrete
backgrounds can
 be potentially
dangerous only in the cases $M_{\Phi^0}\approx M_{Z^0}$ and
$M_{H^\pm}\approx M_{W^\pm}$, while the continuum ones should
have a quite flat distribution in the $M_{b\bar b}$ spectrum, where
 $M_{b\bar b}$ is the invariant mass of the $b\bar b$--pair.
As the aim of a phenomenological analysis is
to finally
select signal candidate events in a window around the Breit--Wigner
resonance of the Higgs bosons, we will ask
that, say, $|M_{\Phi^0(H^\pm)}-M_{b\bar b(\tau\nu_\tau)}|<5$
GeV\footnote{We do not repeat here the considerations
which induced us to adopt a relatively high mass resolution, as
they have been discussed for the case of the \sm\ analysis. For this,
we again refer the reader to ref.~\cite{SMGAL}.}.
If we naively assume that the invariant mass spectra of
the discrete backgrounds are all contained in the regions
 $|M_{Z^0}-M_{b\bar b}|\leq 2\Gamma_{Z^0}=5$ GeV and
 $|M_{W^\pm}-M_{\tau\nu_\tau}|\leq 2\Gamma_{W^\pm}\approx 5$ GeV, then
the fraction of $Z^0/W^\pm$--resonant background events which overlap signal
events is given by \cite{Cheung}
\be\label{peak}
\delta\sigma(Z^0/W^\pm)=\sigma(Z^0/W^\pm)\frac{{\rm{max}}(0, 10~{\rm GeV}-
|M_{\Phi^0/H^\pm}-M_{Z^0/W^\pm}|)}{10~{\rm GeV}},
\ene
for $\Phi^0=H^0,h^0$ and $A^0$.
In using the above equation we tacitly assumed that also the $\Phi^0\ar
b\bar b$ and $H^\pm\ar\tau\nu_\tau$ peaks
are all contained in a region of 10 GeV around the
Higgs--poles\footnote{In fact, the largest Higgs width in the region
of the
parameter space here considered happens for the heavy
scalar $H^0$, at $M_{A^0}=140$ GeV and  for $\tan\beta=30$, giving
$\Gamma_{H^0}\approx 2.9$ GeV.}.\par
In addition, in
case of continuum backgrounds, as these are top--resonant processes
and we are considering hadronic decays of the $W^\pm$'s, in order to
further enhance the signal versus background ratio, we can impose
the veto, say, $|M_{bW\ar b(jj)}-m_t|>15$ GeV. Since by the time
the LEP$\otimes$LHC collider will be operating
the value of the top mass will be well determined, it is quite likely
that the above constrain could reveal very efficient.\par
As criteria for the observability of a signal, we require a rate
of $S\geq6$ events with a significance $S/\sqrt B>4$
for the detection of an isolated Higgs peak, while
for the case of Higgs peaks overlapping with $Z^0$ or $W^\pm$ peaks
we require $S\geq10$ with $S/\sqrt B>6$ \cite{Cheung}. \par
In what follows we will
concentrate only on the regions of parameter space $(\MA,\tb)$ where
we have enough rates to presumably make a statistically significant
analysis: say, at least ${\cal O}(1)$ fb of cross section, and we will
analyse the signatures in eqs.~(\ref{signature1})--(\ref{signature7})
separately.\par

\vskip 1.0cm
\centerline{\underbar{\sl A. Signature $b\bar bb\bar b$.}}
\vskip 0.5cm
\noindent
In this case we have contributions from the signals $H^0A^0X$,
$h^0A^0X$ and $q\bar q\Phi^0$ (here $q=b,t$, with $t\bar t\Phi^0\ar
b\bar b\Phi^0X$, flavours which give the whole of the cross sections
in tabs.~VIIa--b),
and from the backgrounds $Z^0Z^0X$ and $q\bar qZ^0X$, this
latter for $q=b,t$, which yields a cross section of $\approx 110$ fb.\par
Here, the most interesting region in the plane  $(\MA,\tb)$ is
the one with $\tbb$, value for which the combination
$h^0A^0$ seems to be quite promising if $\MA\approx60$ GeV (see
tab.~IVb), whereas the rates for $q\bar q\Phi^0$ are very large
over all the intermediate spectrum of $\MA$, if $\Phi^0=h^0,A^0$.
In the case $q\bar qH^0$ rates are small if $\MA\Ord100$ GeV.
For ${\cal L}=3$
fb$^{-1}$, after a few years of running, it should be possible to
accumulate some tens of $h^0A^0X$ events, practically free from backgrounds,
as both the $A^0$-- and $h^0$--peaks are quite distant from the
$Z^0$--one. The combination $H^0A^0$  is too small for
deserving experimental attention, even it doesn't substantially
contribute in a possible $A^0X$ inclusive analysis.
The cases $q\bar qh^0$ and $q\bar qA^0$ give hundreds or thousands
of events per
year, which should be easily recognised if $\Mh,\MA\ne M_{Z^0}$.
In the case of overlapping $Z^0$ and $h^0/A^0$ peaks, Higgs signal
could be recognised in the form of an excess of $b\bar b$ events at the
$Z^0$ peak. For $q\bar qH^0$, as $\MH-M_{Z^0}>>10$ GeV in the range
where $H^0$--rates are large, there should not be any problem in
selecting the signal.
The case \tba\ seems to be quite discouraging for all the above signals.\par
As this signature involves four $b$--quarks it is crucial that high
$b$--tagging performances can be achieved.
\vskip 1.0cm
\centerline{\underbar{\sl B. Signature $jjb\bar b$.}}
\vskip 0.5cm
\noindent
This channel receives contributions from the signals $W^\pm\Phi^0X$ and
$Z^0\Phi^0X$. The case $q\bar q\Phi^0$ for light flavours $q=u,d,s$
and $c$
practically does not give any event for all $\Phi^0$'s,
as the bulk of the cross
sections come from the subprocesses $q\bar q\Phi^0$ with $q=b,t$, which
give the already considered $4b$--signature. The backgrounds are
$W^\pm Z^0X$, $Z^0Z^0X$, and $q\bar qZ^0X$ for
$q\ne b,t$, which yields a cross section
of $\approx3000$ fb. In addition, the continuum processes $tbX\ar
b\bar bW^\pm X$ and $t\bar tX\ar b\bar bW^\pm X$
enter here as well (with $W^\pm\ar jj$\footnote{It
would be
worth here to consider also $tbX\ar
b\bar bH^\pm X$ and $t\bar tX\ar b\bar b H^\pm X$
as background,
although they contain a \mssm\ charged Higgs. In fact, $H^\pm$'s can
decay to $jj$--pairs. However, as this channel has a small branching
ratio (other than originating from already suppressed $t\ar bH^\pm$ decays,
see tab.~IX)
and as we are tacitly assuming that the two jets in the
signature $jjb\bar b$ reproduce the $W^\pm(Z^0)$--mass
(note that $\MHpm-M_{W^\pm(Z^0)}\OOrd20(10)$ GeV), we can safely
neglect the two above background contributions.}).\par
Due to the small rates, we did not consider here
$A^0$--production at \tba.
Once one multiplies the rates of signal and backgrounds by the BRs
giving the signature $jjb\bar b$, by the yearly luminosity
${\cal L}=3$ fb$^{-1}$ and picks up events in the windows
$|M_{b\bar b}-M_{\Phi^0}|<5$ GeV, it comes out that the only  case which
can give significancies large enough to allow for possible detection
is for $\Phi^0=h^0$ at \tba. The value of $S/\sqrt B$ is approximately
4 over all the range 56 GeV $\Ord\Mh\Ord$ 81 GeV. The case $\Phi^0=H^0$
at \tbb, which has production rates comparable to the ones of the
previous case, is overwhelmed by the $tbX$ and $t\bar tX$ backgrounds
(both productions rates and BRs into $b\bar b$--pairs are in fact smaller with
respect to the light neutral Higgs). Therefore, in the channel
$jjb\bar b$ only the $h^0$ scalar can be detected, and only for large \tb's.

\vskip 1.0cm
\vfill\newpage
\centerline{\underbar{\sl C. Signature $\tau\nu_\tau b\bar b$.}}
\vskip 0.5cm
\noindent
In this case, we have to consider $H^\pm\Phi^0X$,
$tb H^\pm\ar b\bar b H^\pm X$ as signals, and
$W^\pm Z^0X$ and $tb W^\pm\ar  b\bar bW^\pm X$ for backgrounds.
Here, the distinction between signal and background
is subtle, as the final state $tb H^\pm$ enters as signal for
the decay $H^\pm\ar\tau\nu_\tau$ but as background for $\Phi^0\ar  b\bar
b$ because of the top decay $t\ar bX$.
In the signal versus background analysis
 we treated the rates of $tb H^\pm$ exactly on this footing:
when we compute the numbers for the signals separately,
$tb H^\pm$ was considered background to $H^\pm\Phi^0X$, whereas, in the
`inclusive' case $H^\pm X$
(i.e., when we summed up the rates of  $H^\pm\Phi^0X$ and
$tb H^\pm\ar b\bar b H^\pm X$), they contributed to the event rates only.
In computing the
the signal--to--noise ratios we ignored
the case of the $H^0$ in  $H^\pm\Phi^0X$
(whose rates are never greater than $\approx0.4$ fb). \par
For the two
values \tba\ and 30 and in the range 60 GeV $\Ord\MA\Ord$ 140 GeV
the mass of the charged Higgs is always larger than $\approx100$ GeV.
Therefore $W^\pm$-- and $H^\pm$--peaks do not overlap in the spectrum
of the invariant mass $M_{\tau\nu_\tau}$ and charged Higgs signal
should be clearly recognised, whereas the case  $H^\pm\Phi^0X$ is
largely covered by the backgrounds (we found that significancies are always
smaller than 1 after one year of running). So, the signature
$\tau\nu_\tau b\bar b$ definitely gives large chances of charged Higgs
detection (for all masses and \tb's), whereas this latter
 is hopeless in the case
of neutral Higgses.

\vskip 1.0cm
\centerline{\underbar{\sl D. Signature $\tau\nu_\tau\tau\nu_\tau$.}}
\vskip 0.5cm
\noindent
This channel has signal contributions from double charged Higgs
production
$H^+H^-X$ and backgrounds from charged vector boson production
$W^+W^-X$, as well as from neutral production $Z^0Z^0X$ (with one
$Z^0$ decaying to $\tau^+\tau^-$ pairs and the other to neutrinos).
Both the processes with $H^\pm$'s and $W^\pm$'s
benefit from  a large top--resonant
component ($q=b$ in eq.~(\ref{after5})), but only backgrounds
rates have significant contributions for $q\ne b$. The case $Z^0Z^0X$
has a much smaller cross section. A few words are needed here to discuss the
strategy for detecting
Higgs signals, as the presence of two neutrinos should prevent from
reconstructing invariant mass spectra. For example,
one possibility could simply
be the one of looking at the total rates in $\tau\nu_\tau\tau\nu_\tau$
events. An excess of $2\tau2\nu_\tau$ events (i.e., a breaking
of the $\tau$ {\it vs}
$\mu$/$e$ universality), with
respect to the numbers expected from $W^+W^-X$ plus  $Z^0Z^0X$
production, could well be the method of establishing the presence of
$H^\pm$ signals. In that way, these latter should be clearly disentangled
over all the intermediate mass range, for both values of \tba\ and
30, presumably after just one year of running.\par

\vskip 1.0cm
\centerline{\underbar{\sl E. Signature $jj\tau\nu_\tau$.}}
\vskip 0.5cm
\noindent
To this channel there is a signal contribution coming from $q\bar
q'H^\pm X$ when $q\bar q'\ne t\bar b$, whereas backgrounds come from
$W^\pm Z^0X$ (with $Z^0\ar jj$) and $W^+W^- X$ events  (with one
$W^\pm\ar jj$).
As $\MHpm-M_{W^\pm}\OOrd 20$ GeV over the $(\MA,\tb)$ region
here considered, the detection of $H^\pm$--signal should only be a matter of
event rates. For \tba\ numbers are very small, $\Ord{\cal{O}}(10^{-1})$.
In tab.~VIb the cross section of the process $ep\ar
q\bar q' H^\pm$ for $q\bar q'\ne t\bar b$ is
$\approx23(16)[9]\{7\}<4>$ fb,
for $\MA=60(80)[100]\{120\}<140>$ GeV and \tbb.
Therefore, we expect the signature $jj\tau\nu_\tau$ to give further
chances to detect \mssm\ charged Higgses at large \tb, generally over all the
intermediate mass range of the $A^0$.\par
\vskip1.5cm
\noindent
We are aware that, in order to conclude our analysis in a realistic manner,
some additional steps would be necessary now. For example, the gauge bosons
$W^\pm$ and $Z^0$ that we have kinematically
constrained so far to be on--shell should be
allowed to decay. The same should be done for the \mssm\ Higgs bosons
$H^0,h^0,A^0$ and $H^\pm$. In addition, the final state partons should be
evolved into hadrons and reconstructed from the detector acceptances.
Therefore, on the one hand, a clustering scheme of the jets should be
adopted while, on the other hand, information about the detector
design (azimuthal coverage, cell structure, etc ...) and performances
(in particle identification, in microvertex efficiency, etc ...)
should be properly included into the phenomenological simulation.\par
Nevertheless, we have not done all of this.
We have decided not to do that for two substantial reasons,
related to the subject of the kinematical acceptances.
First, doing this would have required a not negligible
computing effort, because of the large numbers of different
processes with different kinematics
here involved (both among signals and backgrounds).
Second, such effort could have risked being finalised in a wrong direction,
in the sense that our choice of kinematical cuts
could have  well been different from the one which will be imposed
by the real detectors.
At present, in fact, the acceptances of the detectors of
the LEP$\otimes$LHC are difficult to predict, as the most recent
and complete studies
on the argument only deal
with simulations done for the LHC (see the
ATLAS {\cite{ATLAS} and CMS \cite{CMS}
Technical Proposals). That is,
we wonder if the detectors designed for a $pp$ machine
will be the same and/or will work in
the same configuration even when they will be set up around a different
kind of machine, an $ep$ collider.
\par
However, in order not to leave this issue completely un--addressed,
we borrow some numbers from ref.~\cite{SMGAL}, where a complete
analysis was attempted. There, the following cuts
\begin{itemize}
\item transverse momentum $p_T^i$ of at least 20 GeV;
\item pseudorapidity $|\eta_i |$ less than 4.5;
\item separation $\Delta
R_{ij}=\sqrt{\Delta\eta_{ij}+\Delta\varphi_{ij}}>1$;
\end{itemize}
were assumed \cite{ZeppenfeldLHC},
for all the $i$--th and $j$--th $b$'s and jets in the
the signature $b\bar b jj$ of the \sm, which would correspond here to the
one obtained in the case of
process (\ref{proc1}) for $\Phi^0=h^0$ and \tba\ (i.e.,
with $\Mh$ between $\approx60$ and
$\approx80$ GeV).
We concentrate only on this case
since this is the one where the effects of the (continuum)
backgrounds are effective but nevertheless still do not prevent
detecting $h^0$--signals.
\par
After applying the above kinematical requirements,
reduction factors of $\approx16-7$ for the signal $W^\pm\phi$,
with $M_\phi=60-140$ GeV\footnote{Here
$\phi$ represents the \sm\ Higgs boson and $M_\phi$ its mass.},
and of $\approx14/11$ for the
$tbX/t\bar tX$ backgrounds were found.
As the only
differences between the \sm\ case of ref.~\cite{SMGAL} and the \mssm\
one studied here (when $\Mh=M_\phi$)
consists in the presence of some angular factors in the
${\cal{SUSY}}$ vertices of reaction (\ref{proc1}) (see
tabs.~A.I--A.IV in the Appendix)
and different (but small) top width effects in the backgrounds (the
substitution $\Gamma_t^{\sm}\ar\Gamma_t^{\mssm}$), the numbers we obtained
there can be safely
used for the present case too.
Therefore, even though the kinematical acceptances
act in the direction of favouring the backgrounds, by reducing
the signal--versus--noise ratio and largely spoiling
the effectiveness of the $M_{bW\ar bjj}$ cut (see ref.~\cite{SMGAL}),
in our opinion
such effects should not have a decisive impact on  the
feasibility of the $h^0$ detection in $jjb\bar b$ events.
We think the same holds also for
the other signatures, especially because there
background events have discrete spectra in the invariant masses
of the Higgs decay products, and the requirements $\MHpm\approx
M_{\tau\nu_\tau}$ and/or $M_{\Phi^0}\approx M_{b\bar b}$ should be generally
sufficient to give large significancies, such that an eventual
reduction due to kinematical cuts shouldn't modify the detection
strategies we indicated.\par
Although our analysis remains partially incomplete,
we believe that the purpose of our study has been reached.
This was in fact to give some hints in the direction of analysing the
impact of using back--scattered photons in $\gamma p$--initiated collisions
at the proposed CERN LEP$\otimes$LHC collider, trying to establish whether
such a machine could give additional informations in the study of
the Higgs sector of the \mssm, once the potential of the two colliders
LEP and LHC (separately operating) was already fully exploited.
This is especially relevant if one considers the
possibility that a long gap in time
between the end of the LEP and LHC era and the beginning of the NLC one
could happen in the future of particle physics.
\par
A brief summary of what we have been doing
and the answers to the above considerations are left in the next section.

\subsection*{Summary and conclusions}

We have studied in this paper some production mechanisms of the Higgs
bosons of the \mssm\ (i.e., $H^0,h^0,A^0$ and $H^\pm$) and of the
possible backgrounds to their signatures at the proposed
LEP$\otimes$LHC $ep$ collider at CERN.\par
Such a machine can be
obtained by crossing an electron/positron beam from LEP with a proton one
from the LHC. It should presumably run with a CM energy at the TeV scale and
with a luminosity between one and ten inverse picobarns per year.
Its discovery/detection potential in the Higgs
sector was already  analysed for the case where the collider
is assumed to operate in the $ep$ mode (i.e., via electron--quark and
electron--gluon scatterings).
Promising results were found for the case of Higgs bosons with intermediate
mass, especially if high $b$--tagging performances can be achieved
in detecting neutral Higgses decaying to $b\bar b$--pairs.
We addressed here the same matter, but assuming the accelerator
working in a possible $\gamma p$ mode, with the incoming photons
produced through
Compton back--scattering of laser light against the electron/positron
beam. This technique has received a lot
of attention in the past few years as a concrete possibility of setting
up real $e\gamma$ and $\gamma\gamma$ interactions at $e^+e^-$ linear
colliders of the next generation. Such photonic interactions
are expected to take place with almost the same
characteristics (in energy of the beams and in integrated luminosity)
as the $e^+e^-$ ones. We studied
the possibility of producing $\gamma p$ interactions at CERN
as we expect the design of the LEP$\otimes$LHC machine not to
prevent the application of the laser back--scattering method.\par
Independently of the fact that ${\cal{SUSY}}$ Higgs bosons
could have already been found either at LEP or LHC, the CERN
``$\gamma p$ machine'' would have
a clear importance on its own, since the fundamental interactions would
take place here via $\gamma$--quark and $\gamma$--gluon scatterings,
these proceeding via a large number of \mssm\ vertices, which can then be
tested. Photons, in fact, directly couple at leading order
to the \mssm\ (charged) Higgs scalars,
whereas electrons/positrons don't (because of the negligible mass of the
electron in the Yukawa couplings).
Therefore, at the NLC, very few Higgs
production mechanisms and a reduced number of
fundamental vertices are involved.\par
Both the high $ep$ energy available at the LEP$\otimes$LHC and the properties
of the back--scattered photons
would make the production of
Higgs events with high rates possible.
Moreover, the absence of strong interactions
from the initial state, which take place at hadron colliders
(via $q\bar q$, $gg$ and $qg$ scatterings), would make the CERN
machine the first TeV environment partially free from the huge QCD
noise typical of the LHC (and of the Tevatron as well).
Finally, both the technology and the
expenses needed in
converting two machines already existing (such as LEP and LHC) and
physically located in the same place (even though maybe not at the
same time) has to be considered, compared to  building
a new one (the NLC): this could make conceivable to expect
the CERN $ep$ accelerator to be operating
well in advance of any
future linear collider.\par
For obvious reasons of space (in reducing the huge amount of
material to a size compatible with a journal publication)
and time (in numerically computing cross sections and distributions
of both signals and backgrounds),
we concentrated only on a limited region of the \mssm\ parameter
space $(\MA,\tb)$. Because of kinematical constrains imposed by the
collider energy and luminosity, we studied Higgs scalars in the intermediate
mass interval
whereas, as example of two opposite situations, we chose two values
at the extremes of the available range of \tb\ (that is, 1.5 and 30).
Our work turns out to be incomplete then.
However, as the discussion of the results has been carried on by
stressing their
dependence on the masses and on the couplings of the
\mssm\ Higgs scalars, we expect our analysis
to be easily translatable to the case
in which different values of \MA\ and/or \tb\ are adopted.
Some remarks are also in order concerning the treatment of the signals, of the
backgrounds and the approach to the kinematical cuts.\par
On the one hand, we assumed a $100\%$ $b$--tagging efficiency, thus neglecting
 considering light quark and gluon jets faking $b$'s in the Higgs
decays $\Phi^0\ar b\bar b$. This is obviously unrealistic but, by the time
of the advent of the LEP$\otimes$LHC, $b$--tagging performances should
be very high, and not too far from the above ideal case.
In addition, signals and discrete backgrounds involving Higgses,
$Z^0$ and $W^\pm$ decaying into $b\bar b$ and $\tau\nu_\tau$
(the signatures of neutral and charged Higgs scalars we have studied
here, respectively) have been computed keeping the bosons
on--shell, and considering the invariant mass of their decay products to
fill a region of only 10 GeV around the corresponding peak.
Such an approximation should be clearly dropped in the end, in order to
predict reliable numbers. However, as we clearly identified as
regions of feasible detection of the \mssm\ Higgs
particles especially the ones well far from the $Z^0$-- and
$W^\pm$--resonances, we expect the inclusion of the tails of the Breit--Wigner
distributions not to substantially modify our conclusions.
In fact, most of the cases in which Higgs and gauge boson peaks
overlap seem to be already out of
the experimental possibilities in the on--shell approximation.\par
On the other hand, a full analysis (including kinematical cuts,
detector efficiencies, hadronization effects, etc ...) was
far beyond our intentions, mainly because a
detailed simulation should necessarily rely on the precise knowledge of
the characteristics of the LEP$\otimes$LHC detectors, which
we cannot have at the moment.
In this respect, a possible way to proceed could well have been, for example,
the one of taking the details needed for this study
from the recent ATLAS \cite{ATLAS} and CMS \cite{CMS}
Technical Proposals for the LHC, which are probably the most complete
and up to date source of useful information.
Nevertheless, we expect that by the time
the CERN $ep$ collider will be operating, both the improvement
in the techniques and the necessity to
adjust the detectors in view of their best performances at a
different kind of machine ($ep$ instead of $pp$), could end up
indicating event selection criteria different from the ones
we could suppose now. What we instead preferred to do here
was to take, as example,
a similar study we performed in the case of the \sm\ in a previous
paper, in order  to
show how in general kinematical cuts should have a decisive
impact on the signal significancies only where these are very small, thus
affecting only restricted regions of the \mssm\ parameter space here
considered. Leaving practically intact in the rest of the cases the
chances of Higgs detections and studies.
\par
Under such premises, we demonstrated the high potential of the
LEP$\otimes$LHC. What we obtained is that in some parts of the
parameter region we studied all the \mssm\ Higgs bosons could be
contemporaneously detected (especially if a high luminosity can be achieved).
However, where this does not happen, at least
two of them are accessible to the experiment and it is never
the case that none of the
 the Higgs scalars can be recognised.
For all the neutral bosons ($H^0,h^0$ and $A^0$) we considered the
$b\bar b$ decay channel, whereas for the charged Higgses ($H^\pm$'s) we
studied the decay mode $\tau\nu_\tau$. The signatures we assumed
are in $b\bar b b\bar b$, $jjb\bar b$, $\tau\nu_\tau b\bar b$, $\tau\nu_\tau
\tau\nu_\tau$ and $jj\tau\nu_\tau$ events.
In detail, the most favourable cases are the following.\par
For the heavy scalar $H^0$ good chances of detection happen when
\tbb\ in the case of the $4b$--signature, via the production subprocess
$g\gamma\ar b\bar bH^0$, if $\MA\OOrd120$ GeV ($\MH\OOrd130$ GeV).
The remaining mass range $\MA\Ord120$ GeV ($\MH\Ord130$ GeV) is quite
difficult to cover, as the only possible way would be via the signature
$jjb\bar b$, through the production processes $q\gamma\ar
q'W^\pm H^0$ and $q\gamma\ar qZ^0H^0$ (the first one mostly), which
have large production rates but small significancies (large continuum
backgrounds). A high luminosity option would be needed in this case
(say, tens of inverse femtobarns per year) to clearly extract $H^0$--signals.
If \tba\ the situation is even less optimistic. Only after a few years
of running at the standard luminosity ${\cal{L}}=3$ fb$^{-1}$ it
should be possible to recognize a few $W^\pm H^0$ events, and these
would not be probably enough for attempting a statistically significant
study. Therefore, we would conclude that for the \mssm\
neutral heavy scalar the parameter region at small \tb\ would remain
practically uncovered, whereas the one at large \tb's should be
accessible by the experiment if $\MH\OOrd130$ GeV (for a standard
${\cal L}$).\par
The light neutral Higgs $h^0$, even with its reduced mass if
compared to \MH, has definitely much more chances to be detected.
The production in events $W^\pm h^0 X$ (giving the signature $jjb\bar
b$) is quite large if \tba. As $M_{Z^0}-\Mh\OOrd10$ GeV over all the
interval 60 GeV $\Ord\MA\Ord$ 140 GeV and the cut in $M_{bW\ar bjj}$
can be successfully exploited in rejecting $tbX$ and $t\bar tX$
events, $h^0$--signals should be
disentangled from the backgrounds up to the maximum value of
$\Mh\approx81$ GeV. A few units of events per year in the above
signature would come from
$Z^0h^0X$ production too. The cases of $H^\pm h^0X$--, $h^0A^0X$-- and
$q\bar qh^0X$--production do not deserve much attention (very small
cross sections). If \tbb, good candidates are $W^\pm h^0X$ events, provided
that $\Mh\OOrd120$ GeV ($\MA\OOrd120$ GeV). The most probable signature
would be again $jjb\bar b$. The case $Z^0h^0X$ at \tbb\ is completely
beyond any experimental possibility. Production events of the type
$H^\pm h^0$ and $h^0A^0X$ contribute by adding some more chances of
$h^0$--detection if \tbb\ only in the case $\Mh\approx\MA\approx60$
GeV (via the signatures $\tau\nu_\tau b\bar b$ and $b\bar b b\bar b$,
respectively). The case where the rewards for $h^0$--detection at high
\tb's are largest is probably via the  subprocess $g\gamma\ar b\bar
bh^0$, if $\Mh\approx\MA\Ord120$ GeV. The production rates are in fact
extremely large  and the $4b$--signature is clean from
backgrounds, provided that high $b$--tagging performances can be
achieved and $M_{h^0}$ is far enough from $M_{Z^0}$.
Therefore, for the \mssm\ neutral light scalar, we conclude that both
the regions \tba\ and 30 are adequately covered, and $h^0$--signals
are observable.\par
The pseudoscalar Higgs $A^0$ is practically uncovered if \tba\ and
$\MA\OOrd80$ GeV. In fact, a few chances at small \tb's occur only when
$\MA\approx60$ GeV, via the signature $\tau\nu_\tau b\bar b$ in $H^\pm
A^0 X$ events and only after a few years of running. The large \tb\ region
case is instead entirely
covered via the $g\gamma\ar b\bar b A^0$ production
mechanism. Even in the case that the final efficiencies
and purities in $b$--tagging are smaller than the ones expected now,
the large production rates should guarantee the
detection of the $A^0$ in the $4b$--mode. In the case of the \mssm\
neutral pseudoscalar Higgs then, only
the large \tb\ region is fully covered, whereas the remaining one is really
difficult, as even the most favourable case $\MA\approx60$ GeV needs a
lot of integrated luminosity.\par
Finally, the case of the charged Higgses. Both single and double
$H^\pm$--production give account of large production rates, at both
\tb's. As
$M_{H^\pm}-M_{W^\pm}\OOrd10$ GeV, backgrounds should be manageable.
Therefore, we
expect that in the intermediate mass range of the $A^0$ (which
correspond to the values 100 GeV $\Ord\MHpm\Ord$ 160 GeV) charged
Higgses should be recognised and detected, both at small and at large
values of \tb.\par
In conclusion then, although we recognise that
a more complete study (especially involving a coverage of the whole \mssm\
parameter space)
 is needed, together with a more refined signal
versus background analysis (once the performances expected from the detectors
of the LEP$\otimes$LHC will become clear), we stress that the possibilities
of the proposed CERN $ep$ collider in testing the Higgs sector of the
\mssm\ are encouraging indeed, with the machine operating
in the $\gamma p$ mode. Therefore, such a project should be
seriously kept into consideration, especially if LEP and LHC together
will not be able to confirm or rule out with certainty the \mssm.

\subsection*{Acknowledgments}

We are grateful to J.B.~Tausk for interesting discussions
and useful suggestions during the early stages of this work, to
T.~Stelzer
for his helpful advice in using MadGraph and, finally, to
W.J.~Stirling for reading the manuscript.

\subsection*{Appendix}

In this section we give the explicit formulae for the helicity amplitudes
of the processes we have studied. Definitions of $S$, $Y$ and $Z$ functions
and of other quantities ($p$, $\lambda$, $\mu$,
$\eta$, etc...), which enter in the following,
can be found in ref.~\cite{ioPR}, with identical notations.\par
Here, we introduce the definitions
\begin{equation}
-b_1=-b_2=b_3=2b_4=2b_5=2b_6=2b_7=1
\end{equation}
for the coefficients of the incoming/outgoing four--momenta,
\begin{equation}
D_{V}(p)={1\over {p^2-M_V^2}},
\quad\quad
D_{q}(p)={1\over {p^2-m_q^2}}
\end{equation}
for the propagators, where $V=W^\pm,H^\pm,Z^0$ or $\gamma$ and $q=u$ or $d$,
\begin{equation}
N_i=[4(p_i\cdot q_i)]^{-1/2},\quad\quad\quad i=1,2
\end{equation}
for the gluon ($i=1$) and the photon ($i=2$) normalisation factor,
where $p_i$($q_i$) is the massless vector four--momentum(any
four--vector not proportional
to $p_i$), with $i=1,2$ \cite{ks}.
The symbols $r_1$ and $r_2$ represent two light--like four--momenta
satisfying the relations
\begin{equation}
r_1^2=r_2^2=0,\quad\quad\quad r_1^\mu+r_2^\mu=p^\mu_4,
\end{equation}
($d\Omega_{r_1(r_2)}$ indicates the solid angle of $r_{1(2)}$ in
the rest frame of $p_4$)
\cite{ks}, $p_6$ and $p_7$ are antispinor four--momenta such that
\begin{equation}
p_6^\mu\equiv p_4^\mu, \quad\quad\quad p_7^\mu\equiv p_5^\mu,
\end{equation}
and
\begin{equation}
\sum_{\lambda=\pm} u(\li)\bar u(\li)=
{p\Dir}_i - m_i,\quad\quad{\rm {with}}~i=6,7,
\end{equation}
while
\begin{equation}
\sum_{\lambda=\pm} u(\li)\bar u(\li)=
{p\Dir}_i + m_i,\quad\quad{\rm {with}}~i=4,5.
\end{equation}
We also define the mass relation
\begin{eqnarray}
\Delta^{V}_{\Phi,\Phi'}& &=\frac{M_{\Phi}^2-M_{\Phi'}^2}{M_{V}^2}
\quad~{\rm if~}\quad \Phi\ne\Phi'~(M_V\ne0), \\ \nonumber
                       & &=0\hskip2.3cm{\rm if~}\quad \Phi=\Phi',
\end{eqnarray}
the additional coefficients
$$
c^V_{\Phi,\Phi';i}=(1-\Delta^{V}_{\Phi,\Phi'})b_i,\quad\quad
{\rm for}~i=4~{\rm and}~6,
$$
\be
c^V_{\Phi,\Phi';i}=-(1+\Delta^{V}_{\Phi,\Phi'})b_i,\quad\quad
{\rm for}~i=5~{\rm and}~7,
\ene
where $V$ and $\Phi,\Phi'$ represent vector and Higgs bosons, respectively,
and the spinor functions\footnote{We adopt the symbol ${\{\lambda\}}$
to
denote a set
of helicities of all external particles in a given reaction,
$\sum_{\{\lambda\}}$ to indicate the usual sum over all their
possible combinations, and the symbol $\sum_{i=j,k,l,...}$
to indicate a sum over $j,k,l,...$ with index $i$.}
$$
{\X}_2=\sum_{\lambda=\pm}\sum_{i=1,3}(-b_i)
Y([2];[i];1,1)
Y([i];(2);1,1),
$$
$$
{\X}_4=\sum_{\lambda=\pm}\sum_{i=5,7}b_i
Y(\{1\};[i];1,1)
Y([i];\{2\};1,1),
$$
$$
{\X}^{qV(')}_{31}=
\sum_{\lambda=\pm}\sum_{i=4,6(5,7)}b_i
Y([3];[i];1,1)
Y([i];[1];c^q_{R_V},c^q_{L_V}),
$$
$$
{\Y}^{(')}_2=\sum_{\lambda=\pm}
\sum_{i=4,6(5,7)}b_i
Y([2];[i];1,1)
Y([i];(2);1,1),
$$
$$
{\Y}_4=\sum_{\lambda=\pm}
Y(\{1\};p_2,\lambda;1,1)
Y(p_2,\lambda;\{2\};1,1),
$$
$$
{\cal F}^{qV}_{31}=
\mu_1\eta_1 Y([3];[1];c^q_{L_V},c^q_{R_V})
-\mu_3\eta_3 Y([3];[1];c^q_{R_V},c^q_{L_V}),
$$
$$
{\Y}^{qV}_{31}=\sum_{\lambda=\pm}
Y([3];p_2,\lambda;1,1)
Y(p_2,\lambda;[1];c^q_{R_V},c^q_{L_V}),
$$
$$
{\tilde{\Y}}^{qV}_{31}={\Y}^{qV}_{31}
-\frac{{\cal F}^{qV}_{31}}{M_V^2}p_2\cdot(p_4+p_5),
$$
$$
{\Z}_{24}=Z([2];(2);\{1\};\{2\};1,1;1,1),
$$
$$
{\Z}_{312}^{qV}=Z([3];[1];
[2];(2);c^q_{R_V},c^q_{L_V};1,1),
$$
$$
{\tilde{\Z}}_{312}^{qV}={\Z}_{312}^{qV}-\frac{{\cal F}^{qV}_{31}}{M_V^2}
({\Y}_2+{\Y}_2'),
$$
$$
{\Z}_{314}^{qV}=Z([3];[1];
\{1\};\{2\};c^q_{R_V},c^q_{L_V};1,1),
$$
\be
{\tilde{\Z}}_{314}^{qV}={\Z}_{314}^{qV}-\frac{{\cal F}^{qV}_{31}}{M_V^2}
({\X}_4-{\Y}_4),
\ene
where $V$ represents a gauge boson $W^\pm$, $Z^0$ or $\gamma$, $q=u$ or
$d$ ($u$-- and $d$--type quarks of arbitrary masses $m_u$ and $m_d$,
respectively), and with the short--hand notations
$[x]=p_{x},\lambda_{x}$ $(x=1,...4)$, $(x)=q_x,\lambda_x$ $(x=1,2)$
and $\{x\}=r_{x},-$ $(x=1,2)$.\par
In the following we adopt $[i]=p_i,\lambda$ and $[j]=p_j,\lambda'$, whereas
the couplings $c_R$, $c_L$ and ${\cal H}$ can be easily
deduced from tabs.~A.I--A.IV. Also, we sometimes make use of the equalities
\be
\Y_2+\Y_2'=\X_2,\quad\quad\quad
\X_{31}^{qV}+\X_{31}^{qV'}=\Y_{31}^{qV}+\F_{31}^{qV}.
\ene
\vskip 1.0cm
\centerline{\sl 1. Process $d\gamma\rightarrow uW^-\Phi^0$.}
\vskip 0.5cm
In order to obtain from fig.~1 the Feynman graphs of  the process
\begin{equation}
d(p_1,\lambda_1) + \gamma (p_2,\lambda_2)
\longrightarrow u (p_3,\lambda_3) + W^- (p_4) + \Phi^0 (p_5),
\end{equation}
where $\Phi^0=H^0,h^0$ or $A^0$, one has to make the following
assignments:
\be\label{ass:WH}
q=d,\quad\quad q'=u,\quad\quad V^{(*)}=W^{\pm(*)},\quad\quad S^*=H^{\pm*}.
\ene
The corresponding matrix element, summed over final spins and averaged
over
initial ones,
is given by
\begin{equation}\label{M2:dph->uWH}
{\left|{\overline M}\right|}=
{e^6\over 4}
N_2^2{3\over {8\pi M^2_{W^\pm}}}
\sum_{\{\lambda\}}
\int d\Omega_{r_1(r_2)}
\sum_{l,m=1}^{17}{T}_l^{\{\lambda\}} T_m^{\{\lambda\}*},
\end{equation}
where
$${\rm i}T_1^{\{\lambda\}}=
D_{u}(p_3+p_5)D_d(p_1+p_2)M_1^{\{\lambda\}}{\cal H}_1,\quad\quad
{\rm i}T_2^{\{\lambda\}}=
D_{d}(p_3+p_4)D_d(p_1+p_2)M_2^{\{\lambda\}}{\cal H}_2,$$
$${\rm i}T_3^{\{\lambda\}}=
D_{d}(p_3+p_4)D_d(p_1-p_5)M_3^{\{\lambda\}}{\cal H}_3,\quad\quad
{\rm i}T_4^{\{\lambda\}}=
D_{u}(p_3+p_5)D_u(p_1-p_4)M_4^{\{\lambda\}}{\cal H}_4,$$
$${\rm i}T_5^{\{\lambda\}}=
D_{u}(p_3-p_2)D_u(p_1-p_4)M_5^{\{\lambda\}}{\cal H}_5,\quad\quad
{\rm i}T_6^{\{\lambda\}}=
D_{u}(p_3-p_2)D_d(p_1-p_5)M_6^{\{\lambda\}}{\cal H}_6,$$
$${\rm i}T_7^{\{\lambda\}}=
D_{W^\pm}(p_4+p_5)D_d(p_1+p_2)M_7^{\{\lambda\}}{\cal H}_7,\quad\quad
{\rm i}T_8^{\{\lambda\}}=
D_{W^\pm}(p_4+p_5)D_u(p_3-p_2)M_8^{\{\lambda\}}{\cal H}_8,$$
$${\rm i}T_9^{\{\lambda\}}=
D_{H^\pm}(p_4+p_5)D_d(p_1+p_2)M_9^{\{\lambda\}}{\cal H}_9,\quad\quad
{\rm i}T_{10}^{\{\lambda\}}=
D_{H^\pm}(p_4+p_5)D_u(p_3-p_2)M_{10}^{\{\lambda\}}{\cal H}_{10},$$
$${\rm i}T_{11}^{\{\lambda\}}=
D_{W^\pm}(p_2-p_4)D_{u}(p_3+p_5)M_{11}^{\{\lambda\}}{\cal H}_{11},\quad\quad
{\rm i}T_{12}^{\{\lambda\}}=
D_{W^\pm}(p_2-p_4)D_{d}(p_1-p_5)M_{12}^{\{\lambda\}}{\cal H}_{12},$$
$${\rm i}T_{13}^{\{\lambda\}}=
D_{W^\pm}(p_1-p_3)D_{W^\pm}(p_4+p_5)M_{13}^{\{\lambda\}}{\cal H}_{13},
\quad\quad
{\rm i}T_{14}^{\{\lambda\}}=
D_{W^\pm}(p_1-p_3)D_{W^\pm}(p_2-p_4)M_{14}^{\{\lambda\}}{\cal H}_{14},$$
$${\rm i}T_{15}^{\{\lambda\}}=
D_{H^\pm}(p_1-p_3)D_{H^\pm}(p_4+p_5)M_{15}^{\{\lambda\}}{\cal H}_{15},
\quad\quad
{\rm i}T_{16}^{\{\lambda\}}=
D_{H^\pm}(p_1-p_3)D_{W^\pm}(p_2-p_4)M_{16}^{\{\lambda\}}{\cal H}_{16},$$
\be\label{T:dph->uWH}
{\rm i}T_{17}^{\{\lambda\}}=
D_{H^\pm}(p_1-p_3)M_{17}^{\{\lambda\}}{\cal H}_{17}.
\ene
We have
$$\hskip-2.0cm
M_{1}^{\{\lambda\}}=
\sum_{\lambda=\pm}\sum_{\lambda'=\pm}
\sum_{i=3,5,7}\sum_{j=1,2}(-b_ib_j)
Y([3];[i];c^{u}_{R_{\Phi^0}},c^{u}_{L_{\Phi^0}})
$$
$$\hskip1.5cm\times
Z([i];[j];
\{1\};\{2\};c_{R_{W^\pm}},c_{L_{W^\pm}};1,1)
Z([j];[1];
[2];(2);c^{d}_{R_{\gamma}},c^{d}_{L_{\gamma}};1,1),
$$
$$\hskip-1.0cm
M_{2}^{\{\lambda\}}=
\sum_{\lambda=\pm}\sum_{\lambda'=\pm}
\sum_{i=3,4,6}\sum_{j=1,2}(-b_ib_j)
Z([3];[i];
\{1\};\{2\};c_{R_{W^\pm}},c_{L_{W^\pm}};1,1)
$$
$$\hskip1.5cm\times
Y([i];[j];c^{d}_{R_{\Phi^0}},c^{d}_{L_{\Phi^0}})
Z([j];[1];
[2];(2);c^d_{R_{\gamma}},c^d_{L_{\gamma}};1,1),
$$
$$\hskip-1.0cm
M_{3}^{\{\lambda\}}=
\sum_{\lambda=\pm}\sum_{\lambda'=\pm}
\sum_{i=3,4,6}\sum_{j=1,5}(-b_ib_j)
Z([3];[i];
\{1\};\{2\};c_{R_{W^\pm}},c_{L_{W^\pm}};1,1)
$$
$$\hskip1.5cm\times
Z([i];[j];
[2];(2);c^d_{R_{\gamma}},c^d_{L_{\gamma}};1,1)
Y([j];[1];c^d_{R_{\Phi^0}},c^d_{L_{\Phi^0}}),
$$
$$\hskip-2.0cm
M_{4}^{\{\lambda\}}=
\sum_{\lambda=\pm}\sum_{\lambda'=\pm}
\sum_{i=3,5,7}\sum_{j=1,4,6}(-b_ib_j)
Y([3];[i];c^u_{R_{\Phi^0}},c^u_{L_{\Phi^0}})
$$
$$\hskip1.5cm\times
Z([i];[j];
[2];(2);c^u_{R_{\gamma}},c^u_{L_{\gamma}};1,1)
Z([j];[1];
\{1\};\{2\};c_{R_{W^\pm}},c_{L_{W^\pm}};1,1),
$$
$$\hskip-2.0cm
M_{5}^{\{\lambda\}}=
\sum_{\lambda=\pm}\sum_{\lambda'=\pm}
\sum_{i=3,2}\sum_{j=1,4,6}(-b_ib_j)
Z([3];[i];
[2];(2);c^u_{R_{\gamma}},c^u_{L_{\gamma}};1,1)
$$
$$\hskip1.5cm\times
Y([i];[j];c^u_{R_{\Phi^0}},c^u_{L_{\Phi^0}})
Z([j];[1];
\{1\};\{2\};c_{R_{W^\pm}},c_{L_{W^\pm}};1,1),
$$
$$\hskip-2.0cm
M_{6}^{\{\lambda\}}=
\sum_{\lambda=\pm}\sum_{\lambda'=\pm}
\sum_{i=3,2}\sum_{j=1,5,7}(-b_ib_j)
Z([3];[i];
[2];(2);c^u_{R_{\gamma}},c^u_{L_{\gamma}};1,1)
$$
$$\hskip1.5cm\times
Z([i];[j];
\{1\};\{2\};c_{R_{W^\pm}},c_{L_{W^\pm}};1,1)
Y([j];[1];c^d_{R_{\Phi^0}},c^d_{L_{\Phi^0}}),
$$
$$\hskip-2.5cm
M_7^{\{\lambda\}}=\sum_{\lambda=\pm}\sum_{i=1,2}
(-b_i)Z([i];[1];[2];(2);c^d_{R_\gamma},c^d_{L_\gamma};1,1)
$$
$$
\times
\{Z([3];[i];\{1\};\{2\};c_{R_{W^\pm}},c_{L_{W^\pm}};1,1)
$$
$$\hskip1.5cm
-{{{\X}_4}\over{M_{W^\pm}^2}}
[\sum_{\lambda'=\pm}\sum_{j=1,2,3}(-b_j)
Y([3];[j];1,1)Y([j];[i];c_{R_{W^\pm}},c_{L_{W^\pm}})]\},
$$
$$\hskip-2.5cm
M_8^{\{\lambda\}}=\sum_{\lambda=\pm}\sum_{i=2,3}
(b_i)Z([3];[i];[2];(2);c^u_{R_\gamma},c^u_{L_\gamma};1,1)
$$
$$
\times
\{Z([i];[1];\{1\};\{2\};c_{R_{W^\pm}},c_{L_{W^\pm}};1,1)
$$
$$\hskip1.5cm
-{{{\X}_4}\over{M_{W^\pm}^2}}
[\sum_{\lambda'=\pm}\sum_{j=1,2,3}(-b_j)
Y([i];[j];1,1)Y([j];[1];c_{R_{W^\pm}},c_{L_{W^\pm}})]\},
$$
$$
M_9^{\{\lambda\}}=\sum_{\lambda=\pm}\sum_{i=1,2}
(-b_i)(-2{\X}_4)
Y([3];[i];c_{R_{H^\pm}},c_{L_{H^\pm}})
Z([i];[1];[2];(2);c^d_{R_\gamma},c^d_{L_\gamma};1,1),
$$
$$
M_{10}^{\{\lambda\}}=\sum_{\lambda=\pm}\sum_{i=2,3}
(b_i)(-2{\X}_4)
Z([3];[i];[2];(2);c^u_{R_\gamma},c^u_{L_\gamma};1,1)
Y([i];[1];c_{R_{H^\pm}},c_{L_{H^\pm}}),
$$
$$\hskip-2.5cm
M_{11}^{\{\lambda\}}=\sum_{\lambda=\pm}\sum_{i=3,5,7}(2b_i)
Y([3];[i];c^u_{R_{\Phi^0}},c^u_{L_{\Phi^0}})
$$
$$\times
[\Z_{24}\sum_{\lambda'=\pm}
Y([i];p_2,\lambda';1,1)Y(p_2,\lambda';[1];c_{R_{W^\pm}},c_{L_{W^\pm}})
$$
$$\hskip.5cm
-\Y_2Z(\{1\};\{2\};[i];[1];1,1;c_{R_{W^\pm}},c_{L_{W^\pm}})
-\Y_4Z([2];(2);[i];[1];1,1;c_{R_{W^\pm}},c_{L_{W^\pm}})],
$$
$$\hskip-2.5cm
M_{12}^{\{\lambda\}}=\sum_{\lambda=\pm}\sum_{i=1,5,7}(-2b_i)
Y([i];[1];c^d_{R_{\Phi^0}},c^d_{L_{\Phi^0}})
$$
$$\times
[\Z_{24}\sum_{\lambda'=\pm}
Y([3];p_2,\lambda';1,1)Y(p_2,\lambda';[i];c_{R_{W^\pm}},c_{L_{W^\pm}})
$$
$$\hskip.5cm
-\Y_2Z(\{1\};\{2\};[3];[i];1,1;c_{R_{W^\pm}},c_{L_{W^\pm}})
-\Y_4Z([2];(2);[3];[i];1,1;c_{R_{W^\pm}},c_{L_{W^\pm}})],
$$
$$\hskip-2.5cm
M_{13}^{\{\lambda\}}=\Z_{24}(\F_{31}^{W^{\pm}}+2\Y_{31}^{W^{\pm}})
-2\X_2\tilde{\Z}_{314}^{W^{\pm}}-(2\Y_4-\X_4)\tilde{\Z}_{312}^{W^{\pm}}
$$
$$
-{1\over M_{W^\pm}^2}
[\X_2\X_4(\Y_{31}^{W^{\pm}}-\X_{31}^{W^{\pm}}-\X_{31}^{W^{\pm{'}}})
+(p_1-p_3)^2(\Z_{24}\F_{31}^{W^{\pm}}+\tilde{\Z}_{312}^{W^{\pm}}\X_4)
+2p_2\cdot(p_1-p_3)\Z_{24}\F_{31}^{W^{\pm}}]
$$
$$\hskip1.5cm
-{1\over M_{W^\pm}^4}
\{[(p_1-p_3)^2+p_2\cdot(p_1-p_3)]
\X_4(\Y_2-\Y_2')\F_{31}^{W^{\pm}}\},
$$
$$
M_{14}^{\{\lambda\}}=2({\tilde{\Y}}^{W^{\pm}}_{31}{\Z}_{24}-
{\Y}_{2}{\tilde{\Z}}^{W^{\pm}}_{314}-{\Y}_{4}{\tilde{\Z}}^{W^{\pm}}_{312}),
$$
$$
M_{15}^{\{\lambda\}}=4Y([3];[1];c_{R_{H^\pm}},c_{L_{H^\pm}})
\X_4(\Y_2+\Y_2'),
$$
$$
M_{16}^{\{\lambda\}}=-2Y([3];[1];c_{R_{H^\pm}},c_{L_{H^\pm}})
[\Z_{24}p_2\cdot(p_4+2p_5)-2\Y_2'\Y_4-2\Y_2\X_4],
$$
\be\label{M:dph->uWH}
M_{17}^{\{\lambda\}}=Y([3];[1];c_{R_{H^\pm}},c_{L_{H^\pm}})
\Z_{24}.
\ene
\vskip 1.0cm
\centerline{\sl 2. Process $d\gamma\rightarrow dZ^0\Phi^0$.}
\vskip 0.5cm
The Feynman graphs of  the process
\begin{equation}
d(p_1,\lambda_1) + \gamma (p_2,\lambda_2)
\longrightarrow d (p_3,\lambda_3) + Z^0 (p_4) + \Phi^0 (p_5),
\end{equation}
where $\Phi^0=H^0,h^0$ or $A^0$, can be obtained from fig.~1 by setting
$$
q=q'=d,\quad\quad\quad V^{(*)}=Z^{0(*)},
$$
$$
M_{i}^{\{\lambda\}}=0,\quad\quad i=11,...17,
$$
\begin{eqnarray}\label{ass:ZH}
(S^*,\Phi^0)& &=(A^{0*},H^0),\\ \nonumber
            & &=(A^{0*},h^0),\\ \nonumber
            & &=(H^{0*}+h^{0*},A^0).
\end{eqnarray}
The formulae for the amplitude squared corresponding to
$\Phi^0=H^0$ and $h^0$ are practically the same as in
the previous section, with the relabeling:
\be\label{ch:dph->dZH}
u\ar d,\quad\quad W^\pm\ar Z^0,\quad\quad H^\pm\ar A^0,
\ene
in eqs.~(\ref{M2:dph->uWH})--(\ref{M:dph->uWH}).
For the case $\Phi^0=A^0$, the same relabeling still hold in
eq.~(\ref{M2:dph->uWH}), whereas in eqs.(\ref{T:dph->uWH})--(\ref{M:dph->uWH})
only for $i=1,...8$.
For diagrams $9$ and $10$, one has to introduce in eqs.~(\ref{T:dph->uWH})
$$
{\rm i}T_9^{\{\lambda\}}=
D_d(p_1+p_2)(D_{H^0}(p_4+p_5)M_{9,H^0}^{\{\lambda\}}{\cal H}_{9,H^0}
            +D_{h^0}(p_4+p_5)M_{9,h^0}^{\{\lambda\}}{\cal H}_{9,H^0}),
$$
\be\label{M910}
{\rm i}T_{10}^{\{\lambda\}}=
D_d(p_3-p_2)(D_{H^0}(p_4+p_5)M_{10,H^0}^{\{\lambda\}}{\cal H}_{10,H^0}
            +D_{h^0}(p_4+p_5)M_{10,h^0}^{\{\lambda\}}{\cal H}_{10,h^0}),
\ene
with
$M_{i,S}^{\{\lambda\}}$, for $i=9,10$ and $S=H^0,h^0$, as given in
eqs.~(\ref{M:dph->uWH}) with the exchanges (\ref{ch:dph->dZH}), where
$A^0\ar H^0,h^0$.
\vskip 1.0cm
\centerline{\sl 3. Process $d\gamma\rightarrow uH^-\Phi^0$.}
\vskip 0.5cm
The Feynman diagrams of  the process
\begin{equation}
d(p_1,\lambda_1) + \gamma (p_2,\lambda_2)
\longrightarrow u (p_3,\lambda_3) + H^- (p_4) + \Phi^0 (p_5),
\end{equation}
where $\Phi^0=H^0,h^0$ or $A^0$, are depicted in fig.~2,
with the assignments:
\be\label{ass:HH1}
q=d,\quad q'=u,\quad S=H^-, \quad(\Phi,\Phi')=(H^-,\Phi^0),
\quad V^{*}=W^{\pm*},\quad
S^*(S^{*'})=H^{\pm*}(H^{\mp*}).
\ene
The amplitude squared, summed over final spins and averaged over initial ones,
is given by
\begin{equation}\label{M2:dph->uHH1}
{\left|{\overline M}\right|}=
{e^6\over 4}
N_2^2
\sum_{\{\lambda\}}
\sum_{l,m=1}^{17}{T}_l^{\{\lambda\}} T_m^{\{\lambda\}*},
\end{equation}
where the $T_i^{\{\lambda\}}$'s, for $i=1,...10$, are the same as in
eqs.~(\ref{T:dph->uWH}) except for a difference in sign, whereas
$${\rm -i}T_{11}^{\{\lambda\}}=
D_{H^\pm}(p_2-p_4)D_{u}(p_3+p_5)M_{11}^{\{\lambda\}}{\cal H}_{11},\quad
{\rm -i}T_{12}^{\{\lambda\}}=
D_{H^\pm}(p_2-p_4)D_{d}(p_1-p_5)M_{12}^{\{\lambda\}}{\cal H}_{12},$$
$${\rm -i}T_{13}^{\{\lambda\}}=
D_{H^\pm}(p_1-p_3)D_{H^\pm}(p_4+p_5)M_{13}^{\{\lambda\}}{\cal H}_{13},\quad
{\rm -i}T_{14}^{\{\lambda\}}=
D_{H^\pm}(p_1-p_3)D_{H^\pm}(p_2-p_4)M_{14}^{\{\lambda\}}{\cal H}_{14},$$
$${\rm -i}T_{15}^{\{\lambda\}}=
D_{W^\pm}(p_1-p_3)D_{W^\pm}(p_4+p_5)M_{15}^{\{\lambda\}}{\cal H}_{15},
{}~{\rm -i}T_{16}^{\{\lambda\}}=
D_{W^\pm}(p_1-p_3)D_{H^\pm}(p_2-p_4)M_{16}^{\{\lambda\}}{\cal H}_{16},$$
\be\label{T:dph->uHH1}
{\rm -i}T_{17}^{\{\lambda\}}=
D_{W^\pm}(p_1-p_3)M_{17}^{\{\lambda\}}{\cal H}_{17}.
\ene
The spinor amplitudes are
$$\hskip-2.0cm
M_{1}^{\{\lambda\}}=
\sum_{\lambda=\pm}\sum_{\lambda'=\pm}
\sum_{i=3,5,7}\sum_{j=1,2}(-b_ib_j)
Y([3];[i];c^{u}_{R_{\Phi^0}},c^{u}_{L_{\Phi^0}})
$$
$$\hskip2.0cm\times
Y([i];[j];c_{R_{H^\pm}},c_{L_{H^\pm}})
Z([j];[1];[2];(2);c^{d}_{R_{\gamma}},c^{d}_{L_{\gamma}};1,1),
$$
$$\hskip-2.0cm
M_{2}^{\{\lambda\}}=
\sum_{\lambda=\pm}\sum_{\lambda'=\pm}
\sum_{i=3,4,6}\sum_{j=1,2}(-b_ib_j)
Y([3];[i];c_{R_{H^\pm}},c_{L_{H^\pm}})
$$
$$\hskip2.0cm\times
Y([i];[j];c^{d}_{R_{\Phi^0}},c^{d}_{L_{\Phi^0}})
Z([j];[1];[2];(2);c^d_{R_{\gamma}},c^d_{L_{\gamma}};1,1),
$$
$$\hskip-2.0cm
M_{3}^{\{\lambda\}}=
\sum_{\lambda=\pm}\sum_{\lambda'=\pm}
\sum_{i=3,4,6}\sum_{j=1,5}(-b_ib_j)
Y([3];[i];c_{R_{H^\pm}},c_{L_{H^\pm}})
$$
$$\hskip2.0cm\times
Z([i];[j];[2];(2);c^d_{R_{\gamma}},c^d_{L_{\gamma}};1,1)
Y([j];[1];c^d_{R_{\Phi^0}},c^d_{L_{\Phi^0}}),
$$
$$\hskip-2.0cm
M_{4}^{\{\lambda\}}=
\sum_{\lambda=\pm}\sum_{\lambda'=\pm}
\sum_{i=3,5,7}\sum_{j=1,4,6}(-b_ib_j)
Y([3];[i];c^u_{R_{\Phi^0}},c^u_{L_{\Phi^0}})
$$
$$\hskip2.0cm\times
Z([i];[j];[2];(2);c^u_{R_{\gamma}},c^u_{L_{\gamma}};1,1)
Y([j];[1];c_{R_{H^\pm}},c_{L_{H^\pm}}),
$$
$$\hskip-1.5cm
M_{5}^{\{\lambda\}}=
\sum_{\lambda=\pm}\sum_{\lambda'=\pm}
\sum_{i=3,2}\sum_{j=1,4,6}(-b_ib_j)
Z([3];[i];[2];(2);c^u_{R_{\gamma}},c^u_{L_{\gamma}};1,1)
$$
$$\hskip2.0cm\times
Y([i];[j];c^u_{R_{\Phi^0}},c^u_{L_{\Phi^0}})
Y([j];[1];c_{R_{H^\pm}},c_{L_{H^\pm}};1,1),
$$
$$\hskip-1.5cm
M_{6}^{\{\lambda\}}=
\sum_{\lambda=\pm}\sum_{\lambda'=\pm}
\sum_{i=3,2}\sum_{j=1,5,7}(-b_ib_j)
Z([3];[i];[2];(2);c^u_{R_{\gamma}},c^u_{L_{\gamma}};1,1)
$$
$$\hskip2.0cm\times
Y([i];[j];c_{R_{H^\pm}},c_{L_{H^\pm}})
Y([j];[1];c^d_{R_{\Phi^0}},c^d_{L_{\Phi^0}}),
$$
$$\hskip-1.0cm
M_7^{\{\lambda\}}=\sum_{\lambda=\pm}\sum_{\lambda'=\pm}
\sum_{i=4,5,6,7}\sum_{j=1,2}
(-c^{W^\pm}_{H^-,\Phi^0;i}b_j)
Y([3];[i];1,1)
Y([i];[j],c_{R_{W^\pm}},c_{L_{W^\pm}})
$$
$$
\hskip3.0cm\times
Z([j];[1];[2];(2);c^d_{R_\gamma},c^d_{L_\gamma};1,1),
$$
$$\hskip-1.5cm
M_8^{\{\lambda\}}=\sum_{\lambda=\pm}\sum_{\lambda'=\pm}
\sum_{i=2,3}\sum_{j=4,5,6,7}
(b_ic^{W^\pm}_{H^-,\Phi^0;j})
Z([3];[i];[2];(2);c^u_{R_\gamma},c^u_{L_\gamma};1,1)
$$
$$
\hskip2.0cm\times
Y([i];[j];1,1)
Y([j];[1];c_{R_{W^\pm}},c_{L_{W^\pm}}),
$$
$$\hskip-1.0cm
M_9^{\{\lambda\}}=\sum_{\lambda=\pm}\sum_{i=1,2}
(-b_i)
Y([3];[i];c_{R_{H^\pm}},c_{L_{H^\pm}})
Z([i];[1];[2];(2);c^d_{R_\gamma},c^d_{L_\gamma};1,1),
$$
$$
M_{10}^{\{\lambda\}}=\sum_{\lambda=\pm}\sum_{i=2,3}
(b_i)
Z([3];[i];[2];(2);c^u_{R_\gamma},c^u_{L_\gamma};1,1)
Y([i];[1];c_{R_{H^\pm}},c_{L_{H^\pm}}),
$$
$$\hskip-2.0cm
M_{11}^{\{\lambda\}}=
\sum_{\lambda=\pm}\sum_{i=3,5,7}(b_i)(-2\Y_2)
Y([3];[i];c^u_{R_{\Phi^0}},c^u_{L_{\Phi^0}})
Y([i];[1];c_{R_{H^\pm}},c_{L_{H^\pm}}),
$$
$$\hskip-2.0cm
M_{12}^{\{\lambda\}}=
\sum_{\lambda=\pm}\sum_{i=1,5,7}(-b_i)(-2\Y_2)
Y([3];[i];c_{R_{H^\pm}},c_{L_{H^\pm}})
Y([i];[1];c^d_{R_{\Phi^0}},c^d_{L_{\Phi^0}}),
$$
$$
M_{13}^{\{\lambda\}}=
(-2\X_2)[2Y([3];[1];c_{R_{H^\pm}},c_{L_{H^\pm}})],
$$
$$
M_{14}^{\{\lambda\}}=
(-2\Y_2)[2Y([3];[1];c_{R_{H^\pm}},c_{L_{H^\pm}})],
$$
$$\hskip-2.0cm
M_{15}^{\{\lambda\}}=
[c^{W^\pm}_{H^-\Phi^0;4}\Y_2+c^{W^\pm}_{H^-\Phi^0;5}\Y_2']
[2\tilde{\Y}_{31}^{W^\pm}+(1-\frac{(p_1-p_3)^2}{M_{W^\pm}^2})
{\F}_{31}^{W^\pm}]
$$
$$
-2\X_2\{[{\X}_{31}^{W^\pm}
      -\frac{{\F}_{31}^{W^\pm}}{M_{W^\pm}^2}p_4\cdot(p_1-p_3)]
      c^{W^\pm}_{H^-\Phi^0;4}
      +[{\X}_{31}^{W^\pm{'}}
      -\frac{{\F}_{31}^{W^\pm}}{M_{W^\pm}^2}p_5\cdot(p_1-p_3)]
      c^{W^\pm}_{H^-\Phi^0;5}\}
$$
$$\hskip2.0cm
+(p_4+p_5-2p_2)\cdot(c^{W^\pm}_{H^-\Phi^0;4}p_4+c^{W^\pm}_{H^-\Phi^0;5}p_5)
\tilde{\Z}_{312}^{W^\pm},
$$
$$
M_{16}^{\{\lambda\}}=
(-2\Y_2)
[\F_{31}^{W^\pm}-2\X_{31}^{W^\pm{'}}
-{(p_1-p_3)\cdot(p_1-p_3-2p_5)\over M_{W^\pm}^2}\F_{31}^{W^\pm}],
$$
\be\label{M:dph->uHH1}
M_{17}^{\{\lambda\}}=
{\tilde{\Z}}_{312}^{W^\pm}
-{\F_{31}^{W^\pm}\X_2\over M_{W^\pm}^2}.
\ene
\vskip 1.0cm
\centerline{\sl 4. Process $d\gamma\rightarrow d\Phi^0\Phi^{0{'}}$.}
\vskip 0.5cm
The Feynman diagrams which describe the reaction
\begin{equation}
d(p_1,\lambda_1) + \gamma (p_2,\lambda_2)
\longrightarrow d (p_3,\lambda_3) + \Phi^0 (p_4) + \Phi^{0{'}} (p_5),
\end{equation}
where $\Phi^0,\Phi^{0{'}}=H^0,h^0$ or $A^0$, are reported in fig.~2,
where
$$
q=q'=d,\quad\quad (\Phi,\Phi')=(\Phi^0,\Phi^{0{'}}),
\quad\quad V^{*}=Z^{0*},
$$
with
$$
M_{i}^{\{\lambda\}}=0,\quad\quad i=11,...17,
$$
and the combinations
\begin{eqnarray}\label{ass:HH2}
(S^*,\Phi^0,\Phi^{0{'}})& &= (H^{0*}+h^{0*},H^{0},H^{0}), \\ \nonumber
                        & &= (H^{0*}+h^{0*},H^{0},h^{0}), \\ \nonumber
                        & &= (A^{0*},H^{0},A^{0}),        \\ \nonumber
                        & &= (H^{0*}+h^{0*},h^{0},h^{0}), \\ \nonumber
                        & &= (A^{0*},h^{0},A^{0}),        \\ \nonumber
                        & &= (H^{0*}+h^{0*},A^{0},A^{0}).
\end{eqnarray}
The amplitude squared is given by a formula identical to
eq.~(\ref{M2:dph->uHH1}).
The expressions for the spinor functions and the propagators
are the same as in eqs.~(\ref{M:dph->uHH1})
for the combinations $(A^{0*},H^0,A^0)$ and $(A^{0*},h^0,A^0)$, after
the exchanges:
\be\label{ch:dph->uHH2}
u\ar d,\quad\quad W^\pm\ar Z^0,\quad\quad H^\pm\ar A^0.
\ene
For the cases in eq.~(\ref{ass:HH2}) with double--flavoured Higgs
propagators, eqs.~(\ref{T:dph->uHH1})--(\ref{M:dph->uHH1}) hold
for the indices $i=1,...8$, while
for diagrams $9$ and $10$, one has to introduce the same equations as in
(\ref{M910}) and the same
$M_{i,S}^{\{\lambda\}}$'s, for $i=9,10$ and $S=H^0,h^0$, as given in
eqs.~(\ref{M:dph->uHH1}) with the exchanges (\ref{ch:dph->uHH2}), where
$A^0\ar H^0,h^0$.
\vskip 1.0cm
\centerline{\sl 5. Process $d\gamma\rightarrow dH^-H^+$.}
\vskip 0.5cm
The Feynman diagrams for
\begin{equation}
d(p_1,\lambda_1) + \gamma (p_2,\lambda_2)
\longrightarrow d (p_3,\lambda_3) + H^- (p_4) + H^+ (p_5),
\end{equation}
are again displayed in fig.~2, where now
$$
q=q'=d,\quad (\Phi,\Phi')=(H^-,H^+),\quad
\quad V^{*}=\gamma^*+Z^{0*},\quad S^*=H^0+h^0,\quad S^{*{'}}=H^{\pm{*}},
$$
\be
M_{i}^{\{\lambda\}}=0,\quad\quad {\rm for}~i=2,3,6,
\ene
and where, moreover, one has to exchange $\Phi\leftrightarrow \Phi'$
in diagram $12$ and replace diagrams $13$ and $15$,
by $14$ and $16$, respectively, but with $\Phi\leftrightarrow \Phi'$.
The matrix element is given by the formula
(\ref{M2:dph->uHH1}), with propagators
as those  obtained for the case $(\Phi,\Phi')=(H^-,\Phi^0)$, except
for
\be\label{ch1:dph->dHH3}
{\rm i}T_5^{\{\lambda\}}=
D_{d}(p_3-p_2)D_u(p_1-p_4)M_5^{\{\lambda\}},
\ene
and with spinor functions as in eqs.~(\ref{M:dph->uHH1}), for $i=1,4,5$,
where
$$
(c^q_{R_{\Phi^0}},c^q_{L_{\Phi^0}})\ar (c_{L_{H^\pm}},c_{R_{H^\pm}}),
\quad\quad q=u,d,
$$
\be\label{ch2:dph->dHH3}
(c^u_{R_{\gamma}},c^u_{L_{\gamma}})\ar (c^d_{R_{\gamma}},c^d_{L_{\gamma}}),
\quad\quad\quad {\rm in}~M_5^{\{\lambda\}}.
\ene
We give explicitly the remaining $T_i^{\{\lambda\}}$'s
and $M_i^{\{\lambda\}}$'s, for $i=7,...17$. They are
$${\rm i}T_7^{\{\lambda\}}=
D_d(p_1+p_2)(D_\gamma(p_4+p_5)M_{7,\gamma}^{\{\lambda\}}{\cal H}_{7,\gamma}
            +D_{Z^0}(p_4+p_5)M_{7,Z^0}^{\{\lambda\}}{\cal H}_{7,Z^0}),
$$
$${\rm i}T_8^{\{\lambda\}}=
D_d(p_3-p_2)(D_\gamma(p_4+p_5)M_{8,\gamma}^{\{\lambda\}}{\cal H}_{8,\gamma}
            +D_{Z^0}(p_4+p_5)M_{8,Z^0}^{\{\lambda\}}{\cal H}_{8,Z^0}),
$$
$$
{\rm i}T_9^{\{\lambda\}}=
D_d(p_1+p_2)(D_{H^0}(p_4+p_5)M_{9,H^0}^{\{\lambda\}}{\cal H}_{9,H^0}
            +D_{h^0}(p_4+p_5)M_{9,h^0}^{\{\lambda\}}{\cal H}_{9,H^0}),
$$
$$
{\rm i}T_{10}^{\{\lambda\}}=
D_d(p_3-p_2)(D_{H^0}(p_4+p_5)M_{10,H^0}^{\{\lambda\}}{\cal H}_{10,H^0}
            +D_{h^0}(p_4+p_5)M_{10,h^0}^{\{\lambda\}}{\cal H}_{10,h^0}),
$$
$${\rm i}T_{11}^{\{\lambda\}}=-
D_{H^\pm}(p_2-p_4)D_{u}(p_3+p_5)M_{11}^{\{\lambda\}}{\cal H}_{11},\quad\quad
{\rm i}T_{12}^{\{\lambda\}}=
D_{H^\pm}(p_2-p_5)D_{d}(p_1-p_4)M_{12}^{\{\lambda\}}{\cal H}_{12},$$
$$
{\rm i}T_{13}^{\{\lambda\}}=-
D_{H^\pm}(p_2-p_5)(D_{H^0}(p_1-p_3)M_{13,H^0}^{\{\lambda\}}{\cal H}_{13,H^0}
                  +D_{h^0}(p_1-p_3)M_{13,h^0}^{\{\lambda\}}{\cal H}_{13,H^0}),
$$
$$
{\rm i}T_{14}^{\{\lambda\}}=
D_{H^\pm}(p_2-p_4)(D_{H^0}(p_1-p_3)M_{14,H^0}^{\{\lambda\}}{\cal H}_{14,H^0}
                  +D_{h^0}(p_1-p_3)M_{14,h^0}^{\{\lambda\}}{\cal H}_{14,H^0}),
$$
$$
{\rm i}T_{15}^{\{\lambda\}}=-
D_{H^\pm}(p_2-p_5)(D_\gamma(p_1-p_3)M_{15,\gamma}^{\{\lambda\}}
{\cal H}_{15,\gamma}
                  +D_{Z^0}(p_1-p_3)M_{15,Z^0}^{\{\lambda\}}{\cal H}_{15,Z^0}),
$$
$$
{\rm i}T_{16}^{\{\lambda\}}=
D_{H^\pm}(p_2-p_4)(D_\gamma(p_1-p_3)M_{16,\gamma}^{\{\lambda\}}
{\cal H}_{16,\gamma}
                  +D_{Z^0}(p_1-p_3)M_{16,Z^0}^{\{\lambda\}}{\cal H}_{16,Z^0}),
$$
\be\label{T:dph->dHH3}
{\rm i}T_{17}^{\{\lambda\}}=
	           D_\gamma(p_1-p_3)M_{17,\gamma}^{\{\lambda\}}
{\cal H}_{17,\gamma}
                  +D_{Z^0}(p_1-p_3)M_{17,Z^0}^{\{\lambda\}}{\cal H}_{17,Z^0},
\ene
and
$$\hskip-1.0cm
M_{7,V}^{\{\lambda\}}=\sum_{\lambda=\pm}\sum_{\lambda'=\pm}
\sum_{i=4,5,6,7}\sum_{j=1,2}
(-c^{V}_{H^-,H^+;i}b_j)
Y([3];[i];1,1)
Y([i];[j];c^d_{R_{V}},c^d_{L_{V}})
$$
$$\hskip2.0cm
\times
Z([j];[1];[2];(2);c^d_{R_\gamma},c^d_{L_\gamma};1,1),
$$
$$\hskip-1.0cm
M_{8,V}^{\{\lambda\}}=\sum_{\lambda=\pm}\sum_{\lambda'=\pm}
\sum_{i=2,3}\sum_{j=4,5,6,7}
(b_ic^{V}_{H^-,H^+;j})Z([3];[i];[2];(2);c^d_{R_\gamma},c^d_{L_\gamma};1,1)
$$
$$\hskip2.0cm
\times
Y([i];[j];1,1)
Y([j];[1];c^d_{R_{V}},c^d_{L_{V}}),
$$
$$\hskip-1.0cm
M_{9,S}^{\{\lambda\}}=\sum_{\lambda=\pm}\sum_{i=1,2}
(-b_i)
Y([3];[i];c^d_{R_{S}},c^d_{L_{S}})
Z([i];[1];[2];(2);c^d_{R_\gamma},c^d_{L_\gamma};1,1),
$$
$$
M_{10,S}^{\{\lambda\}}=\sum_{\lambda=\pm}\sum_{i=2,3}
(b_i)
Z([3];[i];[2];(2);c^u_{R_\gamma},c^u_{L_\gamma};1,1)
Y([i];[1];c^d_{R_{S}},c^d_{L_{S}}),
$$
$$\hskip-2.0cm
M_{11}^{\{\lambda\}}=
\sum_{\lambda=\pm}\sum_{i=3,5,7}(b_i)(-2\Y_2)
Y([3];[i];c_{L_{H^\pm}},c_{R_{H^\pm}})
Y([i];[1];c_{R_{H^\pm}},c_{L_{H^\pm}}),
$$
$$\hskip-2.0cm
M_{12}^{\{\lambda\}}=
\sum_{\lambda=\pm}\sum_{i=1,4,6}(-b_i)(-2\Y_2')
Y([3];[i];c_{L_{H^\pm}},c_{R_{H^\pm}})
Y([i];[1];c_{R_{H^\pm}},c_{L_{H^\pm}}),
$$
$$
M_{13}^{\{\lambda\}}=
(-2\Y_2')[Y([3];[1];c^d_{R_{S}},c^d_{L_{S}})],
$$
$$
M_{14}^{\{\lambda\}}=
(-2\Y_2)[Y([3];[1];c^d_{R_{S}},c^d_{L_{S}})],
$$
$$
M_{15,\gamma}^{\{\lambda\}}=
(-2\Y_2')
[\F_{31}^{\gamma}-2\X_{31}^{\gamma}],
$$
$$
M_{15,Z^0}^{\{\lambda\}}=
(-2\Y_2')
[\F_{31}^{Z^0}-2\X_{31}^{Z^0}
-{(p_1-p_3)\cdot(p_1-p_3-2p_4)\over M_{Z^0}^2}\F_{31}^{Z^0}],
$$
$$
M_{16,\gamma}^{\{\lambda\}}=
(-2\Y_2)
[\F_{31}^{\gamma}-2\X_{31}^{\gamma{'}}],
$$
$$
M_{16,Z^0}^{\{\lambda\}}=
(-2\Y_2)
[\F_{31}^{Z^0}-2\X_{31}^{Z^0{'}}
-{(p_1-p_3)\cdot(p_1-p_3-2p_5)\over M_{Z^0}^2}\F_{31}^{Z^0}],
$$
$$
M_{17,\gamma}^{\{\lambda\}}=
{\tilde{\Z}}_{312}^{d\gamma},
$$
\be\label{M:dph->uHH3}
M_{17,Z^0}^{\{\lambda\}}=
{\tilde{\Z}}_{312}^{d Z^0}
-{\F_{31}^{Z^0}\X_2\over M_{Z^0}^2}.
\ene
\vskip 1.0cm
\centerline{\sl 6. Process $g\gamma\rightarrow u\bar dH^-$.}
\vskip 0.5cm
The Feynman diagrams for
\begin{equation}
g(p_1,\lambda_1) + \gamma (p_2,\lambda_2)
\longrightarrow u (p_3,\lambda_3) + \bar d (p_4,\lambda_4) + H^- (p_5),
\end{equation}
are shown in fig.~3, where
\be\label{ass:gph->udH}
q=u\quad\quad q'=d,\quad \Phi=H^-,\quad
\quad S^{*}=H^{\pm{*}},
\ene
The amplitude squared is
\begin{equation}\label{M2:gph->udH}
{\left|{\overline M}\right|}=
{e^4g_s^2\over 4}
N_1^2N_2^2
\sum_{\{\lambda\}}
\sum_{l,m=1}^{8}{T}_l^{\{\lambda\}} T_m^{\{\lambda\}*}.
\end{equation}
The expressions for the $T_i^{\{\lambda\}}$ 's are
$${\rm -i}T_1^{\{\lambda\}}=
D_d(p_3+p_5)D_d(p_1-p_4)M_{1}^{\{\lambda\}}{\cal H}_1,\quad\quad
{\rm -i}T_2^{\{\lambda\}}=
D_u(p_3-p_2)D_d(p_1-p_4)M_{2}^{\{\lambda\}}{\cal H}_2,$$
$${\rm -i}T_3^{\{\lambda\}}=
D_u(p_3-p_2)D_u(p_4+p_5)M_{3}^{\{\lambda\}}{\cal H}_3,$$
$${\rm -i}T_{i+3}^{\{\lambda\}}={\rm -i}T_i^{\{\lambda\}}
(u\leftrightarrow d;p_3\leftrightarrow p_4),
\quad\quad\quad i=1,...3,$$
\be\label{T:gph->udH}
{\rm -i}T_7^{\{\lambda\}}=
D_{H^\pm}(p_2-p_5)D_d(p_1-p_4)M_{7}^{\{\lambda\}}{\cal H}_7,\quad\quad
{\rm -i}T_8^{\{\lambda\}}={\rm -i}T_7^{\{\lambda\}}
(u\leftrightarrow d;p_3\leftrightarrow p_4),
\ene
while the spinor functions are
$$\hskip-2.0cm
M_{1}^{\{\lambda\}}=
\sum_{\lambda=\pm}\sum_{\lambda'=\pm}
\sum_{i=3,5,7}\sum_{j=1,4}(-b_ib_j)
Y([3];[i];c_{R_{H^\pm}},c_{L_{H^\pm}})
$$
$$\hskip1.5cm\times
Z([i];[j];[2];(2);c^{d}_{R_{\gamma}},c^{d}_{L_{\gamma}};1,1)
Z([j];[4];[1];(1);c^{d}_{R_{g}},c^{d}_{L_{g}};1,1),
$$
$$\hskip-2.0cm
M_{2}^{\{\lambda\}}=
\sum_{\lambda=\pm}\sum_{\lambda'=\pm}
\sum_{i=2,3}\sum_{j=1,4}(-b_ib_j)
Z([3];[i];[2];(2);c^{u}_{R_{\gamma}},c^{u}_{L_{\gamma}};1,1)
$$
$$\hskip1.5cm\times
Y([i];[j];c_{R_{H^\pm}},c_{L_{H^\pm}})
Z([j];[4];[1];(1);c^{d}_{R_{g}},c^{d}_{L_{g}};1,1),
$$
$$\hskip-2.0cm
M_{3}^{\{\lambda\}}=
\sum_{\lambda=\pm}\sum_{\lambda'=\pm}
\sum_{i=2,3}\sum_{j=4,5,7}(-b_ib_j)
Z([3];[i];[2];(2);c^{u}_{R_{\gamma}},c^{u}_{L_{\gamma}};1,1)
$$
$$\hskip1.5cm\times
Z([i];[j];[1];(1);c^{u}_{R_{g}},c^{u}_{L_{g}};1,1)
Y([j];[4];c_{R_{H^\pm}},c_{L_{H^\pm}}),
$$
$$
M_{i+3}^{\{\lambda\}}=M_{i}^{\{\lambda\}}
(u\leftrightarrow d;p_3\leftrightarrow p_4),
\quad\quad\quad i=1,...3,
$$
$$\hskip-2.0cm
M_{7}^{\{\lambda\}}=
\sum_{\lambda=\pm}\sum_{\lambda'=\pm}
\sum_{i=1,4}(-b_i)(-2{\Y_2}')
Y([3];[i];c_{R_{H^\pm}},c_{L_{H^\pm}})
Z([i];[4];[1];(1);c^{d}_{R_{g}},c^{d}_{L_{g}};1,1)
$$
\be\label{M:gph->udH}
M_{8}^{\{\lambda\}}=-M_{7}^{\{\lambda\}}
(u\leftrightarrow d;p_3\leftrightarrow p_4).
\ene
\vskip 1.0cm
\centerline{\sl 7. Process $g\gamma\rightarrow u\bar u\Phi^0$.}
\vskip 0.5cm
The Feynman diagrams for
\begin{equation}
g(p_1,\lambda_1) + \gamma (p_2,\lambda_2)
\longrightarrow u (p_3,\lambda_3) + \bar u (p_4,\lambda_4) + \Phi^0 (p_5),
\end{equation}
with $\Phi^0=H^0,h^0$ or $A^0$, can be obtained from fig.~3 by
$$
q=q'=u,\quad\quad \Phi=\Phi^0,
$$
\be\label{ass:gph->uuH}
M_{i}^{\{\lambda\}}=0,\quad i=7,8.
\ene
With the exchanges
\be\label{ch:gph->uuH}
d\ar u,\quad\quad H^\pm\ar \Phi^0, \quad\quad
(c_{R_{H^\pm}},c_{L_{H^\pm}})\ar(c^u_{R_{\Phi^0}},c^u_{L_{\Phi^0}}),
\ene
in eqs.~(\ref{T:gph->udH})--(\ref{M:gph->udH}), the expressions for
$T_{i}^{\{\lambda\}}$ and $M_{i}^{\{\lambda\}}$ ($i=1,...6$) can be
easily obtained, while eq.~(\ref{M2:gph->udH}) remains the same.
\vskip0.5cm
By trivial relabeling and sign exchanges, it is possible
to obtain from the above formulae the corresponding
ones for the $u$--type quark initiated processes
$$u\gamma\rightarrow dW^+\Phi^0,$$
$$u\gamma\rightarrow uZ^0\Phi^0,$$
$$u\gamma\rightarrow dH^+\Phi^0,$$
$$u\gamma\rightarrow u\Phi^0\Phi^{0'},$$
\be
u\gamma\rightarrow uH^+H^-,
\ene
as for the charge conjugate reactions
$$\bar d\gamma\rightarrow \bar uW^+\Phi^0,$$
$$\bar d\gamma\rightarrow \bar dZ^0\Phi^0,$$
$$\bar d\gamma\rightarrow \bar uH^+\Phi^0,$$
$$\bar d\gamma\rightarrow \bar d\Phi^0\Phi^{0'},$$
\be
\bar d\gamma\rightarrow \bar dH^+H^-,
\ene
and
$$\bar u\gamma\rightarrow \bar dW^-\Phi^0,$$
$$\bar u\gamma\rightarrow \bar uZ^0\Phi^0,$$
$$\bar u\gamma\rightarrow \bar dH^-\Phi^0,$$
$$\bar u\gamma\rightarrow \bar u\Phi^0\Phi^{0'},$$
\be
\bar u\gamma\rightarrow \bar uH^-H^+.
\ene
Finally, the same it can be done for obtaining the helicity amplitudes
for the $g$--initiated processes
$$g\gamma\rightarrow d\bar u H^+,$$
\be
g\gamma\rightarrow d\bar d \Phi^0.
\ene

\newpage
\subsection*{Table Captions}
\begin{description}
\item[table I    ] Cross sections of the processes
$q\gamma\ar q'W^\pm\Phi^0$, where $\Phi^0=H^0,h^0,A^0$,
at $\sqrt s_{ep}=1.36$ TeV, for $M_{A^0}=60,80,100,120,140$ GeV, with
$\tan\beta=1.5$ (a) and 30 (b). The MRS(A) structure functions are used.
The errors are the statistical errors on the numerical calculation.
Entries are in GeV for masses, and in fb for cross sections.

\item[table II   ] Cross sections of the processes
$q\gamma\ar q Z^0\Phi^0$, where $\Phi^0=H^0,h^0,A^0$,
at $\sqrt s_{ep}=1.36$ TeV, for $M_{A^0}=60,80,100,120,140$ GeV, with
$\tan\beta=1.5$ (a) and 30 (b). The MRS(A) structure functions are used.
The errors are the statistical errors on the numerical calculation.
Entries are in GeV for masses, and in fb for cross sections.

\item[table III  ] Cross sections of the processes
$q\gamma\ar q'H^\pm\Phi^0$, where $\Phi^0=H^0,h^0,A^0$,
at $\sqrt s_{ep}=1.36$ TeV, for $M_{A^0}=60,80,100,120,140$ GeV, with
$\tan\beta=1.5$ (a) and 30 (b). The MRS(A) structure functions are used.
The errors are the statistical errors on the numerical calculation.
Entries are in GeV for masses, and in fb for cross sections.

\item[table IV   ] Cross sections of the processes
$q\gamma\ar q \Phi^0\Phi^{0{'}}$, where
$(\Phi^0,\Phi^{0{'}})=(H^0,A^0),(h^0,A^0)$,
at $\sqrt s_{ep}=1.36$ TeV, for $M_{A^0}=60,80,100,120,140$ GeV, with
$\tan\beta=1.5$ (a) and 30 (b). The MRS(A) structure functions are used.
The errors are the statistical errors on the numerical calculation.
Entries are in GeV for masses, and in fb for cross sections.

\item[table V    ] Cross sections of the process
$q\gamma\ar qH^+ H^-$,
at $\sqrt s_{ep}=1.36$ TeV, for $M_{A^0}=60,80,100,120,140$ GeV, with
$\tan\beta=1.5$ (a) and 30 (b). The MRS(A) structure functions are used.
The errors are the statistical errors on the numerical calculation.
Entries are in GeV for masses, and in fb for cross sections.

\item[table VI   ] Cross sections of the process
$g\gamma\ar q\bar q'H^\pm$,
at $\sqrt s_{ep}=1.36$ TeV, for $M_{A^0}=60,80,100,120,140$ GeV, with
$\tan\beta=1.5$ (a) and 30 (b). The MRS(A) structure functions are used.
The errors are the statistical errors on the numerical calculation.
Entries are in GeV for masses, and in fb for cross sections.

\item[table VII  ] Cross sections of the processes
$g\gamma\ar q\bar q\Phi^0$, where $\Phi^0=H^0,h^0,A^0$,
at $\sqrt s_{ep}=1.36$ TeV, for $M_{A^0}=60,80,100,120,140$ GeV, with
$\tan\beta=1.5$ (a) and 30 (b). The MRS(A) structure functions are used.
The errors are the statistical errors on the numerical calculation.
Entries are in GeV for masses, and in fb for cross sections.

\item[table VIII ] Production cross sections for the discrete
and continuum background processes
discussed in the text. Case (a) contains the cross sections which do not have
dependence on the \mssm\ parameters, whereas (b) shows the case in
which resonant $t$--quarks introduce such a dependence through
$\Gamma_t^\mssm$. In (b) the five entries for each process
correspond to the five different values of $\MA=60,80,100,120$
and 140 GeV. Numbers in brackets are for the case \tbb.
The MRS(A) structure functions are used.
The errors are the statistical errors on the numerical calculation.
Entries are in GeV for masses, and in fb for cross sections.

\item[table IX   ] Total top width and BRs of the decay channels
$t\ar bW^\pm$ and $t\ar bH^\pm$ within the \mssm, for \tba\ and 30,
for the different values of $M_{H^\pm}$ corresponding to
$\MA=60,80,100,120$
and 140 GeV.
The total top width in the \sm\ is $\Gamma_t^{\sm}\approx1.57$ GeV.
Entries are in GeV both for masses and widths.

\item[table A.I  ] Neutral ${\cal {MSSM}}$ Higgs boson couplings each other,
to the gauge bosons $W^\pm$, $Z^0$ and $\gamma$,
and to the \mssm\ $H^\pm$'s.

\item[table A.II ] Charged ${\cal {MSSM}}$ Higgs boson couplings
to the gauge bosons $Z^0$ and $\gamma$ (here $c_{2W}\equiv\cos2\theta_W$).

\item[table A.III] \mssm\ right and left handed couplings $(c_R,c_L)$
of $u$-- (upper line) and $d$--type  (lower line) quarks to the neutral
gauge bosons $g$, $\gamma$, $Z^0$
and to the neutral \mssm\ Higgses $H^0$, $h^0$, $A^0$.
We have $g_R^q=-Q^qs^2_W$ and $g_L^q=T^q_3-Q^qs^2_W$ ($q=u,d$), with
$(Q^u, T^u)=(+{2\over 3}, {1\over 2})$ and
$(Q^d, T^d_3)=(-{1\over 3}, -{1\over 2})$ for quark
charges and isospins.

\item[table A.IV ] \mssm\ right and left handed couplings $(c_R,c_L)$
of quarks to the charged gauge bosons $W^\pm$ and
to the charged \mssm\ Higgses $H^\pm$.

\end{description}

\vspace*{\fill}

\subsection*{Figure Captions}
\begin{description}
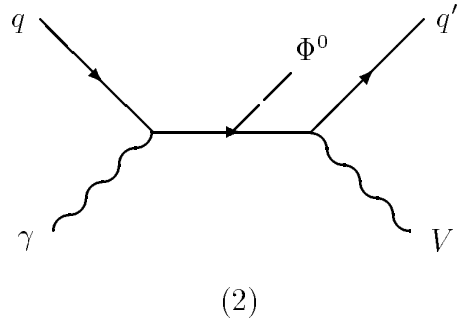
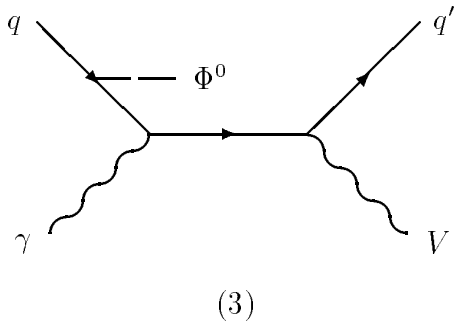
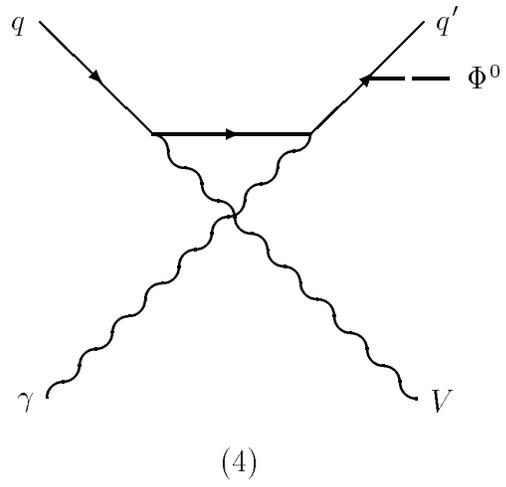
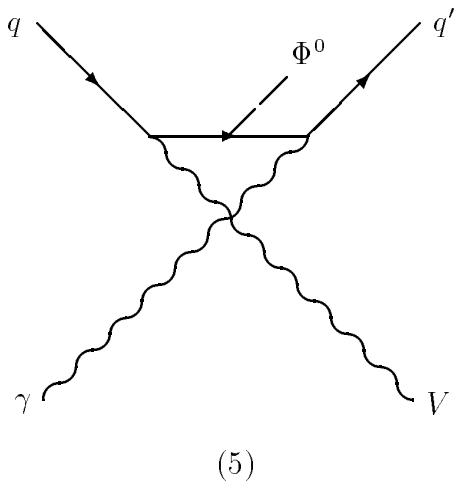
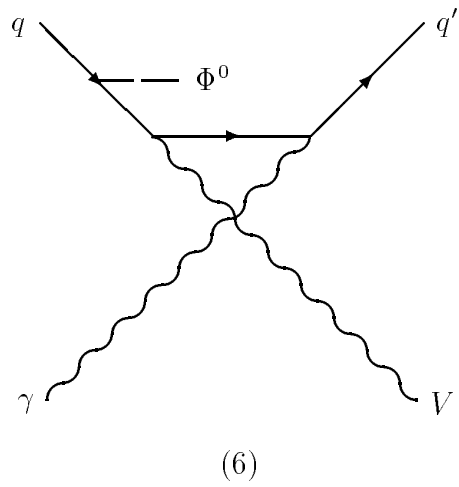
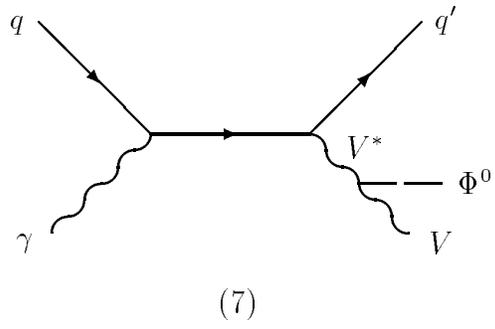
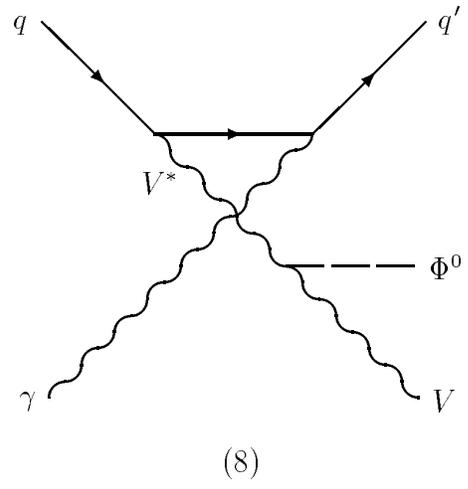
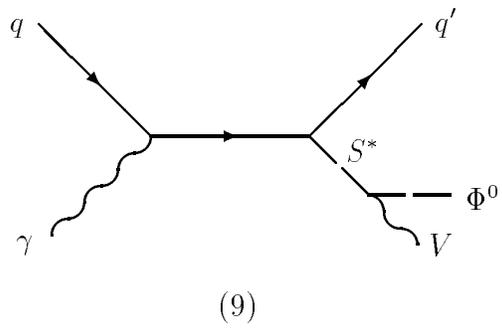
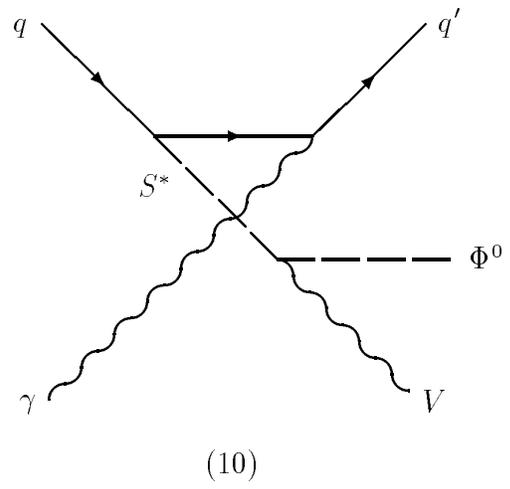
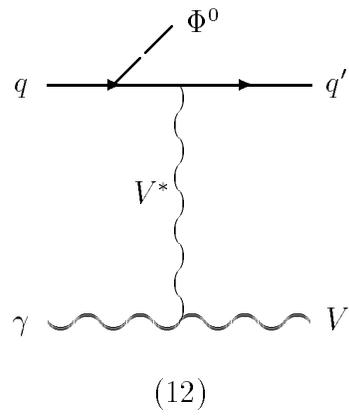
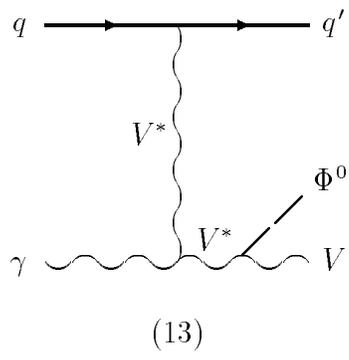
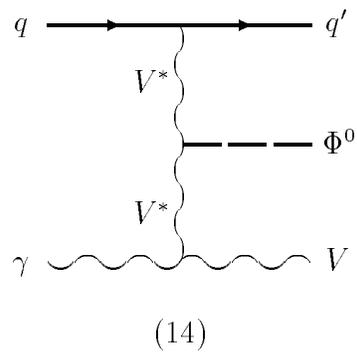
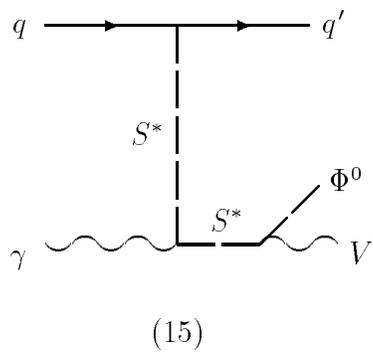
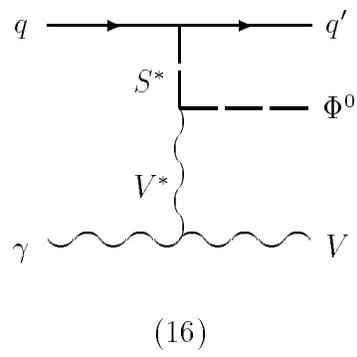
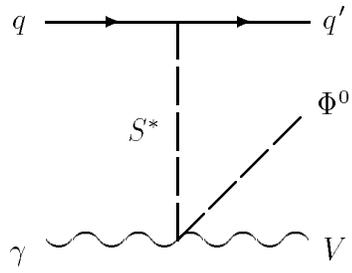
\item[fig.~1 ] Feynman diagrams contributing in the lowest order to
$q\gamma\rightarrow q'V\Phi^0$, where $q(q')$ represents a quark,
$V(V^*)$ an external(internal) vector boson,
$S^*$ an internal scalar Higgs boson and $\Phi^0$ one of
the neutral \mssm\ Higgses, in the unitary gauge.
For the possible combinations of ($q,q',V,V^*,S^*,\Phi^0$) and the
corresponding non--vanishing graphs, see the text.
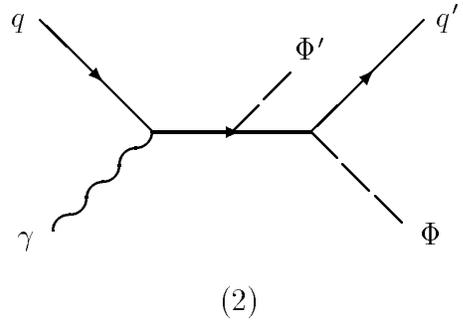
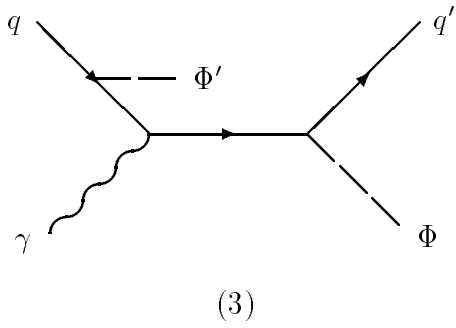
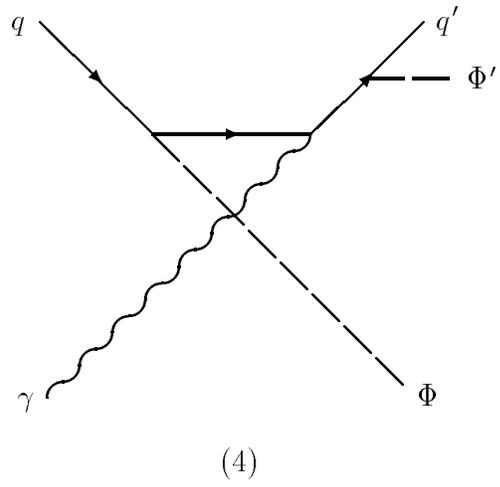
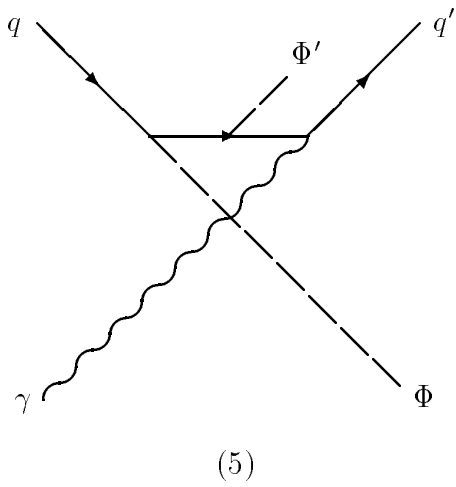
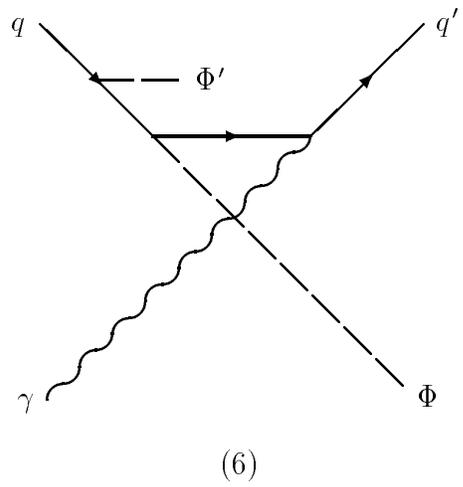
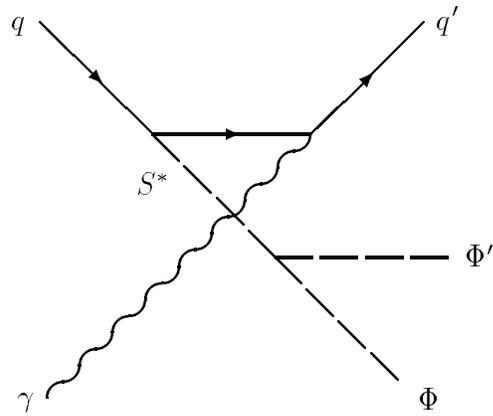
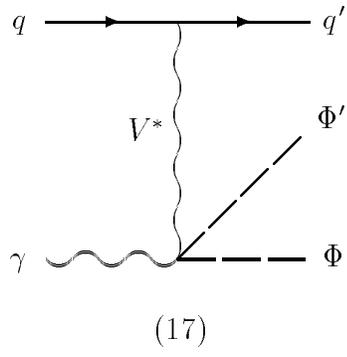
\item[fig.~2 ] Feynman diagrams contributing in the lowest order to
$q\gamma\rightarrow q'\Phi\Phi'$, where $q(q')$ represents a quark,
$V^*$ an internal vector boson,
$S^*$ and $S^{*'}$ internal scalar Higgs bosons and $\Phi$ and
$\Phi'$ both neutral and charged \mssm\ Higgses, in the unitary gauge.
For the possible combinations of ($q,q',V^*,S^*,S^{*'},\Phi,\Phi'$)
and the corresponding non--vanishing  graphs, see the text.
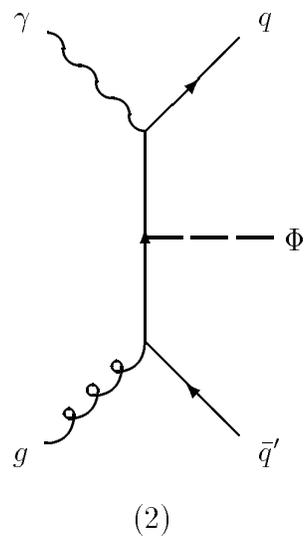
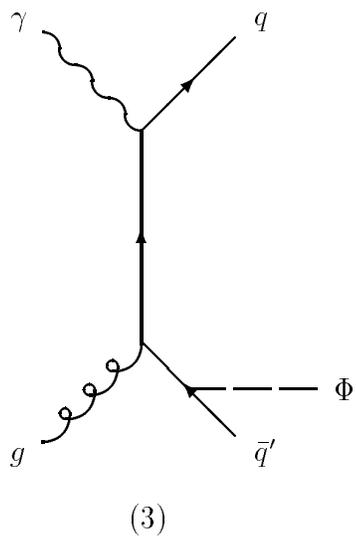
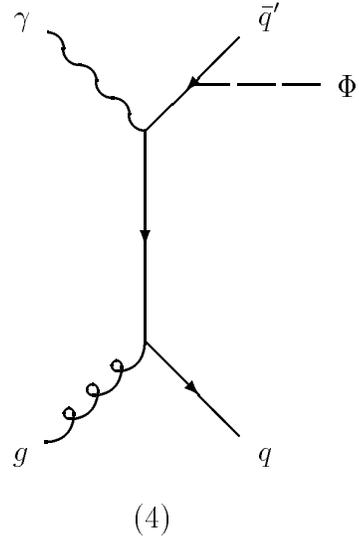
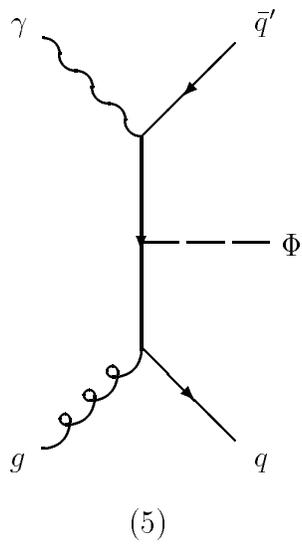
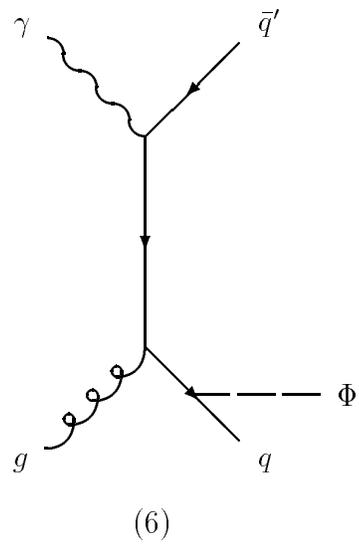
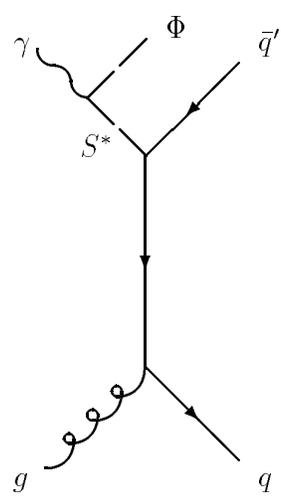
\item[fig.~3 ] Feynman diagrams contributing in the lowest order to
$g\gamma\rightarrow q\bar q'\Phi$, where $q(q')$ represents a quark,
$S^*$ an internal scalar Higgs bosons and $\Phi$
both neutral and charged \mssm\ Higgses, in the unitary gauge.
For the possible combinations of ($q,q',S^*,\Phi$)
and the corresponding non--vanishing graphs, see the text.
\end{description}

\vspace*{\fill}

\vfill
\newpage
\thispagestyle{empty}

\
\vskip4.0cm
\begin{table}
\begin{center}
\begin{tabular}{|c|c|c|c|c|c|}
\hline
\multicolumn{6}{|c|}
{\rule[0cm]{0cm}{0cm}
$\sigma(q\gamma\ar q'W^\pm\Phi^0)$}
\\ \hline
\rule[0cm]{0cm}{0cm}
$M_{H^0}$ &$M_{h^0}$ &$M_{A^0}$ &$H^0$ & $h^0$
&$A^0$   \\ \hline
\rule[0cm]{0cm}{0cm}
$144.4$ & $56.0$ & $60$  & $7.582\pm0.024$ & $37.30\pm0.16$ &
$0.25820\pm0.00090$  \\ \hline
\rule[0cm]{0cm}{0cm}
$150.7$ & $63.7$ & $80$  & $5.767\pm0.019$ & $36.76\pm0.19$ &
$0.18718\pm0.00058$  \\ \hline
\rule[0cm]{0cm}{0cm}
$159.3$ & $70.6$ & $100$ & $3.986\pm0.011$ & $36.80\pm0.17$ &
$0.13096\pm0.00043$  \\ \hline
\rule[0cm]{0cm}{0cm}
$170.1$ & $76.4$ & $120$ & $2.5569\pm0.0069$ & $37.02\pm0.14$ &
$0.09185\pm0.00030$  \\ \hline
\rule[0cm]{0cm}{0cm}
$182.9$ & $80.9$ & $140$ & $1.5431\pm0.0045$ & $37.44\pm0.14$ &
$0.06441\pm0.00020$  \\ \hline
\multicolumn{6}{|c|}
{\rule[0cm]{0cm}{0cm}
$\sqrt s=1.36$ TeV
\quad\quad\quad$
\tan\beta=1.5$
\quad\quad\quad MRS(A)} \\ \hline
\multicolumn{6}{c}
{\rule{0cm}{.9cm}
{\Large Table Ia}}  \\
\multicolumn{6}{c}
{\rule{0cm}{.9cm}}

\end{tabular}
\end{center}
\end{table}


\begin{table}
\begin{center}
\begin{tabular}{|c|c|c|c|c|c|}
\hline
\multicolumn{6}{|c|}
{\rule[0cm]{0cm}{0cm}
$\sigma(q\gamma\ar q'W^\pm\Phi^0)$}
\\ \hline
\rule[0cm]{0cm}{0cm}
$M_{H^0}$ &$M_{h^0}$ &$M_{A^0}$ &$H^0$ & $h^0$
&$A^0$   \\ \hline
\rule[0cm]{0cm}{0cm}
$129.2$ & $59.9$ & $60$  & $24.060\pm0.074$ & $0.8430\pm0.0018$
& $1.5041\pm0.0044$  \\ \hline
\rule[0cm]{0cm}{0cm}
$129.2$ & $79.9$ & $80$  & $23.959\pm0.075$ & $0.6993\pm0.0015$
& $0.9990\pm0.0026$  \\ \hline
\rule[0cm]{0cm}{0cm}
$129.4$ & $99.7$ & $100$ & $23.692\pm0.074$ & $0.8049\pm0.0019$
& $0.6780\pm0.0016$  \\ \hline
\rule[0cm]{0cm}{0cm}
$130.0$ & $119.0$ & $120$ & $21.485\pm0.067$ & $2.9355\pm0.0087$
& $0.4636\pm0.0012$  \\ \hline
\rule[0cm]{0cm}{0cm}
$140.9$ & $128.1$ & $140$ & $1.4487\pm0.0042$ & $22.964\pm0.070$
& $0.31958\pm0.00076$  \\ \hline
\multicolumn{6}{|c|}
{\rule[0cm]{0cm}{0cm}
$\sqrt s=1.36$ TeV
\quad\quad\quad$
\tan\beta=30$
\quad\quad\quad MRS(A)} \\ \hline
\multicolumn{6}{c}
{\rule{0cm}{.9cm}
{\Large Table Ib}}  \\
\multicolumn{6}{c}
{\rule{0cm}{.9cm}}

\end{tabular}
\end{center}
\end{table}

\vfill
\newpage
\thispagestyle{empty}

\
\vskip4.0cm
\begin{table}
\begin{center}
\begin{tabular}{|c|c|c|c|c|c|}
\hline
\multicolumn{6}{|c|}
{\rule[0cm]{0cm}{0cm}
$\sigma(q\gamma\ar q Z^0\Phi^0)$}
\\ \hline
\rule[0cm]{0cm}{0cm}
$M_{H^0}$ &$M_{h^0}$ &$M_{A^0}$ &$H^0$ & $h^0$
&$A^0$   \\ \hline
\rule[0cm]{0cm}{0cm}
$144.4$ & $56.0$ & $60$  & $0.1913\pm0.0031$ & $4.877\pm0.050$
& $(7.962\pm0.023)\times10^{-3}$  \\ \hline
\rule[0cm]{0cm}{0cm}
$150.7$ & $63.7$ & $80$  & $0.1283\pm0.0014$ & $3.941\pm0.045$
& $(5.186\pm0.015)\times10^{-3}$  \\ \hline
\rule[0cm]{0cm}{0cm}
$159.3$ & $70.6$ & $100$ & $0.0803\pm0.0011$ & $3.260\pm0.065$
& $(3.551\pm0.010)\times10^{-3}$  \\ \hline
\rule[0cm]{0cm}{0cm}
$170.1$ & $76.4$ & $120$ & $0.0419\pm0.0014$ & $2.998\pm0.037$
& $(2.4628\pm0.0078)\times10^{-3}$  \\ \hline
\rule[0cm]{0cm}{0cm}
$182.9$ & $80.9$ & $140$ & $0.02421\pm0.00028$ & $2.705\pm0.037$
& $(1.7317\pm0.0051)\times10^{-3}$  \\ \hline
\multicolumn{6}{|c|}
{\rule[0cm]{0cm}{0cm}
$\sqrt s=1.36$ TeV
\quad\quad\quad$
\tan\beta=1.5$
\quad\quad\quad MRS(A)} \\ \hline
\multicolumn{6}{c}
{\rule{0cm}{.9cm}
{\Large Table IIa}}  \\
\multicolumn{6}{c}
{\rule{0cm}{.9cm}}

\end{tabular}
\end{center}
\end{table}


\begin{table}
\begin{center}
\begin{tabular}{|c|c|c|c|c|c|}
\hline
\multicolumn{6}{|c|}
{\rule[0cm]{0cm}{0cm}
$\sigma(q\gamma\ar q Z^0\Phi^0)$}
\\ \hline
\rule[0cm]{0cm}{0cm}
$M_{H^0}$ &$M_{h^0}$ &$M_{A^0}$ &$H^0$ & $h^0$
&$A^0$   \\ \hline
\rule[0cm]{0cm}{0cm}
$129.2$ & $59.9$ & $60$  & $0.7443\pm0.0085$ & $1.3557\pm0.0047$ &
$2.2500\pm0.0080$  \\ \hline
\rule[0cm]{0cm}{0cm}
$129.2$ & $79.9$ & $80$  & $0.7406\pm0.0073$ & $0.9509\pm0.0032$ &
$1.5374\pm0.0052$  \\ \hline
\rule[0cm]{0cm}{0cm}
$129.4$ & $99.7$ & $100$ & $0.7474\pm0.0078$ & $0.6819\pm0.0023$ &
$1.0753\pm0.0040$  \\ \hline
\rule[0cm]{0cm}{0cm}
$130.0$ & $119.0$ & $120$ & $0.677\pm0.015$ & $0.5162\pm0.0022$ &
$0.7622\pm0.0031$  \\ \hline
\rule[0cm]{0cm}{0cm}
$140.9$ & $128.1$ & $140$ & $0.3412\pm0.0011$ & $0.770\pm0.017$ &
$0.5376\pm0.0018$  \\ \hline
\multicolumn{6}{|c|}
{\rule[0cm]{0cm}{0cm}
$\sqrt s=1.36$ TeV
\quad\quad\quad$
\tan\beta=30$
\quad\quad\quad MRS(A)} \\ \hline
\multicolumn{6}{c}
{\rule{0cm}{.9cm}
{\Large Table IIb}}  \\
\multicolumn{6}{c}
{\rule{0cm}{.9cm}}

\end{tabular}
\end{center}
\end{table}

\vfill
\newpage
\thispagestyle{empty}

\
\vskip4.0cm
\begin{table}
\begin{center}
\begin{tabular}{|c|c|c|c|c|c|c|}
\hline
\multicolumn{7}{|c|}
{\rule[0cm]{0cm}{0cm}
$\sigma(q\gamma\ar q'H^\pm\Phi^0)$}
\\ \hline
\rule[0cm]{0cm}{0cm}
$M_{H^0}$ &$M_{h^0}$ &$M_{A^0}$ &$M_{H^\pm}$
&$H^0$ & $h^0$ &$A^0$   \\ \hline
\rule[0cm]{0cm}{0cm}
$144.4$ & $56.0$ & $60$  & $100.0$ & $0.3621\pm0.0019$
& $1.1599\pm0.0085$ & $2.834\pm0.021$  \\ \hline
\rule[0cm]{0cm}{0cm}
$150.7$ & $63.7$ & $80$  & $113.1$ & $0.2683\pm0.0017$
& $0.5857\pm0.0037$ & $1.3684\pm0.0072$  \\ \hline
\rule[0cm]{0cm}{0cm}
$159.3$ & $70.6$ & $100$ & $128.1$ & $0.1944\pm0.0011$
& $0.2866\pm0.0021$ & $0.6745\pm0.0037$  \\ \hline
\rule[0cm]{0cm}{0cm}
$170.1$ & $76.4$ & $120$ & $144.2$ & $0.1334\pm0.00081$
& $0.1352\pm0.0013$ & $0.3518\pm0.0021$  \\ \hline
\rule[0cm]{0cm}{0cm}
$182.9$ & $80.9$ & $140$ & $161.2$ & $0.08182\pm0.00059$
& $0.06572\pm0.00031$ & $0.1858\pm0.0011$  \\ \hline
\multicolumn{7}{|c|}
{\rule[0cm]{0cm}{0cm}
$\sqrt s=1.36$ TeV
\quad\quad\quad$
\tan\beta=1.5$
\quad\quad\quad MRS(A)} \\ \hline
\multicolumn{7}{c}
{\rule{0cm}{.9cm}
{\Large Table IIIa}}  \\
\multicolumn{7}{c}
{\rule{0cm}{.9cm}}

\end{tabular}
\end{center}
\end{table}


\begin{table}
\begin{center}
\begin{tabular}{|c|c|c|c|c|c|c|}
\hline
\multicolumn{7}{|c|}
{\rule[0cm]{0cm}{0cm}
$\sigma(q\gamma\ar q'H^\pm\Phi^0)$}
\\ \hline
\rule[0cm]{0cm}{0cm}
$M_{H^0}$ &$M_{h^0}$ &$M_{A^0}$ &$M_{H^\pm}$
&$H^0$ & $h^0$ &$A^0$   \\ \hline
\rule[0cm]{0cm}{0cm}
$129.2$ & $59.9$ & $60$  & $100.0$ & $(6.833\pm0.019)\times10^{-3}$
& $2.8210\pm0.019$ & $2.833\pm0.020$  \\ \hline
\rule[0cm]{0cm}{0cm}
$129.2$ & $79.9$ & $80$  & $113.1$ & $(6.527\pm0.029)\times10^{-3}$
& $1.3660\pm0.0081$ & $1.3697\pm0.0072$  \\ \hline
\rule[0cm]{0cm}{0cm}
$129.4$ & $99.7$ & $100$ & $128.1$ & $(8.344\pm0.044)\times10^{-3}$
& $0.6732\pm0.0056$ & $0.6758\pm0.0037$  \\ \hline
\rule[0cm]{0cm}{0cm}
$130.0$ & $119.0$ & $120$ & $144.2$ & $(31.52\pm0.24)\times10^{-3}$
& $0.3224\pm0.0019$ & $0.3523\pm0.0021$  \\ \hline
\rule[0cm]{0cm}{0cm}
$140.9$ & $128.1$ & $140$ & $161.2$ & $(171.3\pm1.0)\times10^{-3}$
& $0.015369\pm0.000073$ & $0.1860\pm0.0011$  \\ \hline
\multicolumn{7}{|c|}
{\rule[0cm]{0cm}{0cm}
$\sqrt s=1.36$ TeV
\quad\quad\quad$
\tan\beta=30$
\quad\quad\quad MRS(A)} \\ \hline
\multicolumn{7}{c}
{\rule{0cm}{.9cm}
{\Large Table IIIb}}  \\
\multicolumn{7}{c}
{\rule{0cm}{.9cm}}

\end{tabular}
\end{center}
\end{table}

\vfill
\newpage
\thispagestyle{empty}

\
\vskip4.0cm
\begin{table}
\begin{center}
\begin{tabular}{|c|c|c|c|c|}
\hline
\multicolumn{5}{|c|}
{\rule[0cm]{0cm}{0cm}
$\sigma(q\gamma\ar q \Phi^0\Phi^{0{'}})$}
\\ \hline
\rule[0cm]{0cm}{0cm}
$M_{H^0}$ &$M_{h^0}$ &$M_{A^0}$ &$H^0A^0$
&$h^0A^0$   \\ \hline
\rule[0cm]{0cm}{0cm}
$144.4$ & $56.0$ & $60$  & $0.1468\pm0.0023$ & $1.0889\pm0.0092$   \\ \hline
\rule[0cm]{0cm}{0cm}
$150.7$ & $63.7$ & $80$  & $0.1004\pm0.0014$ & $0.3226\pm0.0041$   \\ \hline
\rule[0cm]{0cm}{0cm}
$159.3$ & $70.6$ & $100$ & $0.0715\pm0.0013$ & $0.1183\pm0.0018$   \\ \hline
\rule[0cm]{0cm}{0cm}
$170.1$ & $76.4$ & $120$ & $0.0409\pm0.00056$ & $0.0437\pm0.00075$   \\ \hline
\rule[0cm]{0cm}{0cm}
$182.9$ & $80.9$ & $140$ & $0.02656\pm0.00049$ & $0.0203\pm0.00025$   \\ \hline
\multicolumn{5}{|c|}
{\rule[0cm]{0cm}{0cm}
$\sqrt s=1.36$ TeV
\quad\quad\quad$
\tan\beta=1.5$
\quad\quad\quad MRS(A)} \\ \hline
\multicolumn{5}{c}
{\rule{0cm}{.9cm}
{\Large Table IVa}}  \\
\multicolumn{5}{c}
{\rule{0cm}{.9cm}}

\end{tabular}
\end{center}
\end{table}


\begin{table}
\begin{center}
\begin{tabular}{|c|c|c|c|c|}
\hline
\multicolumn{5}{|c|}
{\rule[0cm]{0cm}{0cm}
$\sigma(q\gamma\ar q \Phi^0\Phi^{0{'}})$}
\\ \hline
\rule[0cm]{0cm}{0cm}
$M_{H^0}$ &$M_{h^0}$ &$M_{A^0}$ &$H^0A^0$
&$h^0A^0$   \\ \hline
\rule[0cm]{0cm}{0cm}
$129.2$ & $59.9$ & $60$  & $(6.072\pm0.024)\times10^{-3}$
& $4.002\pm0.039$   \\ \hline
\rule[0cm]{0cm}{0cm}
$129.2$ & $79.9$ & $80$  & $(4.279\pm0.026)\times10^{-3}$
& $1.1668\pm0.0098$   \\ \hline
\rule[0cm]{0cm}{0cm}
$129.4$ & $99.7$ & $100$ & $(4.489\pm0.057)\times10^{-3}$
& $0.4270\pm0.0038$   \\ \hline
\rule[0cm]{0cm}{0cm}
$130.0$ & $119.0$ & $120$ & $(16.05\pm0.18)\times10^{-3}$
& $0.1746\pm0.0021$   \\ \hline
\rule[0cm]{0cm}{0cm}
$140.9$ & $128.1$ & $140$ & $(79.9\pm1.0)\times10^{-3}$
& $0.00802\pm0.00013$   \\ \hline
\multicolumn{5}{|c|}
{\rule[0cm]{0cm}{0cm}
$\sqrt s=1.36$ TeV
\quad\quad\quad$
\tan\beta=30$
\quad\quad\quad MRS(A)} \\ \hline
\multicolumn{5}{c}
{\rule{0cm}{.9cm}
{\Large Table IVb}}  \\
\multicolumn{5}{c}
{\rule{0cm}{.9cm}}

\end{tabular}
\end{center}
\end{table}

\vfill
\newpage
\thispagestyle{empty}

\
\vskip4.0cm
\begin{table}
\begin{center}
\begin{tabular}{|c|c|}
\hline
\multicolumn{2}{|c|}
{\rule[0cm]{0cm}{0cm}
$\sigma(q\gamma\ar qH^+ H^-)$}
\\ \hline
\rule[0cm]{0cm}{0cm}
$M_{H^\pm}$ &$H^+H^-$   \\ \hline
\rule[0cm]{0cm}{0cm}
$100.0$ & $18.18\pm0.42$   \\ \hline
\rule[0cm]{0cm}{0cm}
$113.1$ & $10.96\pm0.16$   \\ \hline
\rule[0cm]{0cm}{0cm}
$128.1$ & $6.06\pm0.16$   \\ \hline
\rule[0cm]{0cm}{0cm}
$144.2$ & $2.991\pm0.064$   \\ \hline
\rule[0cm]{0cm}{0cm}
$161.2$ & $1.577\pm0.034$   \\ \hline
\multicolumn{2}{|c|}
{\rule[0cm]{0cm}{0cm}
$\sqrt s=1.36$ TeV
\quad$
\tan\beta=1.5$
\quad MRS(A)} \\ \hline
\multicolumn{2}{c}
{\rule{0cm}{.9cm}
{\Large Table Va}}  \\
\multicolumn{2}{c}
{\rule{0cm}{.9cm}}

\end{tabular}
\end{center}
\end{table}


\begin{table}
\begin{center}
\begin{tabular}{|c|c|}
\hline
\multicolumn{2}{|c|}
{\rule[0cm]{0cm}{0cm}
$\sigma(q\gamma\ar qH^+ H^-)$}
\\ \hline
\rule[0cm]{0cm}{0cm}
$M_{H^\pm}$ &$H^+H^-$   \\ \hline
\rule[0cm]{0cm}{0cm}
$100.0$ & $28.13\pm0.40$   \\ \hline
\rule[0cm]{0cm}{0cm}
$113.1$ & $16.52\pm0.21$   \\ \hline
\rule[0cm]{0cm}{0cm}
$128.1$ & $9.52\pm0.14$   \\ \hline
\rule[0cm]{0cm}{0cm}
$144.2$ & $4.244\pm0.056$   \\ \hline
\rule[0cm]{0cm}{0cm}
$161.2$ & $1.867\pm0.071$   \\ \hline
\multicolumn{2}{|c|}
{\rule[0cm]{0cm}{0cm}
$\sqrt s=1.36$ TeV
\quad$
\tan\beta=30$
\quad MRS(A)} \\ \hline
\multicolumn{2}{c}
{\rule{0cm}{.9cm}
{\Large Table Vb}}  \\
\multicolumn{2}{c}
{\rule{0cm}{.9cm}}

\end{tabular}
\end{center}
\end{table}

\vfill
\newpage
\thispagestyle{empty}

\
\vskip4.0cm
\begin{table}
\begin{center}
\begin{tabular}{|c|c|}
\hline
\multicolumn{2}{|c|}
{\rule[0cm]{0cm}{0cm}
$\sigma(g\gamma\ar q\bar q'H^\pm)$}
\\ \hline
\rule[0cm]{0cm}{0cm}
$M_{H^\pm}$ &$H^\pm$   \\ \hline
\rule[0cm]{0cm}{0cm}
$100.0$ & $367.3\pm2.7$   \\ \hline
\rule[0cm]{0cm}{0cm}
$113.1$ & $270.5\pm3.8$   \\ \hline
\rule[0cm]{0cm}{0cm}
$128.1$ & $174.3\pm1.0$   \\ \hline
\rule[0cm]{0cm}{0cm}
$144.2$ & $84.78\pm0.48$   \\ \hline
\rule[0cm]{0cm}{0cm}
$161.2$ & $22.78\pm0.15$   \\ \hline
\multicolumn{2}{|c|}
{\rule[0cm]{0cm}{0cm}
$\sqrt s=1.36$ TeV
\quad$
\tan\beta=1.5$
\quad MRS(A)} \\ \hline
\multicolumn{2}{c}
{\rule{0cm}{.9cm}
{\Large Table VIa}}  \\
\multicolumn{2}{c}
{\rule{0cm}{.9cm}}

\end{tabular}
\end{center}
\end{table}


\begin{table}
\begin{center}
\begin{tabular}{|c|c|}
\hline
\multicolumn{2}{|c|}
{\rule[0cm]{0cm}{0cm}
$\sigma(g\gamma\ar q\bar q'H^\pm)$}
\\ \hline
\rule[0cm]{0cm}{0cm}
$M_{H^\pm}$ &$H^\pm$   \\ \hline
\rule[0cm]{0cm}{0cm}
$100.0$ & $621.7\pm4.8$   \\ \hline
\rule[0cm]{0cm}{0cm}
$113.1$ & $460.6\pm6.0$   \\ \hline
\rule[0cm]{0cm}{0cm}
$128.1$ & $291.9\pm1.7$   \\ \hline
\rule[0cm]{0cm}{0cm}
$144.2$ & $142.2\pm1.7$   \\ \hline
\rule[0cm]{0cm}{0cm}
$161.2$ & $40.65\pm0.31$   \\ \hline
\multicolumn{2}{|c|}
{\rule[0cm]{0cm}{0cm}
$\sqrt s=1.36$ TeV
\quad$
\tan\beta=30$
\quad MRS(A)} \\ \hline
\multicolumn{2}{c}
{\rule{0cm}{.9cm}
{\Large Table VIb}}  \\
\multicolumn{2}{c}
{\rule{0cm}{.9cm}}

\end{tabular}
\end{center}
\end{table}

\vfill
\newpage
\thispagestyle{empty}

\
\vskip4.0cm
\begin{table}
\begin{center}
\begin{tabular}{|c|c|c|c|c|c|}
\hline
\multicolumn{6}{|c|}
{\rule[0cm]{0cm}{0cm}
$\sigma(g\gamma\ar q\bar q\Phi^0)$}
\\ \hline
\rule[0cm]{0cm}{0cm}
$M_{H^0}$ &$M_{h^0}$ &$M_{A^0}$ &$H^0$ & $h^0$
&$A^0$   \\ \hline
\rule[0cm]{0cm}{0cm}
$144.4$ & $56.0$ & $60$  & $0.1914\pm0.0047$ & $2.1015\pm0.0059$
& $1.4169\pm0.0041$  \\ \hline
\rule[0cm]{0cm}{0cm}
$150.7$ & $63.7$ & $80$  & $0.1541\pm0.0020$ & $1.6574\pm0.0051$
& $0.6810\pm0.0022$  \\ \hline
\rule[0cm]{0cm}{0cm}
$159.3$ & $70.6$ & $100$ & $0.1174\pm0.0029$ & $1.3948\pm0.0036$
& $0.3640\pm0.0019$  \\ \hline
\rule[0cm]{0cm}{0cm}
$170.1$ & $76.4$ & $120$ & $0.0844\pm0.0025$ & $1.2121\pm0.0030$
& $0.2081\pm0.0010$  \\ \hline
\rule[0cm]{0cm}{0cm}
$182.9$ & $80.9$ & $140$ & $0.0588\pm0.0014$ & $1.1043\pm0.0025$
& $0.1253\pm0.00092$  \\ \hline
\multicolumn{6}{|c|}
{\rule[0cm]{0cm}{0cm}
$\sqrt s=1.36$ TeV
\quad\quad\quad$
\tan\beta=1.5$
\quad\quad\quad MRS(A)} \\ \hline
\multicolumn{6}{c}
{\rule{0cm}{.9cm}
{\Large Table VIIa}}  \\
\multicolumn{6}{c}
{\rule{0cm}{.9cm}}

\end{tabular}
\end{center}
\end{table}


\begin{table}
\begin{center}
\begin{tabular}{|c|c|c|c|c|c|}
\hline
\multicolumn{6}{|c|}
{\rule[0cm]{0cm}{0cm}
$\sigma(g\gamma\ar q\bar q\Phi^0)$}
\\ \hline
\rule[0cm]{0cm}{0cm}
$M_{H^0}$ &$M_{h^0}$ &$M_{A^0}$ &$H^0$ & $h^0$
&$A^0$   \\ \hline
\rule[0cm]{0cm}{0cm}
$129.2$ & $59.9$ & $60$  & $0.2743\pm0.0010$ & $428.3\pm1.7$
& $449.9\pm1.6$  \\ \hline
\rule[0cm]{0cm}{0cm}
$129.2$ & $79.9$ & $80$  & $0.3432\pm0.0011$ & $209.13\pm0.83$
& $218.87\pm0.85$  \\ \hline
\rule[0cm]{0cm}{0cm}
$129.4$ & $99.7$ & $100$ & $0.6488\pm0.0020$ & $115.30\pm0.55$
& $117.97\pm0.53$  \\ \hline
\rule[0cm]{0cm}{0cm}
$130.0$ & $119.0$ & $120$ & $8.649\pm0.027$ & $62.86\pm0.28$
& $67.66\pm0.31$  \\ \hline
\rule[0cm]{0cm}{0cm}
$140.9$ & $128.1$ & $140$ & $36.38\pm0.17$ & $4.514\pm0.020$
& $40.76\pm0.19$  \\ \hline
\multicolumn{6}{|c|}
{\rule[0cm]{0cm}{0cm}
$\sqrt s=1.36$ TeV
\quad\quad\quad$
\tan\beta=30$
\quad\quad\quad MRS(A)} \\ \hline
\multicolumn{6}{c}
{\rule{0cm}{.9cm}
{\Large Table VIIb}}  \\
\multicolumn{6}{c}
{\rule{0cm}{.9cm}}

\end{tabular}
\end{center}
\end{table}

\vfill
\newpage
\thispagestyle{empty}

\
\vskip4.0cm
\begin{table}
\begin{center}
\begin{tabular}{|c|c|}
\hline
\rule[0cm]{0cm}{0cm}
{Background} &  $\sigma$  \\ \hline
\rule[0cm]{0cm}{0cm}
$ep\ar W^\pm Z^0 X$ & $219.8\pm3.2$  \\ \hline
\rule[0cm]{0cm}{0cm}
$ep \ar Z^0 Z^0 X$  & $10.98\pm0.60$  \\ \hline
\rule[0cm]{0cm}{0cm}
$ep\ar q\bar q Z^0X$ & $3139\pm49$ \\ \hline
\rule[0cm]{0cm}{0cm}
$ep\ar W^+W^-X$ & $1805\pm55$ \\ \hline
\rule[0cm]{0cm}{0cm}
$ep\ar q\bar q' W^\pm X$ & $17114\pm150$ \\ \hline
\multicolumn{2}{|c|}
{\rule[0cm]{0cm}{0cm}
$\sqrt s=1.36$ TeV\quad\quad MRS(A)}
 \\ \hline
\multicolumn{2}{c}
{\rule{0cm}{.9cm}
{\Large Table VIIIa}}  \\
\multicolumn{2}{c}
{\rule{0cm}{.9cm}}

\end{tabular}
\end{center}
\end{table}


\begin{table}
\begin{center}
\begin{tabular}{|c|c|c|c|}
\hline
\rule[0cm]{0cm}{0cm}
$W^+W^-X$ ($t$--res.) & $tbW^\pm X$
& $tbX\ar b\bar bW^+W^- X$
& $t\bar tX\ar b\bar bW^+W^- X$ \\ \hline
\rule[0cm]{0cm}{0cm}
$809\pm20(707\pm11)$ & $1590.0\pm6.6(1406.9\pm8.5)$
& $291\pm16(262\pm18)$ & $532.6\pm2.0(423.4\pm1.0)$   \\ \hline
\rule[0cm]{0cm}{0cm}
$758\pm26(704\pm11)$ & $1586.7\pm7.5(1401.8\pm6.2)$
& $305\pm22(280\pm20)$ & $587.2\pm1.0(489.5\pm1.1)$   \\ \hline
\rule[0cm]{0cm}{0cm}
$783\pm23(714\pm11)$ & $1593.6\pm6.8(1397.3\pm6.3)$
& $323\pm20(306\pm21)$ & $656.2\pm1.1(579.2\pm1.1)$   \\ \hline
\rule[0cm]{0cm}{0cm}
$783\pm22(705\pm10)$ & $1576.2\pm7.8(1401.3\pm6.1)$
& $341\pm34(331\pm22)$ & $730.2\pm1.4(685.6\pm1.1)$   \\ \hline
\rule[0cm]{0cm}{0cm}
$789\pm29(708\pm11)$ & $1569.3\pm7.7(1389.8\pm6.4)$
& $356\pm36(352\pm26)$ & $791.2\pm1.0(780.5\pm1.2)$   \\ \hline
\multicolumn{4}{|c|}
{\rule[0cm]{0cm}{0cm}
$\sqrt s=1.36$ TeV
\quad\quad\quad\quad $\tba(30)$
\quad\quad\quad\quad MRS(A)} \\ \hline
\multicolumn{4}{c}
{\rule{0cm}{.9cm}
{\Large Table VIIIb}}  \\
\multicolumn{4}{c}
{\rule{0cm}{.9cm}}

\end{tabular}
\end{center}
\end{table}


\begin{table}
\begin{center}
\begin{tabular}{|c|c|c|c|}
\hline
\rule[0cm]{0cm}{0cm}
$M_{H^\pm}$ & BR$(t\ar bW^\pm)$ & BR$(t\ar bH^\pm)$
& $\Gamma_t^{\mssm}$\\ \hline
\rule[0cm]{0cm}{0cm}
$100.0$ & $0.81(0.73)$ & $0.19(0.27)$ &  $1.94(2.17)$\\ \hline
\rule[0cm]{0cm}{0cm}
$113.1$ & $0.85(0.78)$ & $0.15(0.22)$ &  $1.84(2.02)$\\ \hline
\rule[0cm]{0cm}{0cm}
$128.1$ & $0.90(0.85)$ & $0.10(0.15)$ &  $1.75(1.86)$\\ \hline
\rule[0cm]{0cm}{0cm}
$144.2$ & $0.95(0.92)$ & $0.05(0.08)$ &  $1.66(1.71)$\\ \hline
\rule[0cm]{0cm}{0cm}
$161.2$ & $0.99(0.98)$ & $0.01(0.02)$ &  $1.59(1.60)$ \\ \hline
\multicolumn{4}{|c|}
{\rule[0cm]{0cm}{0cm}
$M_{W^\pm}\approx80$ GeV
\quad$
\tan\beta=1.5(30)$
\quad
$m_t=175$ GeV} \\ \hline
\multicolumn{4}{c}
{\rule{0cm}{.9cm}
{\Large Table IX}}  \\
\multicolumn{4}{c}
{\rule{0cm}{.9cm}}

\end{tabular}
\end{center}
\end{table}

\vfill
\newpage
\thispagestyle{empty}
\begin{table}
\begin{center}
\begin{tabular}{|c|c|c|c|}     \hline
\rule[0cm]{0cm}{0cm}
$\;\;\;\;\;$                     &
$H^0$                            &
$h^0$                            &
$A^0$                            \\ \hline
\rule[0cm]{0cm}{0cm}
$W^\pm W^\mp$                                          &
${M_{W^\pm}c_{\beta\alpha}\over{s_W}}$                &
${M_{W^\pm}s_{\beta\alpha}\over{s_W}}$                &
$0$                                                    \\ 
\rule[0cm]{0cm}{0cm}
$H^\pm H^\mp$                                                           &
$\frac{M_{W^\pm}}{s_W}(c_{\beta\alpha}-
{1\over {2c^2_W}}c_{2\beta}c_{\alpha\beta})$ &
$\frac{M_{W^\pm}}{s_W}(s_{\beta\alpha}+
{1\over {2c^2_W}}c_{2\beta}s_{\alpha\beta})$ &
$0$                                            \\ 
\rule[0cm]{0cm}{0cm}
$W^{\pm} H^\mp(\gamma)$                       &
${s_{\beta\alpha}\over{2s_W}}$                &
$-{c_{\beta\alpha}\over{2s_W}}$               &
$-\frac{\rm{i}}{2s_W}$                          \\ 
\rule[0cm]{0cm}{0cm}
$Z^0Z^0$                                               &
${M_{W^\pm}\over{s_Wc^2_W}}c_{\beta\alpha}$                &
${M_{W^\pm}\over{s_Wc^2_W}}s_{\beta\alpha}$                &
$0$                                                    \\ 
\rule[0cm]{0cm}{0cm}
$Z^0A^0$                                               &
$\frac{\rm{i}}{2s_Wc_W}s_{\beta\alpha}$                 &
$-\frac{\rm{i}}{2s_Wc_W}c_{\beta\alpha}$                &
$0$                                                    \\ 
\rule[0cm]{0cm}{0cm}
$H^0H^0$
&
${3M_{W^\pm}\over {2s_Wc^2_W}}c_{2\alpha}c_{\alpha\beta}$                &
$-{M_{W^\pm}\over {2s_Wc^2_W}}(2s_{2\alpha}c_{\alpha\beta}+
c_{2\alpha}s_{\alpha\beta})$                &
$0$                                                      \\ 
\rule[0cm]{0cm}{0cm}
$H^0h^0$                                                &
$-{M_{W^\pm}\over {2s_Wc^2_W}}(2s_{2\alpha}c_{\alpha\beta}
+c_{2\alpha}s_{\alpha\beta})$                &
${M_{W^\pm}\over {2s_Wc^2_W}}(2s_{2\alpha}s_{\alpha\beta}
-c_{2\alpha}c_{\alpha\beta})$                &
$0$                                        \\ 
\rule[0cm]{0cm}{0cm}
$H^0A^0$                                              &
$0$                                                   &
$0$                                                   &
$-{M_{W^\pm}\over {2s_Wc^2_W}}c_{2\beta}c_{\alpha\beta}$    \\ 
\rule[0cm]{0cm}{0cm}
$h^0h^0$                                              &
${M_{W^\pm}\over {2s_Wc^2_W}}(2s_{2\alpha}s_{\alpha\beta}-c_{2\alpha}
c_{\alpha\beta})$                &
${3M_{W^\pm}\over {2s_Wc^2_W}}c_{2\alpha}s_{\alpha\beta}$   &
$0$                                                   \\ 
\rule[0cm]{0cm}{0cm}
$h^0A^0$                                              &
$0$                                                   &
$0$                                                   &
${M_{W^\pm}\over {2s_Wc^2_W}}c_{2\beta}s_{\alpha\beta}$     \\ 
\rule[0cm]{0cm}{0cm}
$A^0A^0$                                              &
$-{M_{W^\pm}\over {2s_Wc^2_W}}c_{2\beta}c_{\alpha\beta}$    &
${M_{W^\pm}\over {2s_Wc^2_W}}c_{2\beta}s_{\alpha\beta}$     &
$0$                                                   \\ \hline
\end{tabular}
\end{center}
\centerline{\Large Table A.I}
\end{table}
\vfill
\newpage
\thispagestyle{empty}
\begin{table}
\begin{center}
\begin{tabular}{|c|c|}     \hline
\rule[0cm]{0cm}{0cm}
$\;\;\;\;\;$                     &
$H^\pm H^\mp$                    \\ \hline
\rule[0cm]{0cm}{0cm}
$\gamma$                             &
$1$                                  \\ 
\rule[0cm]{0cm}{0cm}
$Z^0$                                &
$\frac{1}{2s_Wc_W}c_{2W}$            \\ 
\rule[0cm]{0cm}{0cm}
$\gamma\gamma$                       &
$-2$                                 \\ 
\rule[0cm]{0cm}{0cm}
$\gamma Z^0$                             &
$-\frac{c_{2W}}{s_Wc_W}$             \\ \hline
\end{tabular}
\end{center}
\centerline{\Large Table A.II}
\end{table}
\vfill
\newpage
\pagestyle{empty}
\begin{table}
\begin{center}
\begin{tabular}{|c|c|c|c|c|c|}     \hline
\rule[0cm]{0cm}{0cm}
$g$          &
$\gamma$     &
$Z^0$        &
$H^0$        &
$h^0$        &
$A^0$        \\ \hline
\rule[0cm]{0cm}{0cm}
$(1,1)$                                                  &
$Q^u(1,1)$                                               &
${1\over{s_Wc_W}}(g_R^u,g_L^u)$                          &
$\frac{m_u}{2M_{W^\pm}s_W}\frac{s_\alpha}{s_\beta}(1,1)$ &
$\frac{m_u}{2M_{W^\pm}s_W}\frac{c_\alpha}{s_\beta}(1,1)$ &
$-{\rm{i}}\frac{m_u}{2M_{W^\pm}s_W}{1\over t_\beta}(1,-1)$         \\
\rule[0cm]{0cm}{0cm}
$(1,1)$                                                   &
$Q^d(1,1)$                                                &
${1\over{s_Wc_W}}(g_R^d,g_L^d)$                           &
$\frac{m_d}{2M_{W^\pm}s_W}\frac{c_\alpha}{c_\beta}(1,1)$  &
$-\frac{m_d}{2M_{W^\pm}s_W}\frac{s_\alpha}{c_\beta}(1,1)$ &
$-{\rm{i}}\frac{m_d}{2M_{W^\pm}s_W}t_\beta(1,-1)$ \\ \hline
\end{tabular}
\end{center}
\centerline{\Large Table A.III}
\end{table}
\vfill
\newpage
\pagestyle{empty}
\begin{table}
\begin{center}
\begin{tabular}{|c|c|}     \hline
\rule[0cm]{0cm}{0cm}
$W^\pm$      &
$H^\pm$      \\ \hline
\rule[0cm]{0cm}{0cm}
${1\over{\sqrt2 s_W}}(0,1)$                             &
$-{1\over{2\sqrt2M_{W^\pm}s_W}}(m_dt_\beta,m_u/t_\beta)$ \\ \hline
\end{tabular}
\end{center}
\centerline{\Large Table A.IV}
\end{table}
\vfill
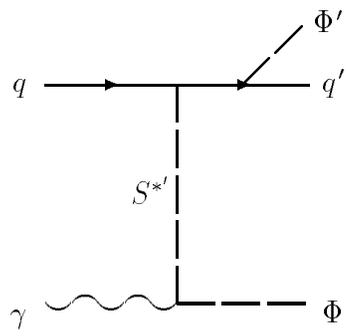
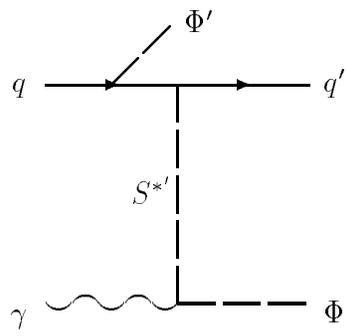
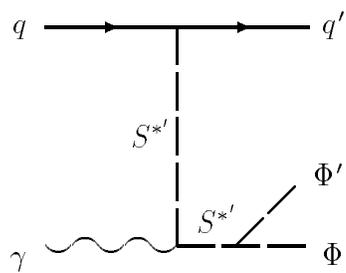
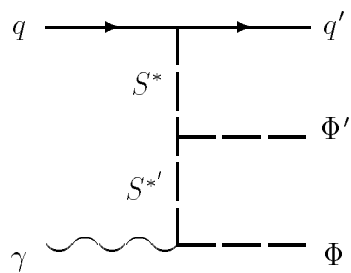
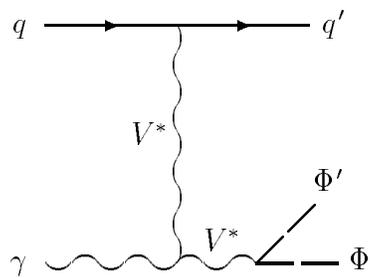
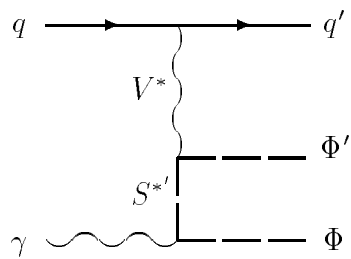
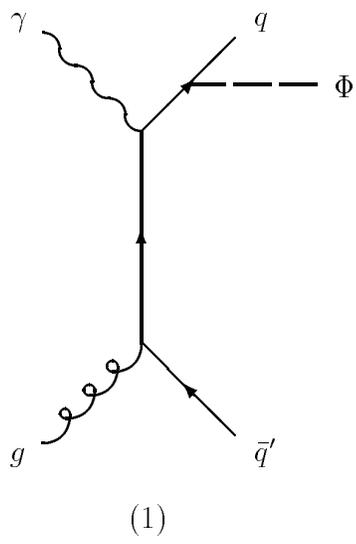
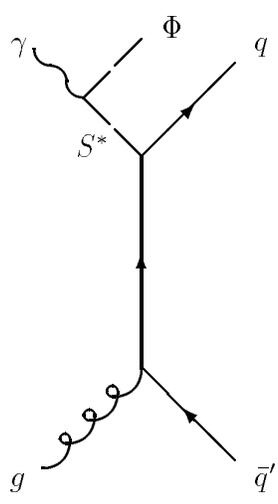
\end{document}